\documentclass[a4paper,11pt]{article}
\pdfoutput=1 

\usepackage{jheppub} 

\usepackage[T1]{fontenc} 
\usepackage{amsmath}
\usepackage{xcolor}
\usepackage{mathtools}
\usepackage{subcaption}
\usepackage{multirow}
\newcommand{\ga}{\alpha}
\newcommand{\gb}{\beta}

\newcommand{\gl}{\lambda}
\newcommand{\gs}{\sigma}
\newcommand{\go}{\omega}

\newcommand{\gD}{\Delta}

\newcommand{\gL}{\Lambda}

\newcommand{\Dtilde}{\tilde{D}}

\newcommand{\cG}{\mathcal G}

\newcommand{\cN}{\mathcal N}
\newcommand{\cO}{\mathcal O}

\newcommand{\cS}{\mathcal S}

\newcommand{\cV}{\mathcal V}

\newcommand{\bC}{\mathbb C}

\newcommand{\bZ}{\mathbb Z}

\renewcommand{\Im}{\mbox{Im}}

\renewcommand{\Re}{\mbox{Re}}
\newcommand{\Res}{\mbox{Res}}

\newcommand{\be}{\begin{equation}}
\newcommand{\bea}{\begin{eqnarray}}

\newcommand{\ee}{\end{equation}}
\newcommand{\eea}{\end{eqnarray}}
\newcommand{\half}{\frac{1}{2}}

\newcommand{\ret}{\nonumber \\}
\newcommand{\sk}{\vspace{1 em} \noindent}

\newcommand{\mS}{\mathfrak S}

\newtheorem{prop}{Proposition}[section]

\newcommand{\rnc}[1]
    {\text{\MakeUppercase{\romannumeral #1}}}
\newcommand*\diff{\mathop{}\!\mathrm{d}} 
\title{\boldmath Painlevé I and exact WKB: Stokes phenomenon for two-parameter transseries}

\author{Alexander van Spaendonck}
\author{and Marcel Vonk}

\affiliation{Institute for Theoretical Physics, University of Amsterdam,
Science Park 904, 1090 GL Amsterdam, The Netherlands}

\emailAdd{a.b.n.vanspaendonck@uva.nl}
\emailAdd{m.l.vonk@uva.nl}

\abstract{
For more than a century, the Painlevé I equation has played an important role in both physics and mathematics. Its two-parameter family of solutions was studied in many different ways, yet still leads to new surprises and discoveries. Two popular tools in these studies are the theory of isomonodromic deformation that uses the exact WKB method, and the asymptotic description of transcendents in terms of two-parameter transseries. Combining methods from both schools of thought, and following work by Takei and collaborators, we formulate complete, two-parameter connection formulae for solutions when they cross arbitrary Stokes lines in the complex plane. These formulae allow us to study Stokes phenomenon for the full two-parameter family of transseries solutions. In particular, we recover the exact expressions for the Stokes data that were recently found by Baldino, Schwick, Schiappa and Vega and compare our connection formulae to theirs. We also explain several ambiguities in relating transseries parameter choices to actual Painlevé transcendents, study the monodromy of formal solutions, and provide high-precision numerical tests of our results.
}

\begin{document} 
\maketitle
\flushbottom

\section{Introduction}
Ever since their discovery in the early 20th century \cite{PAINLEVE}, the six Painlevé equations have played an important role in many branches of physics and mathematics. These second order ordinary differential equations, whose special properties we shall describe in more detail in section \ref{sec:transP1}, pop up in areas from fluid mechanics to quantum gravity and from random matrix theory to conformal field theory.

\sk
Even the simplest of the six equations, the {\em Painlevé I equation}
\be
 \label{eq:PIintro}
 u(z)^2-\frac{1}{6}\frac{d^2 u(z)}{dz^2} =z
\ee
has been an intriguing object of study for more than a century now, and several questions about the equation and its solutions $u(z)$, the {\em Painlevé I transcendents}, are still partially or entirely unanswered. In this paper, one of these partially answered questions will be our main subject. It concerns the relation between formal solutions to the first Painlevé equation and its functional solutions, the transcendents. 

\subsection{From formal solutions to transcendents}
The simplest formal solution to (\ref{eq:PIintro}) is a power series in the complex variable $z$, or more precisely, in $z^{-5/2}$. Being a series in a negative power of $z$, it is clear what this solution describes: it portrays the asymptotic behavior of a transcendent as $z \to \infty$. Indeed, there exist many Painlevé I transcendents that -- at least in some directions in the $z$-plane -- have the required asymptotics. In Boutroux' classification that stems from 1913 \cite{BOUTROUX}, these are the so-called {\em tritronquée} and {\em tronquée} solutions that we shall also describe in some more detail in section \ref{sec:transP1}.

\sk
However, this relation between a single formal solution and a subclass of transcendents cannot be the whole story. Clearly, the relation is not one-to-one, as the power series solution is essentially unique, whereas -- since (\ref{eq:PIintro}) is a second order ODE -- the transcendents form a two-parameter family.

\sk
It is well-known how to obtain a two-parameter family of solutions also on the formal side \cite{Yos, AKT3, Marino:2008ya, Marino:2008vx, GIKM, ASV1, Shim, Iwa1}. To achieve this, one extends the concept of a power series to that of a {\em transseries} (see e.g.\ \cite{Edgar} for an introduction), a formal expansion not only in $z$, but also in other {\em transmonomials} such as $e^{-A/z^c}, e^{+A/z^c}$ (where $A$ and $c$ are constants) and $\log(z)$. Using these building blocks, it was shown in \cite{ASV1}, building on pioneering work in \cite{GIKM}, that one can construct a complete, 2-parameter family of transseries solutions to the Painlevé I equation. These 2-parameter transseries solutions have also appeared in many different guises in other work, e.g.\ \cite{AKT3, Shim, Iwa1, Bonelli:2016qwg, Lis1, Compere:2021zfj}.

\sk
An issue that all members of the family of transseries solutions share, is that whenever a power series in $z^{-\ga}$ for some $\ga$ appears (in practice, for Painlevé I, $\ga$ is always 5/2 or 5/4), no matter which other transmonomials it multiplies, it is always an {\em asymptotic} series, with coefficients that grow factorially. Therefore, these series do not converge for any value of $z$, and so one cannot simply sum them to obtain a functional, transcendent solution. Thus, one is left no choice but to interpret the formal transseries solutions in terms of (hyper-) asymptotics.

\sk
Of course, this is not to say that there are not ways to turn asymptotic series into functions. Émile Borel, a contemporary of Paul Painlevé, developed the procedure of what we now know as `Borel summation' precisely to achieve this. Thus, one may wonder if we cannot simply Borel sum every power series inside a transseries solution (a procedure known as Borel-Écalle summation) to find a transcendent.  At first sight, this may seem to be an impossible task: there are at least three scales in the story, roughly $z$, $e^{-A/z^c}$ and $e^{+A/z^c}$, and at least one of them will always grow without bounds when $z \to \infty$ in some direction in the complex plane. However, as we shall see in detail in this paper, if we keep $z$ {\em fixed} and the parameters of the problem are small enough, one can still Borel-Écalle sum every asymptotic series in the problem (in numerical practice with some cutoffs, of course) and obtain actual transcendents.

\sk
While the answer to the question of summation is thus in principle `yes', the procedure is not unambiguous. Often, the power series one encounters are `non Borel-summable', meaning that one must make a specific choice of {\em lateral Borel summation} to sum them. More importantly, the result one then finds at best corresponds to a given transcendent in a specific wedge-shaped sector of the complex $z$-plane -- a result of the occurence of {\em Stokes phenomenon}.

\subsection{Stokes phenomenon and isomonodromic deformation}
Just like the power series solution to Painlevé I, the transseries solutions, when appropriately summed, describe the asymptotic behaviour of transcendents. Ever since the work of Stokes \cite{STOKES} it has been known, however, that beyond the leading power series description, the asymptotic behaviour of complex functions can suddenly change at special loci -- {\em Stokes lines} -- in the complex plane. It is known, as we shall review in detail in section \ref{sec:transP1}, that for the Painlevé I transseries solutions this means that at the Stokes lines, the values of the two free parameters jump.

\sk
This in particular means that the map between two-parameter transseries and two-parameter transcendents is a very intricate one, and it is this map that we aim to understand better in this paper. On the transcendent side, one picks the parameters once (for example by choosing boundary conditions $u(z_0)$ and $u'(z_0)$ at some point $z_0 \in \bC$) which fixes the entire transcendent. On the transseries side, one can choose $z_0$ and fix the two transseries parameters, but these parameters are then only piecewise constant as one changes $z_0$, and will jump at the Stokes lines. As a result, one must first understand the {\em Stokes automorphism} acting on the transseries parameters at every Stokes line, before even being able to map from transseries to transcendents.

\sk
The quest to fully understand the Stokes automorphisms for the two-parameter transseries of Painlevé I has been a long one, starting with \cite{ASV1} where it was shown that there are actually {\em two} automorphisms that need to be understood for two types of Stokes lines, that each of these automorphisms can be encoded in a seemingly infinite number of {\em Stokes constants}, and that these Stokes constants have several intriguing algebraic relations among themselves. In \cite{Aniceto:2013fka} the problem was studied further, and recently in \cite{BSSV} the research culminated when Baldino, Schwick, Schiappa and Vega showed that even more relations existed. These relations allowed them to relate all Stokes constants to two numbers, one of which is the known Stokes constant of Painlevé I, and a second one for which they were then able to give a closed form expression, thereby matching all the previous numerical results up to a large number of decimal places.

\sk
This result was an enormous step forward from the previous state of the art, since before it only a single Stokes constant had been computed exactly \cite{Kap1, Kap2, Dav1, Tak1} (see also \cite{Cos4, Cos5} for an interesting approach). Intriguingly, the new method of computation, based on alien calculus and Écalle's theory of resurgence \cite{ECALLE} (see e.\ g.\ \cite{Sauzin, Dor1, ABS1} for reviews), differs completely from how the value of the original Stokes constant had been found. Notably, in work by Takei \cite{Tak1, Tak4}, the first Stokes constant was computed by applying the theory of isomonodromic deformation. The question that sparked the research in this paper was: could one use similar methods to Takei's, and in this way recover and verify the results of \cite{BSSV} for all Stokes constants of the full two-parameter transeries?

\sk
As we describe in detail in what follows, the answer to that question is `yes'. One can set up a similar isomonodromic analysis: relate the Painlevé I equation to an associated linear problem, solve that problem using the methods of exact WKB analysis, and study the monodromy of solutions to the problem -- a monodromy that is only consistent if the parameters in the new problem satisfy the original Painlevé I equation. By requiring these parameters to be described by full two-parameter transseries, and finding the connection formulae for the associated linear problem in terms of these\footnote{In \cite{Tak4, KT3}, Y. Takei and T. Kawai already described connection formulae for two-parameter transseries across a Stokes line. Similar ideas involving isomonodromic deformations are also present in both published \cite{Iwa1} and unpublished work of K. Iwaki on 2-parameter tau functions. Here we rederive those formulae with a few tweaks needed to obtain the precise Stokes data for the ``stringy conventions'' of (\ref{eq:PIintro}).}, one can indeed find the complete Stokes automorphisms, including the exact values of all the Stokes constants.

\subsection{Results}
In this paper, we provide the above-mentioned connection formulae for all the different Stokes lines that a transseries solution needs to cross before completing its full monodromy. As expected, these transitions alternate between the two types that were studied in \cite{ASV1} and follow-up work. From the connection formulae, we can find a closed form for all ten Stokes automorphisms -- though in practice the form of the expressions is ugly and it is much nicer to write them down in a more implicit form. When expanded in alien derivatives, the Stokes automorphisms indeed give back the exact form of all the Stokes constants that were found in \cite{BSSV}.

\sk
Having the Stokes automorphisms at our disposal, we are then able to derive several other interesting results. One is about the uniqueness of transseries expansions when a specific transcendent solution is given. It is clear that, even in a given sector in the complex plane, such an expansion can {\em not} be unique, as there are also many different ways to start from an expansion and Borel-Écalle sum it to a transcendent -- choosing different paths of integration in the inverse Borel transform. 

\sk
Here, we make this non-uniqueness precise: we show that first of all there is a choice of branch in expressions like $z^{-5/4}$ and (roughly) $\log z$ that leads to an infinite number of possible transseries expansions for a given transcendent in a given direction. The values of the transseries parameters for these different transseries can all be related by a simple mapping that we denote by $T$. Secondly, in mapping between the transseries parameters and the actual monodromy data of the associated linear problem, another logarithmic map appears, which allows us to essentially shift the {\em product} of the two parameters by $\frac12 \bZ$ without changing the transcendent that is being described. Together, these two ambiguities show us how to reduce the `covering space' of all transseries parameter values to the underlying `monodromy data space' that describes all different transcendents. We pay some special attention to how all of this works for the special loci in monodromy data space that describe Boutroux' special {\em tronquée} and {\em tritronqué} solutions.

\sk
Finally, we perform some high-precision numerical tests on our results. These tests not only help us confirm the validity of our connection formulae and Stokes transitions; they also clarify that one can indeed Borel-Écalle sum {\em two-parameter} transseries in a consistent way, and that these formal transseries -- albeit in a many-to-one way, and only after a specific Borel summation has been chosen -- indeed describe well-defined transcendent solutions.

\subsection{Outline}
Our hope is that this work appeals to members of two communities -- roughly the `isomonodromic deformation community' and the `transseries community'. For this reason, we have tried to write this paper in a self-contained way, explaining some background from both schools of thought in sections \ref{sec:transP1} and \ref{sec:isodef} before moving on to our new results in sections \ref{sec:Stok2par} and especially \ref{sec:results}. 

\sk
We start in section \ref{sec:transP1} by summarising some of the basic features of the Painlevé I equation and what so far has been known about its transseries solutions. We recall some of the basic techniques that were used in \cite{ASV1} to study these transseries, and in particular review how Stokes phenomenon was described in that work and its follow-ups, culminating in \cite{BSSV}. Section \ref{sec:isodef} is mostly a review of work of Takei, Aoki and Kawai. We switch to their alternative formulation of the Painlevé I equation that carries an additional large parameter $\eta$, and following \cite{AKT3} we construct a two-parameter transseries for this equation, which we show to be equivalent to the one constructed in \cite{ASV1}. Subsequently, we follow \cite{Tak1} and study the associated linear problem  and compute all the relevant linear Stokes transitions that allow us to examine the monodromy of solutions around the irregular singularity at infinity. 

\sk
In section \ref{sec:Stok2par}, we establish explicit relations between the monodromy data of the linear problem and the two parameters of the Painlevé I transseries solutions. These relations constitute a connection formula, and although we only perform the computation explicitly for two non-linear Stokes transitions, we show how the result can easily be generalised to arbitrary Stokes transitions. In section \ref{sec:results} we discuss some interesting properties of our formulae and shed further light on the relation between the Painlevé I transcendents and their transseries representations. Moreover, we show how one can rederive all the Stokes constants of the Painlevé I equation up to arbitrary order, finding complete agreement with \cite{BSSV}. Finally, we perform numerical computations to check the connection formulae and the maps between two-parameter transseries and Painlevé I transcendents. In four appendices, we provide some further background and details to the computations in the main text.

\section{Transseries expansions of Painlevé I solutions}
\label{sec:transP1}
In this section, we review some properties of the Painlevé I equation and its solutions -- both formal and `functional'. For the formal, transseries solutions, we use notation and results from \cite{ASV1}, where many more details can be found.

\subsection{Painlevé I equation and Boutroux classification}
\label{sec:P1BC}
Painlevé transcendents form a special class of functions that occur extensively in mathematics and physics. The Painlevé transcendents are solutions to second order, nonlinear, ordinary differential equations, with the defining property that all moveable singularities are poles. That is: any singularities whose location in the complex plane is determined by the two boundary conditions of the ODE are poles: no moveable branch points or other singularities appear.

\sk
Paul Painlevé and his successors studied and classified functions with those properties \cite{PAINLEVE}, culminating in a classification of all second order ODEs that have solutions of this type. The classification consists of six families of equations, labeled Painlevé I through Painlevé VI. The Painlevé VI equation is the most generic one: it has four parameters, and all other Painlevé equations can be obtained from it using successive scaling (`coalescence') limits. 

\sk
The simplest of all Painlevé equations, one of the end points of the cascade of limits, is the parameter-free Painlevé I equation -- see e.\ g.\ \cite{Del} for a review about this equation in the context that we are interested in. By changing variables the equation can be written in several different forms. For now, we write it in one of the forms that is popular in the physics (and especially string theory) literature:
\begin{equation}
    u(z)^2-\frac{1}{6}\frac{d^2 u(z)}{dz^2} =z.
    \label{eq:P1}
\end{equation}
Being a second order ODE, the Painlevé I equation has a two-parameter family of solutions. The solutions $u(z)$ of the equation were studied and classified in detail by Pierre Boutroux in the early 20th century \cite{BOUTROUX}. An important fact in the study of these solutions is that the equation (\ref{eq:P1}) has the $\bZ_5$-symmetry
\be
 z\to \go \; z, \qquad u \to \go^3 \; u ,
\ee
with $\go^5 = 1$. This symmetry maps solutions to new `rotated' solutions, and special solutions exist that are themselves symmetric.

\begin{figure}
 \centering
 \includegraphics[width=0.3\textwidth]{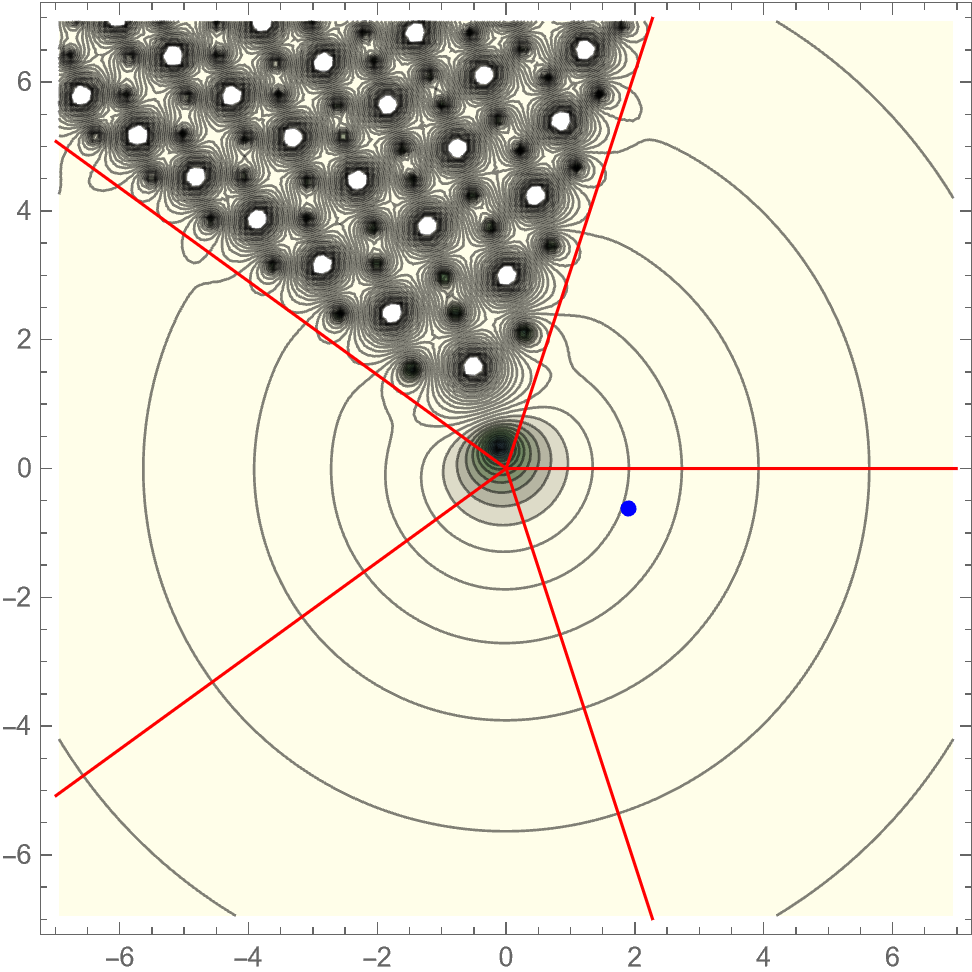}\hspace{1em}
 \includegraphics[width=0.3\textwidth]{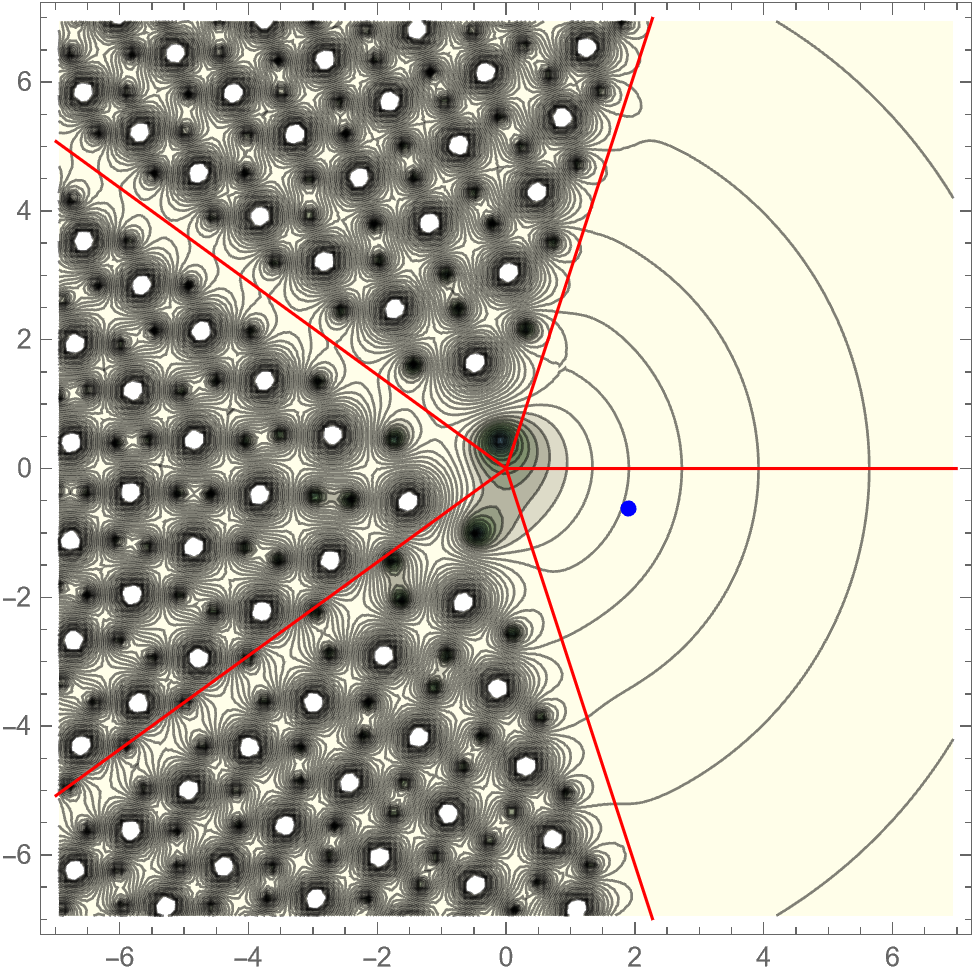}\hspace{1em}
 \includegraphics[width=0.3\textwidth]{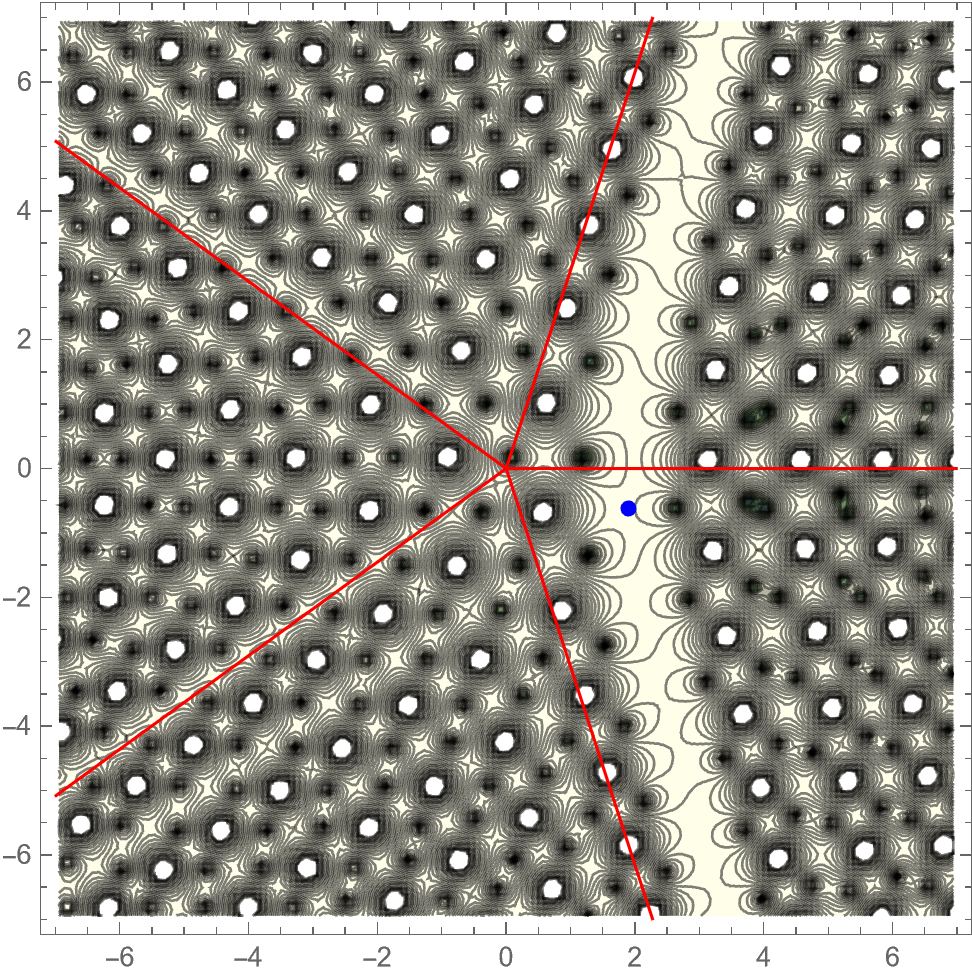} \\
 (a)\hspace{0.3\textwidth}(b)\hspace{0.3\textwidth}(c)
     \caption{Contour plots of (a) a tritronquée solution, (b) a tronquée solution and (c) a general solution. We plot the absolute value of $u(z)$ in the complex $z$-plane; the poles of the solutions are clearly visible as dots in the image. The red lines indicate the boundaries of the five asymptotic sectors. The blue dot marks the value $z_0=2e^{-i\pi/10}$ where the boundary conditions $u(z_0)$ and $u'(z_0)$ are fed to the numerical solver. See section \ref{sec:test} for further details on how these plots are made. }
     \label{fig:boutroux}
\end{figure}

\sk
More importantly, the $\bZ_5$ structure shows up in the possible locations of the moveable poles. In particular, Boutroux' classification shows that all solutions have a single, non-moveable essential singularity at $z=\infty$ and have infinitely many moveable second order poles. For generic members of the two-parameter family of solutions, these poles occur all throughout the complex plane, in a pattern which at infinity becomes $\bZ_5$ symmetric -- see figure \ref{fig:boutroux}c. Closer to the origin in the complex $z$-plane, the poles fill out sectors that asymptote to wedges of opening angle $2\pi/5$. 

\sk
Boutroux showed that moreover there are five one-parameter subfamilies of {\em tronquée} solutions (related to each other by the $\bZ_5$ symmetry) where the poles only appear in three sectors of the complex plane, as in figure \ref{fig:boutroux}b. Within these one-parameter subfamilies, there are special isolated solutions -- five in total, again related by the symmetry -- where the poles only occur in one sector, the so-called {\em tritronquée} solutions (figure \ref{fig:boutroux}a).

\subsection{Series and transseries solutions}
\label{sec:uTS}
The tronquée and tritronquée solutions form the basis for an asymptotic analysis of the Painlevé I solutions, since in the pole-free sectors they have a nice asymptotic behaviour as $z \to \infty$. Since in these sectors the solution is varying slowly, the second derivative term in (\ref{eq:P1}) becomes subdminant as $z \to \infty$, and therefore the solutions behave asymptotically as $u(z) \sim \sqrt{z}$.

\sk
It is straightforward to extend this description of the asymptotic behaviour and include subleading terms. The easiest way to do so is to plug a power series {\em ansatz} starting with $\sqrt{z}$ into (\ref{eq:P1}); one then finds the solution
\be
 \label{eq:pertsol}
 u(z) \sim \sqrt{z} \left( 1 - \frac{1}{48} z^{-5/2} - \frac{49}{4608} z^{-5} - \frac{1225}{55296} z^{-15/2} - \ldots \right).
\ee
However, while this power series describes the large $z$ behaviour in the pole-free sectors very well, it is an {\em asymptotic} series whose coefficients grow factorially -- that is, it does not converge for any large but finite fixed value of $z$. A second, related problem is that the power series solution around $\sqrt{z}$ is unique: it has no free parameters, despite the fact that the differential equation (\ref{eq:P1}) is of second order and therefore should have a two-parameter family of solutions.

\sk
To address the problem of convergence first, it is well known how to deal with asymptotic series such as (\ref{eq:pertsol}) and obtain actual values for $u(z)$ out of them. To begin with, one can try to use Borel summation on the power series. (See e.g. section 2 of \cite{Dor1} for an introduction to Borel summation.) As we shall see in a moment, for the inclusion of nonperturbative corrections it is most natural to do the Borel transformation with respect to the variable $z^{-5/4}$, not with respect to $z^{-5/2}$. Doing so, one finds that the Borel transform $\hat{u}(s)$ has two singularities, at $s = \pm A$ in the Borel plane, with
\be
 A = \frac{8\sqrt{3}}{5},
 \label{eq:instantonaction}
\ee
making the series (\ref{eq:pertsol}) non-Borel summable when $z^{-5/4}$ is a positive or negative real number.

\sk
The terminology `non-Borel summable' is somewhat confusing, however: as is well known, one can still Borel sum the series using a {\em lateral Borel summation}, involving an integration path in the complex Borel plane that avoids the singularity either to the left or to the right. This simply means that there is no longer a natural unique answer coming out of the Borel procedure, but of course, in the end there should not be -- we are expecting a full two-parameter family of solutions!

\sk
Assuming for the moment that we want to do a lateral Borel summation close the positive real $s$-axis, and given the singularity at $s=A$, one easily sees that the ambiguity in the Borel summed answer is of order $e^{-A z^{5/4}}$. This hints at a solution to our second problem, the fact that there are no degrees of freedom in our solution: one could try to construct further formal asymptotic solutions as double expansions of the form
\be
 u(z) \sim \sqrt{z} \sum_{n=0}^\infty \gs_1^n e^{-nAz^{5/4}} u^{(n)}(z),
\ee
where all the $u^{(n)}(z)$ are themselves perturbative expansions in some power of $z$. The leading $u^{(0)}(z)$ is then simply the power series appearing in (\ref{eq:pertsol}). We shall discuss the meaning of the new variable $\gs_1$ in a moment.

\sk
A formal solution of the above type indeed exists, as one can easily check by plugging the {\em ansatz} into the Painlevé I equation. Indeed, up to a choice of sign one finds back the same value (\ref{eq:instantonaction}) for the `instanton action' $A$, and discovers that the $u^{(n)}(z)$ for $n>0$ are uniquely determined\footnote{The exception to this uniqueness is the value of $u^{(1)}_0$, which can be absorbed in the definition of $\gs_1$ and can therefore freely be chosen to equal e.g. $u^{(1)}_0=1$.} series of the form
\be
 u^{(n)} = z^{-5n/4} \sum_{g=0}^\infty u^{(n)}_g z^{-5g/4}.
\ee
As we alluded to before, contrary to the $n=0$ series, the series for $n>0$ turn out to be expansions in $z^{-5/4}$. This fact was given a string theory interpretation in \cite{ASV1}, where the $n=0$ series was identified with a closed string sector with expansion variable $g_s^2 = z^{-5/2}$ whereas the $n>0$ sectors should be thought of as open string sectors with expansion variable $g_s=z^{-5/4}$. Another important point is that one finds solutions for {\em any} value of $\gs_1$. That is: moving from a single expansion in $z^{-5/4}$ to a double expansion, also in powers of $e^{-Az^{5/4}}$ (which is nonperturbative as a function of $z^{-5/4}$), changes our fixed solution into a one-parameter family of solutions!

\sk
Expansions in multiple `transmonomials' such as the one above are called {\em transseries} solutions -- see e.g. \cite{Edgar} for a review. The formal one-parameter transseries solution that we have found can again be thought of as a description of the asymptotic (or rather: {\em hyperasymptotic} \cite{BerryHowls90}) behaviour of a one-parameter family of `true', function solutions. In fact, one can turn a transseries solution into a function solution by choosing an integration path in the Borel plane, Borel summing all of the $u^{(n)}(z)$ along this path, and finally summing over $n$, which for $\gs_1$ small enough now turns out to be a {\em convergent} expansion. This procedure is known as Borel-Écalle summation -- see e.g. \cite{ABS1} for a review.

\sk
Of course, one may now wonder about the next step: can we also construct the full {\em two}-parameter family of solutions in this way? From the formal transseries perspective, this question was addressed in \cite{ASV1}, following the pioneering work by Garoufalidis, Its, Kapaev and Mariño in \cite{GIKM}, and it was found that indeed a two-parameter transseries solution exists. This solution has the form of a {\em triple} expansion,
\be
 \label{eq:ufull}
 u(z) = \sum_{n,m=0}^\infty \gs_1^n \gs_2^m \; e^{-(n-m) A z^{5/4}} u^{(n|m)}(z),
\ee
with $u^{(n|m)}(z)$ again perturbative expansions in $z^{-5/4}$ with a somewhat awkward prefactor:
\be
 \label{eq:unm}
 u^{(n|m)} (z) = z^{-\frac58 \gb_{nm} + \frac12 + \frac{5}{2\sqrt{3}}(n-m)\gs_1 \gs_2} \sum_{h=0}^\infty u^{(n|m)}_h z^{-5h/4}.
\ee
Here, we used the shorthand
\be
 \gb_{nm} =
 \begin{cases}
  2n & \mbox{if~} n=m, \\
  n + (m\mod 2) & \mbox{if~} n>m, \\
  m + (n\mod 2) & \mbox{if~} m>n.
 \end{cases}
\ee
Importantly, in (\ref{eq:ufull}) we now also see the other possible value for the instanton action, $-A$, appearing, and the transseries now has two free parameters, $\gs_1$ and $\gs_2$. Note that the power in the prefactor of the perturbative series not only depends on $n$ and $m$, but also on $\gs_1$ and $\gs_2$. This dependence can be removed by Taylor expanding the prefactor in $\gs_1 \gs_2$, at the cost of introducing a finite, fourth expansion, this time in powers of $\log z$ \cite{ASV1}.

\sk
The transseries (\ref{eq:ufull}) gives us a two-parameter family of formal solutions, but it is not straightforward to interpret an expression of this form as a (hyper-) asymptotic expansion of a function solution, since in almost every direction in the complex plane, either $e^{+Az^{5/4}}$ or $e^{-Az^{5/4}}$ grows without bounds. This fact is of course related to the fact that, as we mentioned earlier, there are no pole-free sectors in a generic two-parameter solution, and so we should not really expect a good interpretation of the two-parameter transseries in terms of asymptotics\footnote{This statement is true if we think of asymptotic expansions in terms of functions that decay at infinity. One can however describe the large $z$ behaviour of generic Painlevé I solutions in terms of Weierstrass elliptic functions, as was already shown by Boutroux \cite{BOUTROUX}. For a derivation of this elliptic behaviour see e.g.\ \cite{JK1} and for the interplay with Stokes' phenomenon \cite{KK}.}. 

\sk
This by no means implies that the two-parameter transseries solution is useless, though. In fact, one can still convert solutions of the form (\ref{eq:ufull}) into functions by applying the exact same procedure as for the one-parameter transseries: first {\em fix} a value of $z$, as well as a path in the Borel plane, then Borel sum all $u^{(n|m)}(z)$, and finally sum over $n$ and $m$. For $\gs_1$ and $\gs_2$ small enough, the latter sums are again convergent and so we obtain a true function for such small values -- which can then at least in principle (though this is quite difficult numerically) be analytically continued to larger values of $\gs_1$ and $\gs_2$.

\subsection{Stokes phenomenon}
\label{sec:Stokes}
When relating formal solutions to function solutions (i.e.\ transcendents), {\em Stokes phenomenon} \cite{STOKES} plays an important role. The phenomenon was originally discovered as a discrete change in asymptotic behavior of certain functions -- the Airy function being a prime example -- as one changes the direction in the complex plane of the variable in which the asymptotics is described. A modern description of the phenomenon for a function $u(z)$ is as follows:
\begin{itemize}
    \item When crossing {\bf \em Stokes lines} running from $0$ to $\infty$ in some direction in the complex $z$-plane, the (hyper-) asymptotic behavior of the function picks up new, exponentially small contributions.
    \item At {\bf \em anti-Stokes lines} in the complex $z$-plane, these nonperturbative contributions become of the same order of magnitude as the terms in the perturbative expansion, and therefore the asymptotic behaviour of the function becomes very different.
\end{itemize}
In the world of formal solutions, the transseries turns out to be an ideal tool to describe Stokes phenomenon. To start with the {\em anti-Stokes lines}: exponential terms of the form $e^{\pm A z^{5/4}}$ are already present in the transseries description, and therefore we can describe the anti-Stokes lines as loci where $A z^{5/4}$ becomes purely imaginary. Note that this requires a choice of branch for $z^{5/4}$, and as a result what is an anti-Stokes line for one choice of branch may not be one for another.

\sk
When crossing a {\em Stokes line}, something different happens to the transseries: not only is $A z^{5/4}$ now purely real, but here the asymptotic behavior of the function solutions beyond the leading power series changes abruptly, and therefore one should choose different parameters $\gs_1, \gs_2$ to describe the same Painlevé I transcendent.

\sk
The Stokes phenomenon for the 2-parameter transseries solution to the Painlevé I equation was studied in \cite{ASV1,Aniceto:2013fka,BSSV} using Écalle's theory of resurgence \cite{ECALLE}. It was shown that the five `special' lines that separate the sectors in the complex $z$-plane (see again figure \ref{fig:boutroux}), alternatingly play the role of anti-Stokes lines (when the branch of $z^{5/4}$ is appropriate) and of Stokes lines. At the Stokes lines, a {\em Stokes automorphism} $\mS$ maps the values of $\gs_1, \gs_2$ to new values, appropriate for describing the asymptotics in the next sector.

\sk
Écalle's theory of resurgence allows one to build up each Stokes automorphism out of operators known as {\em alien derivatives}, see e.g. \cite{ABS1} for a review. Using these techniques, \cite{ASV1} found results that can be summarized as follows:
\begin{itemize}
    \item In terms of the Écalle time $x=z^{-5/4}$, for Painlevé I transseries there are essentially two different Stokes phenomena at play -- denoted in \cite{ASV1} by $\underline{\mS}_0$ and $\underline{\mS}_\pi$.
    \item $\underline{\mS}_0$ occurs whenever in the $z$-plane a Stokes line is crossed that corresponds to the {\em positive} real axis in the complex $x$-plane. $\underline{\mS}_\pi$ occurs when the Stokes line corresponds to the {\em negative} real $x$-axis.
    \item In the case of the one-parameter transseries for which $\gs_2=0$, the automorphism $\underline{\mS}_0$ has a `classical', simple form: it simply corresponds to mapping
    \be
     \gs_1 \to \gs_1 + S_1
     \ee
     and keeping $\gs_2=0$ also after the transition. Here
     \be
      S_1=-i\frac{3^{1/4}}{2\sqrt{\pi}}
     \ee
    is the (first) Stokes constant for Painlevé I.
    \item For the same one-parameter transseries, the automorphism $\underline{\mS}_\pi$ is much more complicated. Expanding in alien derivatives, \cite{ASV1} gave expressions for the automorphism in terms of a seemingly infinite number of other Stokes constants labelled $S_{-k}$.
    \item For the full two-parameter transseries, both $\underline{\mS}_0$ and $\underline{\mS}_\pi$ are complicated nonlinear maps. Expanding in alien derivatives, these automorphisms can be encoded in two infinite arrays of Stokes constants, denoted $S_{\pm k}^{(\ell)}$ and $\tilde{S}_{\pm k}^{(\ell)}$ in \cite{ASV1}.
    \item Finally, in \cite{ASV1}, several of the above-mentioned Stokes constant were computed numerically up to very high precision. While at the time it turned out impossible to find closed expressions for any of these numbers (apart from the already known $S_1$ and a number that up to high precision equalled $i/2$), it was shown that there are several simple algebraic relations between the constants.
\end{itemize}
The last point of course left two interesting questions: can we find further closed expressions for the Stokes constants, and can we find further relations between them? As for relations, more were found and explained in \cite{Aniceto:2013fka}, and then recently the full structure of relations between Stokes constants was uncovered in \cite{BSSV}. For the values themselves, for a long time only the value of $S_1$ had been known analytically from the work of \cite{Kap1, Tak1, Kap2}. This also changed with the recent work of \cite{BSSV}, whose relations reduced all the unknown Stokes constants to two numbers, one of which is the known constant $S_1$ and they other which in that paper was matched to an exact expression as well as confirmed up to very high numerical precision.

\sk
The method to relate the different Stokes constants that was used by \cite{BSSV}, based on a deep understanding of the alien calculus, is very different from the methods used to compute the Stokes constant $S_1$ by Kapaev in \cite{Kap1} or Takei in \cite{Tak1}. These latter two approaches used the theory of isomonodromic deformation to relate the Painlevé I equation to an associated linear problem. In the approach of Takei and his collaborators, the linear problem was massaged into a single second order ODE whose monodromy data gave way to the computation of the Stokes constant $S_1$. The main question that led to the current paper was: is it possible to extend Takei's methods to also compute the higher Stokes constants? If this would be possible in closed form, it could not only reproduce and verify the results of \cite{BSSV}, but might also provide workable (though complicated) closed forms for the full Stokes automorphisms $\underline{\mS}_0$ and $\underline{\mS}_\pi$, that could then be used to better understand the relation between asymptotic transseries expansions and transcendents.

\sk
It is precisely this question that we answer in an affirmative way in this paper. Before writing down the full answer, though, to be self-contained we must first review the method of isomonodromic deformation that was used by Takei and collaborators. It is to this topic that we turn next.

\section{Isomonodromic deformation}
\label{sec:isodef}
In order to understand the Stokes phenomenon of transseries solutions for the Painlevé I equation, we now turn to a review of the method of isomonodromic deformation. We consider its incarnation as developed over the years by T. Aoki, T. Kawai and Y. Takei \cite{AKT2, AKT3, Tak1}. The method involves an alternative formulation of the Painlevé I equation which is related to an associated linear problem, which in fact turns out to be a Schrödinger equation. The Painlevé I equation acts as a \textit{condition for isomonodromic deformation} on this linear problem, which means that its {\em monodromy data}, that we define shortly, are preserved if a parameter $\lambda(t)$ appearing in the linear problem is a solution to the Painlevé I equation. 

\sk
Before studying this linear problem, we first construct the transseries solutions to this alternative Painlevé I equation in subsection \ref{sec3.1} and show how they relate to the original transseries solutions that we have described in the previous section. In section \ref{sec:EWKBmon} we then review the exact WKB approach by Takei \cite{Tak1}, which employs techniques due to Voros \cite{Vor1}, and compute all the necessary ingredients for establishing explicit connection formulae for the two-parameter transseries solutions to the Painlevé I equation. We try to stay close to Takei's derivation and conventions, but will mention it when we sporadically deviate from those. 

\sk
For our purposes, the ``string theory conventions'' of writing the Painlevé I equation in the form (\ref{eq:P1}) are not very convenient. In fact, to be able to apply the exact WKB formalism, it is useful to introduce an additional parameter in the equation. We therefore switch to the conventions of the \textit{Painlevé I equation with a large parameter}:
\begin{equation}
    \frac{d^2 \lambda}{dt^2} = \eta^2 (6\lambda^2+t).
    \label{eq:P1L}
\end{equation}
The parameter $\eta$ is considered to be large; we shall see shortly that the solutions we consider are expansions in powers of $\eta^{-1/2}$. Using the right change of variables and fixing $\eta$, we can scale this equation to any of its different forms in the literature. In particular, setting $\eta = \frac{1}{6}$  and identifying $z = -t/6$ we find back the form $u^2-\frac{1}{6}u_{zz}  = z$ of (\ref{eq:P1}). For comparison with the literature we shall transform our results back to this string theory convention later, but for the time being we keep $\eta$ unfixed and use it to describe the asymptotics of solutions to (\ref{eq:P1L}). 

\sk
As is well known, the Painlevé I equation with a large parameter is equivalent to the Hamiltonian system \cite{JM, Oka1}
\begin{equation}
    \begin{split}
        \frac{d \lambda}{d t } &=  \eta \frac{\partial H}{\partial\nu} \\
         \frac{d \nu}{d t } &= - \eta\frac{\partial H}{\partial\lambda},
    \end{split}
    \label{eq:ham}
\end{equation}
with Hamiltonian
\be
 H = \nu^2/2-(2\lambda^3+t\lambda),
\ee
as can easily be checked. Note that there are two differences with the classical Hamiltonian systems one often encounters in physics: first of all the Hamiltonian depends on $t$ explicitly, and secondly there are explicit factors of the parameter $\eta$, which we shall soon identify with the inverse of Planck's constant. If we want, we can absorb those factors by thinking of $\eta t$ as the `physical time', while we think of $t$ as a rescaled `quantum time' that is usually used in the mathematics literature. 

\sk
The nonlinear classical Hamiltonian system (\ref{eq:ham}) is not the formulation we are most interested in in this paper. It is well known that there is an associated linear quantum system that is much better suited for our purposes. In fact, the system (\ref{eq:ham}) acts as a condition of integrability for the following second-order differential equation:
\begin{equation}
   \left( -\frac{\partial^2}{\partial x^2}+\eta^2 Q(x)\right)\psi(x) = 0,
   \label{eq:SL1}
\end{equation}
where $Q(x)$ is defined as follows:
\begin{equation}
    Q(x) = 4x^3+2tx+2H-\eta^{-1} \frac{\nu}{x-\lambda}+\eta^{-2} \frac{3}{4(x-\lambda)^2}.
    \label{eq:SL2}
\end{equation}
In appendix \ref{app:isomon}, we remind the reader of the relation between the nonlinear and the linear systems. Clearly, equation (\ref{eq:SL1}) is a Schrödinger type equation with $\hbar = \eta^{-1}$ and a quantum-corrected potential energy that has terms that explicitly depend on this $\hbar$. One can study this equation extensively using the tools of the exact WKB method that we briefly review in appendix \ref{app:exactWKB}.

\sk
Let us now sketch the strategy to address the non-linear Stokes phenomenon of the Painlevé I equation. The Schrödinger equation (\ref{eq:SL1}) above has an irregular singularity at infinity around which its basis of solutions has a non-trivial monodromy, defined by the {\em monodromy data} that we shall specify later. The theory of isomonodromic deformations tells us that solutions to the {\em nonlinear} Hamiltonian system (\ref{eq:ham}), which are in fact solutions to the first Painlevé equation, preserve the monodromy data of the {\em linear} quantum problem (\ref{eq:SL1}). Turning this logic around we can construct solutions $\lambda(t)$ to the first Painlevé equation by fixing the monodromy data. Even better, we can consider an asymptotic solution -- in this paper a two-parameter transseries solution -- and study it on either side of a Stokes line of $\lambda(t)$. Imposing that the monodromy data remains fixed across the Stokes line allows us to relate the transseries asymptotics of a single Painlevé I transcendent on either side of the Stokes line and thereby establish a connection formula. The main technical hurdle is obtaining explicit relations between the Painlevé I asymptotics and the monodromy data, which can be achieved by means of the exact WKB method

\subsection{Another transseries}
\label{sec3.1}
Before we can start examining the linear problem (\ref{eq:SL1}), we must first address the asymptotics of the solution $\lambda(t)$ to (\ref{eq:P1L}). As it is a solution to the Painlevé I equation with a large parameter, we expect that, just as the solution to the original equation  (\ref{eq:P1}), this solution admits a two-parameter asymptotic expansion. We briefly check that this is the case by constructing a transseries following the multiple-scales approach of e.g.\ \cite{AKT3}. Subsequently, we show that with the identifications mentioned in the previous subsection, this transseries reduces to the one discussed in section \ref{sec:uTS}.

\sk
We start by defining $F(\lambda, t) = 6\lambda^2+t$ so that the first Painlevé equation reads
\begin{equation}
    \frac{d^2\lambda}{dt^2} = \eta^2 F(\lambda(t), t).
    \label{eq:F1}
\end{equation}
We then write $\lambda(t)$ as a perturbation around the solution of $F(\lambda_0, t) = 0$:
\begin{equation}
    \lambda(t) = \lambda_0(t)+\eta^{-1/2} \Lambda(\eta, t)
    \label{eq:lambdaLambda}
\end{equation}
where the factor $\eta^{-1/2}$ is introduced for future convenience and $\gl_0(t)=\sqrt{-t/6}$. This of course requires a choice of branch for the square root, but since we are interested in asymptotics in certain sectors of the complex $t$-plane only, we can simply choose a branch and stick with it. We also introduce a new scale \cite{AKT3}
\begin{equation}
    \tau = \tau(t, \eta) = \eta \int^t \sqrt{\frac{\diff F}{\diff \lambda}(\lambda_0, t')} \; dt'
\end{equation}
which, as we shall see in a moment, is the exponent of the instanton transmonomial of the transseries we are constructing. The goal is now to find a solution $\Lambda= \Lambda(\eta, t, \tau)$, which we assume to be a formal series in $\eta^{-1/2}$, such that $\lambda(t)$ in (\ref{eq:lambdaLambda}) is a solution of the first Painlevé equation. First of all, we can differentiate $\Lambda(\eta, t, \tau)$ with respect to $t$ twice and obtain
\begin{equation}
    \frac{\diff ^2\Lambda}{\diff t^2} = \left[ \frac{\partial^2}{\partial t^2}+\left(\frac{\diff \tau}{\diff t}\right)^2 \frac{\partial^2}{\partial \tau^2}+\frac{\diff ^2\tau}{\diff t^2} \frac{\partial}{\partial \tau}+2\frac{\diff \tau}{\diff t}\frac{\partial^2}{\partial t \partial \tau}\right]\Lambda(\eta, t, \tau).
\end{equation}
Then, on the left hand side of (\ref{eq:F1}) we can insert this expression, while on the right hand side we can expand the function $F(\lambda, t)$ around the solution $\lambda_0$:
\begin{equation}
    \begin{split}
        \frac{d^2\lambda_0}{dt^2} + \eta^{-1/2}\left[ \frac{\partial^2}{\partial t^2}+\eta^2 F^{(1)} \frac{\partial^2}{\partial \tau^2}+\eta\left(\frac{\diff }{\diff t} \sqrt{F^{(1)}}\right) \frac{\partial}{\partial \tau}+2\eta\sqrt{F^{(1)}}\frac{\partial^2}{\partial t \partial \tau}\right]\Lambda\\
        = \eta^2 \left( F^{(0)} + F^{(1)} \eta^{-1/2}\Lambda +  \frac{1}{2}F^{(2)} \eta^{-1}\Lambda^2+ \frac{1}{6}F^{(3)} \eta^{-3/2}\Lambda^3+\frac{1}{24}F^{(4)} \eta^{-2}\Lambda^4\right)\\+\mathcal{O}(\eta^{-1/2}).
    \end{split}
\end{equation}
Here, we have denoted $F^{(n)} = \frac{\diff ^n F}{\diff \lambda^n}(\lambda_0(t), t)$, so in the case of Painlevé I the $F^{(3)}$ and $F^{(4)}$ terms are actually zero. Then we plug a formal power series ansatz,
\be
 \Lambda = \sum_{n=0}^\infty \Lambda_{n/2}(t, \tau) \eta^{-n/2},
\ee
into the expression above and obtain a series of equations order by order in $\eta^{-1/2}$, of which the first one at order $\eta^2$ is vacuous and the next three read
\begin{equation}
    \begin{split}
        &\left(\frac{\partial^2}{\partial \tau^2}-1\right)\Lambda_0 = 0,\\
        &\left(\frac{\partial^2}{\partial \tau^2}-1\right)\Lambda_{1/2} = \frac{F^{(2)}}{2F^{(1)}} \Lambda_0^2,\\
        &\left(\frac{\partial^2}{\partial \tau^2}-1\right)\Lambda_1 = \frac{F^{(2)}}{F^{(1)}} \Lambda_0\Lambda_{1/2}-\frac{2}{\sqrt{F^{(1)}}}\frac{\partial^2 \Lambda_0}{\partial t \partial \tau}-\frac{\frac{d}{dt}F^{(1)}}{2(F^{(1)})^{3/2}}\frac{\partial \Lambda_0}{\partial \tau},
    \end{split}
    \label{Lsystem}
\end{equation}
where we now have used $F^{(3)}=F^{(4)}=0$. These equations are then solved recursively by using the ansatz
\begin{equation}
    \Lambda_{n/2} = \sum_{m=-(n+1)/2}^{(n+1)/2} a^{(n)}_m(t) \; e^{2m \tau} 
\end{equation}
and imposing non-secularity conditions as is standard in multiple scale analysis. Once the dust settles, for all $\Lambda_{n/2}$, we find a dependence on only two free parameters, that we shall call $\alpha$ and $\beta$. Details can again be found in \cite{AKT3}; the first three terms that one finds are
\bea
     \Lambda_0 \;\;\; & = & (12\lambda_0)^{-1/4}\left(\alpha \lambda_0^{5\alpha\beta}e^{\tau}+\beta \lambda_0^{-5\alpha\beta}e^{-\tau}\right) \ret
     \Lambda_{1/2} & = & 2(12\lambda_0)^{-3/2}\left(\alpha^2 \lambda_0^{10\alpha\beta}e^{2\tau}-6\alpha\beta +\beta^2 \lambda_0^{-10\alpha\beta}e^{-2\tau}\right) \ret
     \Lambda_{1} \;\;\; & = & (12\lambda_0)^{-11/4}\left(3\alpha^3 \lambda_0^{15\alpha\beta}e^{3\tau} +\left(-\frac{15}{4}\alpha+22\alpha^2\beta-282 \alpha^3\beta^2\right)\lambda_0^{5\alpha\beta}e^\tau \right. \ret
       && \quad +\left. \left(\frac{15}{4}\beta+22\alpha\beta^2+282 \alpha^2\beta^3\right)\lambda_0^{-5\alpha\beta}e^{-\tau} +3\beta^3 \lambda_0^{-15\alpha\beta}e^{-3\tau}\right).
    \label{eq:AKTtrans}
\eea
These expressions now constitute a transseries. We recognize the `instanton transmonomial factors' of the form $e^{(n-m) \tau}$ -- for $n$ and $m$ integer -- accompanying products of the free parameters, $\alpha^n \beta^m$, in the familiar way. 

\sk
Note that a single instanton transmonomial appears in multiple $\Lambda_{n/2}$'s, each time with a different power of $\lambda_0$, and therefore of $t$. All such terms belong to the same transseries sector with labels $(n, m)$. For instance, in $\Lambda_{0}$ and $\Lambda_{1}$, we find terms that scale with $\alpha \cdot e^\tau$, which therefore must be terms in the $(1,0)$ transseries sector. More perturbative fluctuations in this sector, as well as new terms that appear in higher order instanton sectors, appear at $\Lambda_{n/2}$ for higher $n$. 

\sk
Of course, this transseries, obtained using the multiple-scales analysis, should reproduce the transseries that we discussed in the `stringy' conventions in section \ref{sec:uTS}. As a warmup for our later computation when we want to relate the Stokes data in the two conventions, let us check that indeed the transseries match. As mentioned before, with $\lambda_0 = \sqrt{-t/6} = \sqrt{z} $ and setting $\eta=1/6$, we recover the form of the first Painlevé equation studied in \cite{ASV1}. We can first check the instanton transmonomial $e^{\pm\tau}$, for which we find
\begin{equation}
    \tau =\eta \int^t \sqrt{12 \lambda_0(t')} \; \diff t' =- \frac{8\sqrt{3}}{5}\left(-\frac{t}{6}\right)^{5/4} =- \frac{8\sqrt{3}}{5}z^{5/4},
\end{equation}
which indeed carries the instanton action $A$ from  (\ref{eq:instantonaction}). 

\sk
Next, we can check some of the leading coefficients in the transseries expansion. Our complete transseries solution of the form $\lambda = \lambda_0+\sum_{n=0}^\infty \Lambda_{n/2} \eta^{-(1+n)/2}$ has components 
\begin{equation}
    \begin{split}
        \lambda_0 & \; = \; \sqrt{z}\\
        \eta^{-1/2}\Lambda_0 & \; = \; z^{-1/8}\left(\sigma_1 e^I+\sigma_2 e^{-I}\right)\\
        \eta^{-1}\Lambda_{1/2} & \; = \; z^{-3/4}\left(\frac{1}{6}\sigma_1^2 e^{2I}-\sigma_1 \sigma_2+\frac{1}{6}\sigma_2^2 e^{-2I}\right) \\ 
        \eta^{-3/2}\Lambda_1 & \; = \; z^{-11/8}\Bigg(\frac{1}{48}\sigma_1^3 e^{3I}+\left(-\frac{5}{64\sqrt{3}}\sigma_1+ \frac{11}{72}\sigma_1^2\sigma_2- \frac{47}{24\sqrt{3}}\sigma_1^3\sigma_2^2\right)e^I \\ 
        & \qquad \qquad +\left(\frac{5}{64\sqrt{3}}\sigma_2+ \frac{11}{72}\sigma_1\sigma_2^2+ \frac{47}{24\sqrt{3}}\sigma_1^2\sigma_2^3\right)e^{-I}+\frac{1}{48} \sigma_2^3e^{-3I}\Bigg),
    \end{split}
    \label{eq:AKTtransseries}
\end{equation}
where we have rescaled the transseries parameters $(\alpha, \beta)=-3^{-1/4}(\sigma_1, \sigma_2)$ and inserted the value of $\eta^{-1/2}=-\sqrt{6}$, which is consistent with the identification $\eta = 1/6$ that we dicussed at the start of this section, and turns out to be the sign choice that gives us back the conventions of \cite{ASV1}. Moreover, in the above expression we have used the shorthand
\begin{equation}
    e^I = z^{\frac{5}{2\sqrt{3}}\sigma_1\sigma_2 } e^{-\frac{8\sqrt{3}}{5}z^{5/4}}.
\end{equation}
We see that the terms on the right-hand side of (\ref{eq:AKTtransseries}) indeed match those found for the two-parameter transseries in \cite{ASV1}, now organized by the power of $z$ with which they appear. Additionally, we can expand the first factor in $e^I$ as
\begin{equation}
    z^{\frac{5}{2\sqrt{3}}\sigma_1\sigma_2} = \exp\left(\frac{5}{2\sqrt{3}}\sigma_1\sigma_2\log(z)\right) = \sum_{k=0}^\infty \frac{1}{k!}\left(\frac{5}{2\sqrt{3}}\right)^k (\sigma_1 \sigma_2)^k \log(z)^k
    \label{eq:translogexp}
\end{equation}
which, as was observed in \cite{ASV1}, leads to the additional \textit{logarithmic sectors} that occur because the Painlevé I equation has a resonant transseries solution.

\sk
From (\ref{eq:AKTtransseries}) we see that for the Painlevé I equation\footnote{For the higher Painlevé equations, one can equivalently introduce a large parameter $\eta$ to play the role of the inverse Planck's constant, but in those cases the $\eta$-expansion is truly different from the large $z$ expansion -- see e.g. \cite{Gregori:2021tvs}. We thank Ricardo Schiappa for pointing this out to us.}, $\eta$ is nothing but a convenient bookkeeping device that allows us to make an $\eta$-expansion equivalent to the large $z$ expansion: if we ignore the overall factor of $\sqrt{z}$, then we see that all terms in (\ref{eq:AKTtransseries}) scale with $\left(\eta^{-1/2}z^{-5/8}\right)^{n+m}$.

\subsection{Exact WKB and the monodromy in the linear problem}
\label{sec:EWKBmon}

\subsubsection{General aspects of the monodromy}
\label{sec:mono}
By now, we have mentioned the importance of the monodromy data of our problem several times. Before studying the specific linear problem introduced in equations (\ref{eq:SL1}) and (\ref{eq:SL2}), we would first like to explain what exactly this monodromy data is, and how it relates to the linear Stokes geometry of the Schrödinger problem. Some relevant background on the exact WKB method and Stokes phenomenon can be found in appendix \ref{app:exactWKB}.

\sk
The leading term $Q_0(x)$ of the potential $Q(x)$ in (\ref{eq:SL2}) forms a cubic polynomial in $x$, which implies that the corresponding Stokes graph has five asympotic directions going to infinity. Since our problem is a second order linear ODE, there will be a two-dimensional basis of solutions that undergoes a linear Stokes phenomenon across each of these asymptotic directions. These Stokes phenomena can be encoded in triangular two-by-two matrices of the form
\begin{equation}
            M_{2i} = \begin{pmatrix}
            1& m_{2i} \\
            0& 1
            \end{pmatrix} \hspace{1cm} \text{and}\hspace{1cm}
             M_{2i-1} = \begin{pmatrix}
            1& 0 \\
            m_{2i-1}& 1
            \end{pmatrix},
            \label{eq:G1}
\end{equation}
where $i = 1 .... 5$. Note that there are {\em ten} Stokes matrices, two for each asymptotic Stokes line, since we are working on a Stokes curve that lives on a  double cover of the complex plane. Whenever an asympototic solution expressed in the basis $(\psi_+, \psi_-)^T$ crosses an asymptotic direction, we multiply it by the corresponding Stokes matrix. Here, as usual, we denote by $\psi_+$ the basis solution that grows exponentially towards infinity along the positive real axis on the principal sheet, and by $\psi_-$ the one that decays exponentially. We call the matrices $M_j$ \textit{Stokes matrices} and the corresponding numbers $m_j$  \textit{Stokes multipliers} -- the latter should not be confused with the Stokes constants of the Painlevé I equation. The Stokes multipliers determine the monodromy of solutions around the irregular singular point at infinity in the complex $x$-plane, and hence we say that these multipliers define the \textit{monodromy data} of the linear problem \cite{JMU1}. Besides the five asymptotic directions that the Stokes graph has, it also contains a square root branch cut originating from the simple turning point. The branch cut of course can lie in any convenient direction we choose. Crossing this cut in the counterclockwise direction is equivalent to multiplying our set of solutions with 
\be
    B = \begin{pmatrix}
         0 & -i \\
         -i & 0
    \end{pmatrix},
\ee
as follows from the WKB ansatz (\ref{eq:WKBsol3}). Thus, if we make a round trip, we find the following constraint:
\begin{equation}
    B M_{j+4} M_{j+3} M_{j+2} M_{j+1} M_{j} = I
\end{equation}
for any $j$, where the $M_k$ are defined cyclically whenever $k>10$ or $k<1$. One can express this contraint in terms of the Stokes multipliers to find
\begin{equation}
    \begin{split}
        m_{j+3} &= i(1+m_j m_{j+1}), \\
        m_{j+5} &= m_j.
    \end{split}
    \label{eq:mult}
\end{equation}
We see that the Stokes multipliers are not all independent and reduce to essentially two degrees of freedom. The fact that our transseries also has two parameters is no coincidence, and it will be our goal to relate the monodromy data of the linear problem, encoded in the Stokes multipliers $m_i$, to the two free parameters of the Painlevé I transseries solution. However, we shall see that this relation is not one-to-one and that therefore only the Stokes multipliers truly parametrize the different Painlevé I transcendents. We come back to this point in section \ref{sec:TPS} when we study the Stokes phenomena of tronquée and tritronquée solutions.

\sk
Let us now turn to the problem at hand. Recall that $\lambda_0(t)$ is a solution to $F(\lambda_0(t), t) = 0$. We consider a perturbation around this solution\footnote{Note that the fact that $\nu(t)$ starts at order $\eta^{-1/2}$ is not inconsistent with the Hamiltonian relation $\nu = \eta^{-1} \gl'(t)$. The reason is that due to the instanton transmonomials, $\gL'(t)$ in fact is of order $\eta^{+1}$.},
\begin{equation}
    \begin{split}
        \lambda(t)& = \lambda_0(t)+\eta^{-1/2}\Lambda, \\
        \nu(t) &= \eta^{-1/2} \mathcal{N},
    \end{split}
\end{equation}
where again $\Lambda = \sum_{n=0}\Lambda_{n/2}\eta^{-n/2}$ and $\cN = \sum_{n=0}\cN_{n/2}\eta^{-n/2}$ are formal power series in $\eta^{-1/2}$. The above ansatz allows us to rewrite the potential (\ref{eq:SL2}) as
\begin{equation}
    Q(x)  = 4(x-\lambda_0)^2(x+2\lambda_0) + \eta^{-1}(\mathcal{N}-12\lambda_0\Lambda^2)+\mathcal{O}(\eta^{-3/2}),
\end{equation}
where $\lambda_0 = \sqrt{z}=\sqrt{-t/6}$. We have factorized the leading term $Q_0$, commonly known as the \textit{principal term} (see also appendix \ref{app:exactWKB}), from which we see that the Stokes graph constructed out of $Q_0$ has two turning points: one simple turning point at $x=-2\lambda_0$ and one double turning point at $x=\lambda_0$.  

\begin{figure}
     \centering
     \includegraphics[width=1 \textwidth]{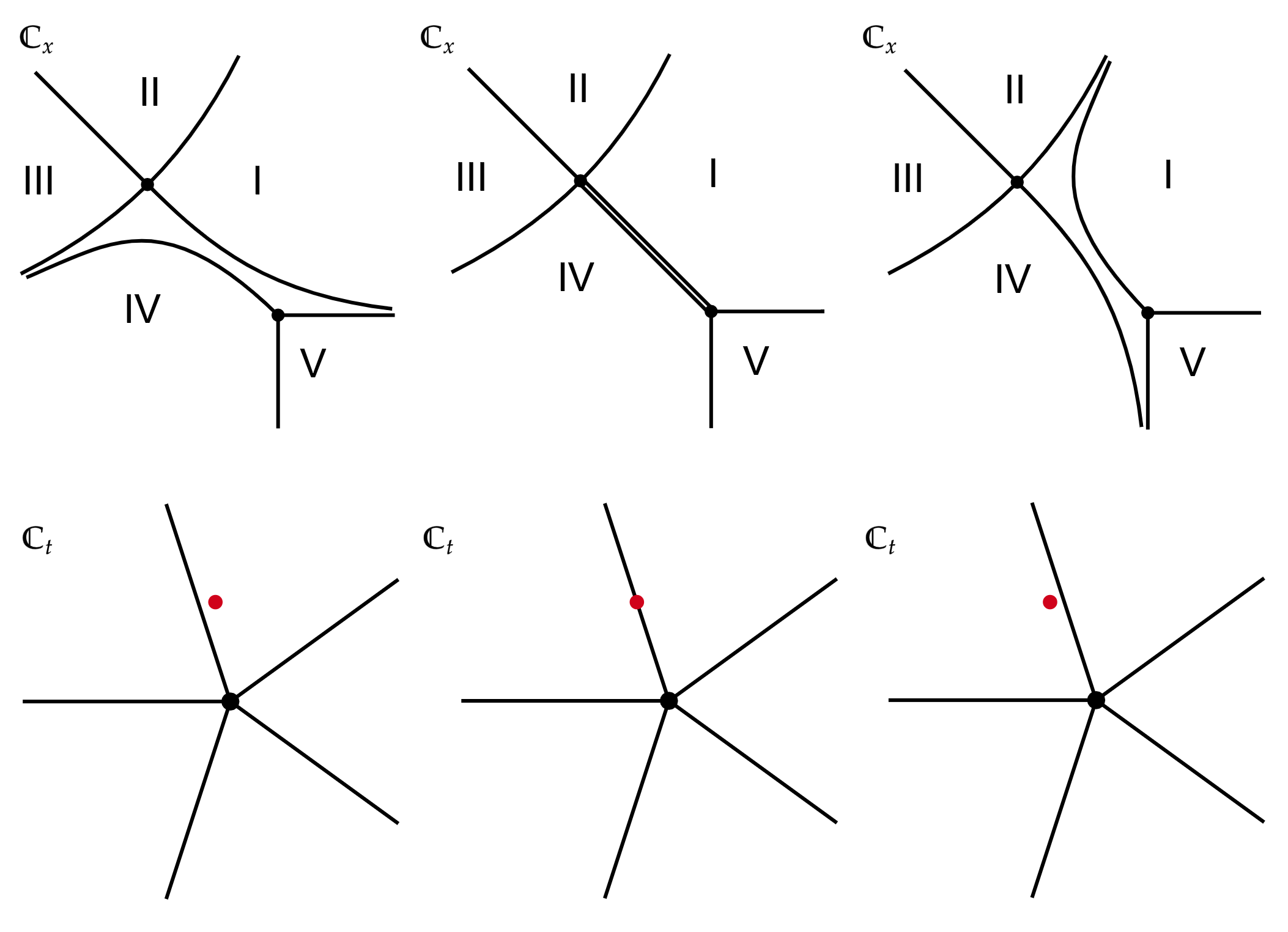}
     \caption{Three Stokes graphs in the $\bC_x$-plane before, at, and after the non-linear Stokes transition. The top three panels show the Stokes graphs of the linear problem that correspond to Painlevé I solutions $\lambda(t)$ that have a value of $t = -6z = -6\lambda_0^2$ displayed in the bottom three graphs respectively. The double line in the upper middle panel denotes a saddle trajectory ending at two different turning points. The roman numerals denote the different asymptotic regions.}
     \label{fig:Wallcrossing}
 \end{figure}

\sk
Examples of generic Stokes graphs for such a potential are displayed in the upper left and upper right panels of figure \ref{fig:Wallcrossing}. These graphs have three Stokes lines emerging from the simple turning point and four coming out of the double turning point. All these lines approach infinity along one of the five asymptotic directions, separated evenly by angles of $2\pi/5$. We see that two pieces of information are needed to determine the monodromy data of our problem: 1) the exact linear Stokes phenomenon across the Stokes lines emanating from both simple and double turning points, and 2) the topology of the Stokes graph, i.e.\ we need to know which Stokes line approaches which asymptotic direction. Let us address these two points in some more detail, starting with the latter.

\paragraph{Topology of the Stokes graph.}

In $Q_0(x)$, $t$ plays the role of an external parameter determining the precise shape of the Stokes graph. For generic infinitesimal changes in $t$, we observe no change in the topology of the graph. However, when $t$ crosses what is called a {\em non-linear Stokes line} in the $\mathbb{C}_t$ plane (see figure \ref{fig:Wallcrossing}), a change does occur\footnote{We want to stress again that in this problem there are two types of Stokes phenomena: the linear Stokes phenomenon that solutions $\psi_\pm(x)$ of the Schrödinger equation undergo, and the non-linear Stokes phenomenon that affects solutions $\lambda(t)$ of the Painlevé I equation. We use the terminology of linear and non-linear Stokes lines to make this distinction extra clear -- but note the perhaps confusing fact that the non-linear Stokes lines are straight lines, whereas the linear Stokes lines are not.}. First of all, note that something interesting happens when $t$ lies on top of the non-linear Stokes line: there, two of the linear Stokes lines (in $\mathbb{C}_x$) merge to form a \textit{saddle connection}, which is a single Stokes line connecting two turning points. This occurs if and only if
\begin{equation}
    \operatorname{Im}\left(\int_{-2\lambda_0}^{\lambda_0}\sqrt{Q_0(x')}\diff x'\right) =  \operatorname{Im}\left(-\frac{24\sqrt{3}}{5}\left(-\frac{t}{6}\right)^{5/4}\right) = 0,
    \label{eq:saddle}
\end{equation}
which produces the five rays in the $\mathbb{C}_t$ plane that represent the nonlinear Stokes lines of the Painlevé I equation. 

\sk
If we now consider three values of $t$ close to one another, one exactly on the non-linear Stokes line and two on either side -- see the bottom half of figure \ref{fig:Wallcrossing} -- then we find the three topologically distinct Stokes graphs depicted in the top of that figure. From these graphs we can read off which linear Stokes lines approach a certain asymptotic direction and contribute to the corresponding Stokes matrix. Therefore, the topology of the Stokes graph will determine the explicit expressions for the monodromy data in terms of parameters $t$, $\lambda(t)$ and $\nu(t)$. This topology abruptly changes when $t$ crosses the locus defined by (\ref{eq:saddle}), and thus new expressions for the Stokes data emerge after crossing those rays. In the end, to describe a single solution to the Painlevé I equation, we fix the monodromy data, i.e.\ we require that the Stokes matrices remain fixed when crossing each non-linear Stokes line, which as we shall see induces a change in the asymptotics of $\lambda(t)$. As we show in this paper, this change is exactly the \textit{non-linear} Stokes phenomenon of the full two-parameter solution to the first Painlevé equation.

\paragraph{Linear Stokes phenomenon}

Besides the topology of the Stokes graph we need to know the Stokes phenomenon across all linear Stokes lines emanating from the simple and double turning points. What happens at the transitions across lines emanating from the simple turning point is well known: locally, the simple turning point looks like that of the Airy equation, for which the Stokes phenomenon is well understood. The Stokes phenomenon around the double turning point, however, is more challenging. Using a series of coordinate transformations \cite{Tak1} one can locally describe the double turning point in terms of a Weber equation. The solutions to this Weber equation are well understood as they are parabolic cylinder functions and so one can extract their Stokes phenomenon and translate it back to the global picture with the full potential $Q(x)$. We will review this procedure in the next subsection, with some further details spelled out in appendix \ref{app:weber}.

\subsubsection{Stokes phenomenon of the double turning point}
In order to obtain explicit expressions for the Stokes multipliers $m_i$ in (\ref{eq:G1}), we need to understand how to analytically continue solutions to the Schrödinger equation (\ref{eq:SL1}) across the Stokes lines that emanate from the double turning point. In this subsection we review how this can be done by a series of transformations \cite{Tak1}.

\sk
The first of these transformations is provided by the \textit{local transformation theorem} -- proved to all orders of $\eta^{-1/2}$ in \cite{AKT3}. The idea is to switch from the coordinate $x$ to a new coordinate $\zeta(x, t, \eta)$, which is an expansion in $\eta^{-1/2}$ and which also depends on $t$, so that we can locally map the potential (\ref{eq:SL2}) to a potential that to leading order is quadratic in the new coordinate $\zeta$ -- i.e.\ any corrections to this potential will be suppressed by higher powers of $\eta^{-1/2}$. Let us paraphrase this proposition that was formulated in \cite{Tak1}:

\begin{prop}[Aoki, Kawai, Takei]  There exists a neighbourhood $U$ of the double turning point $x=\lambda_0$, and a formal series
\begin{equation}
    \zeta(x, t, \eta) = \sum_{j =0}^{\infty}\zeta_{j/2}(x, t)\eta^{-j/2}
    \label{eq:zetaexp}
\end{equation}
whose coefficients $\zeta_{j/2}(x, t)$ are holomorphic in $x\in U$, as well as formal series\footnote{Actually, the coefficients $\zeta_{n/2}$, $E_{n/2}$ and $\rho_{n/2}$ do still depend on $\eta$ through instanton terms $e^{\pm\tau(\eta, t)}$, but this subtlety plays no role in our inquiry.}
\begin{equation}
    \begin{split}
        E(t,\eta)& = \sum_{j=0}^\infty E_{j/2}(t)\eta^{-j/2}\\
        \rho(t,\eta) &= \sum_{j=0}^\infty \rho_{j/2}(t)\eta^{-j/2}
    \end{split}
\end{equation}
 such that the following conditions are satisfied:
\begin{equation}
    \begin{split}
        &\frac{\partial\zeta_0}{\partial x} \neq 0 \\
        &\zeta_0(x, t)\Big|_{x=\lambda_0} = 0\\
        &\zeta_{1/2} = 0,
        \end{split}
    \label{eq:localconditions}
\end{equation}
and such that the potential $Q$ can be expanded as
\be
        Q(x, t, \eta) = \left(\frac{\partial\zeta}{\partial x}\right)^2 \left[ 4\zeta^2+\eta^{-1}E+\frac{\eta^{-3/2}\rho}{\zeta-\eta^{-1/2}\xi}+\frac{3\eta^{-2}}{4(\zeta-\eta^{-1/2}\xi)^2}\right]-\frac{1}{2}\eta^{-2}\{\zeta ; x\}.
        \label{eq:Qdef}
\ee
Here $\{\zeta; x\}$ denotes the usual Schwarzian derivative:
\begin{equation}
    \{\zeta ; x\} = \left(\frac{\frac{\partial^3\zeta}{\partial x^3}}{\frac{\partial\zeta}{\partial x}}\right) -\frac{3}{2}\left(\frac{\frac{\partial^2\zeta}{\partial x^2}}{\frac{\partial \zeta}{\partial x}}\right)^2,
\end{equation}
and we have defined\footnote{Note that $\xi$ is of order $\eta^0$ because $\zeta_0(x)$ evaluated at $x=\lambda_0$ vanishes.} $\xi = \eta^{1/2}\zeta(\lambda(t), t, \eta)$. 
\end{prop}
Using this proposition one can straightforwardly check that a new Schrödinger equation can be constructed for $\varphi \simeq (\frac{\partial\zeta}{\partial x})^{1/2}\psi$ which must satisfy
\begin{equation}
    \left( -\frac{\partial^2}{\partial \zeta^2}+\eta^2 \tilde Q(\zeta)\right) \varphi(\zeta) = 0,
    \label{eq:S1}
\end{equation}
with
\begin{equation}
    \tilde Q(\zeta) =  4\zeta^2+\eta^{-1}E+\frac{\eta^{-3/2}\rho}{\zeta-\eta^{-1/2}\xi}+\frac{3\eta^{-2}}{4(\zeta-\eta^{-1/2}\xi)^2}.
    \label{eq:Qnew}
\end{equation}
We see that the new potential is quadratic with a correction $E$ appearing at order $\eta^{-1} = \hbar$, thus mimicking the harmonic oscillator and its energy.

\sk
One can show \cite{AKT3}  that the series $E(t, \eta)$ is related to the solution $S_{\text{odd}}$ of the Riccati equation associated to the original Schrödinger problem (\ref{eq:SL2}) via
\begin{equation}
    E(t, \eta) = 4 \Res_{x=\lambda_0}S_{odd}.
    \label{eq:Eres}
\end{equation}
Computing the expansions of $E, \rho$ and $\xi$ in $\eta^{-1/2}$ is now a tedious but straightforward exercise. In \cite{AKT3} the following leading terms were computed:
\bea
    E_0 & = & \phantom{+} (3\lambda_0)^{-1/2}(\cN_0^2-12\lambda_0\Lambda_0^2) \ret
    \rho_0 & = & -(3\lambda_0)^{-1/4}\cN_0 \ret
    \xi_0 & = & \phantom{+} (3\lambda_0)^{1/4}\Lambda_0.
    \label{eq:ers}
\eea
For the new differential equation (\ref{eq:S1}), describing the solution locally near the double turning point, we can write out the WKB solutions which, when expanded in $\eta^{-1/2}$, are of the form
\begin{equation}
    \varphi_\pm(\zeta) = \exp(\pm\eta \zeta^2) \; \zeta^{-1/2\pm E/4} \; \Big(1+\mathcal{O}(\eta^{-1/2})\Big).
    \label{eq:localWKBsol}
\end{equation}
Now there is a second transformation that maps the Schrödinger equation (\ref{eq:S1}) to the {\em Weber equation},
\begin{equation}
    \left(\frac{\partial^2 }{\partial y^2}+\kappa+\frac{1}{2}-\frac{y^2}{4}\right)w(y) = 0,
    \label{eq:W1}
\end{equation}
where $\kappa = -E/4-1$. This mapping essentially removes any higher order corrections and is described in more detail in appendix \ref{app:weber}.

\sk
Combining these two transformations, we have managed to rewrite our Schrödinger problem in the form we need. As mentioned before, solutions to (\ref{eq:W1}) are parabolic cylinder functions, whose asymptotics are well known. Hence, we can map their asympotics back to solutions of (\ref{eq:S1}) in order to extract their Stokes phenomenon in the original coordinates. The resulting Stokes transformations for the $\varphi$-solutions are then:
\begin{figure}[ht]
    \centering
    \includegraphics[width=0.7\textwidth]{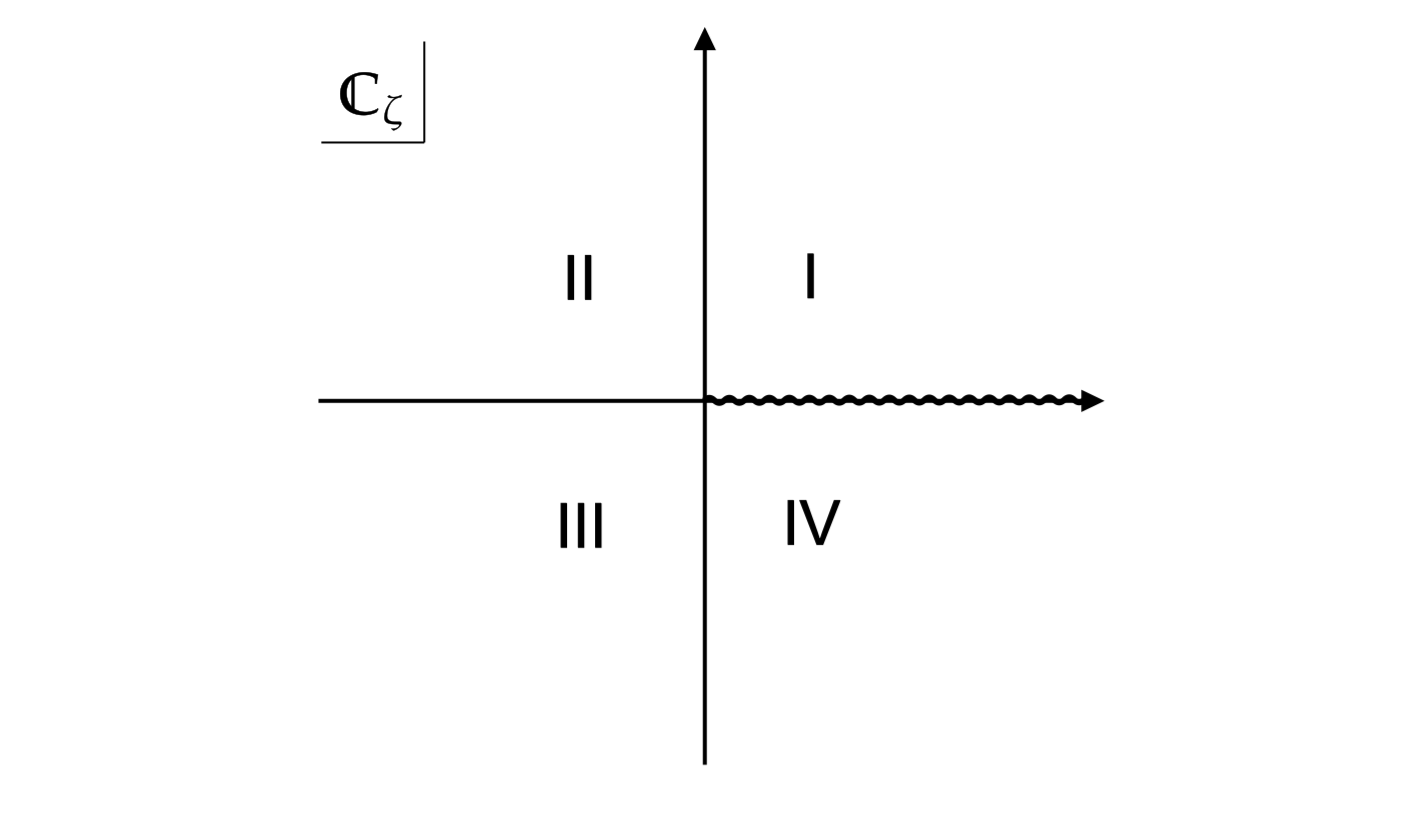}
    \caption{The local Stokes graph for the system (\ref{eq:S1})-(\ref{eq:Qnew}).}
    \label{fig:Weberrnc}
\end{figure}
\begin{equation}
\begin{split}
&\begin{cases}
    \varphi_+^{\rnc{1}} &= \varphi_+^\rnc{2}\\
    \varphi_-^\rnc{1} &= \varphi_-^\rnc{2} +\frac{\rho-2\xi}{2}\frac{\sqrt{2\pi}}{\Gamma(E/4+1)}e^{-i\pi E/4}(2\sqrt{\eta})^{E/2} \; \varphi_+^\rnc{2}
\end{cases} \\
&\begin{cases}
    \varphi_+^\rnc{2} &= \varphi_+^\rnc{3} +\frac{2}{\rho-2\xi}\frac{\sqrt{2\pi}}{\Gamma(-E/4)}e^{i\pi (E+1)/2}(2\sqrt{\eta})^{-E/2} \; \varphi_-^\rnc{3}\\
    \varphi_-^\rnc{2} &= \varphi_- ^\rnc{3}
\end{cases}\\
&\begin{cases}
    \varphi_+^\rnc{3} &= \varphi_+^\rnc{4}\\
    \varphi_-^\rnc{3} &= \varphi_-^\rnc{4} -\frac{\rho-2\xi}{2}\frac{\sqrt{2\pi}}{\Gamma(E/4+1)}e^{-3i\pi E/4}(2\sqrt{\eta})^{E/2} \; \varphi_+^\rnc{4},
\end{cases}
\end{split}
\label{eq:phiStokes}
\end{equation}
where a branchcut lies on top of the axis separating regions $\rnc{4}$ and $\rnc{1}$. These solutions $\varphi_\pm(\zeta)$ are then connected to the solutions $\psi_\pm(x)$ of (\ref{eq:SL1}) in the following way:
\begin{equation}
    \psi_\pm(x) = C_\pm(t, \eta) \left(\frac{\partial \zeta}{\partial x}\right)^{-1/2}\varphi_\pm(\zeta),
    \label{eq:psiphiconnection}
\end{equation}
where the $C_\pm(t, \eta)$ are $x$-independent formal series in $\eta^{-1/2}$ that can be calculated recursively. Explicitly writing out both solutions $\psi_\pm(x)$ and $\varphi_\pm(\zeta)$ and taking into account the expansion for $\zeta(x)$ from (\ref{eq:zetaexp}), we can find the leading terms
\begin{equation}
    C_\pm(t, \eta)= \exp\left(\mp \frac{24\sqrt{3}}{5}\lambda_0^{5/2} \eta \right) \left(4(3\lambda_0)^{5/4}\right)^{\mp E_0/4}e^{\mp i \pi E_0/4}(1+\mathcal{O}(\eta^{-1/2})).
    \label{eq:Cpm}
\end{equation}
Note that this expression differs from \cite{Tak1} by the factor $\exp(\mp i\pi E_0/4)$. In Appendix \ref{app:matching} we give a derivation of the above result for $C_\pm(t,\eta)$ and clarify this discrepancy. 

\sk
Finally, we have reached the point where we can present the Stokes phenomenen around the double turning point. Transforming back the result from (\ref{eq:phiStokes}) into expressions for $\psi_\pm(x)$, we find
 \begin{equation}
     \begin{split}
     &\begin{cases}
          \psi_+^{\rnc{1}} &= \psi_+^\rnc{2}\\
    \psi_-^\rnc{1} &= \psi_-^\rnc{2} +a_{12}\psi_+^\rnc{2}
    \end{cases}\\
    &\begin{cases}
    \psi_+^{\rnc{2}} &= \psi_+^\rnc{3} +a_{23}\psi_+^\rnc{3}\\
    \psi_-^\rnc{2} &= \psi_-^\rnc{3} 
    \end{cases}\\
    &\begin{cases}
    \psi_+^{\rnc{3}} &= \psi_+^\rnc{4}\\
    \psi_-^\rnc{3} &= \psi_-^\rnc{4} +a_{34}\psi_+^\rnc{4}
    \end{cases}
     \end{split}
     \label{eq:G2}
 \end{equation}
 where
 \begin{equation}
    \begin{split}
        a_{12} &= \phantom{-} \frac{C_-(t,\eta)}{C_+(t,\eta)} \frac{\rho-2\xi}{2}\frac{\sqrt{2\pi}}{\Gamma(1+E/4)} e^{-i\pi E/4}(2\sqrt{\eta})^{E/2}\\
        a_{23} &= \phantom{-} \frac{C_+(t,\eta)}{C_-(t,\eta)} \frac{\rho+2\xi}{2}\frac{i\sqrt{2\pi}}{\Gamma(1-E/4)} e^{i\pi E/2}(2\sqrt{\eta})^{-E/2}\\
        a_{34} &= -\frac{C_-(t,\eta)}{C_+(t,\eta)} \frac{\rho-2\xi}{2}\frac{\sqrt{2\pi}}{\Gamma(1+E/4)} e^{-3i\pi E/4}(2\sqrt{\eta})^{E/2}.
    \end{split}
    \label{eq:aij}
\end{equation}
We call these $a_{ij}$ \textit{connection coefficients} and they will play an important role in fully expressing the monodromy data in terms of the Painlevé I solution, as we do in the next section.

\subsubsection{Stokes multipliers}

\begin{figure}[ht]
     \centering
     \includegraphics[width=1 \textwidth]{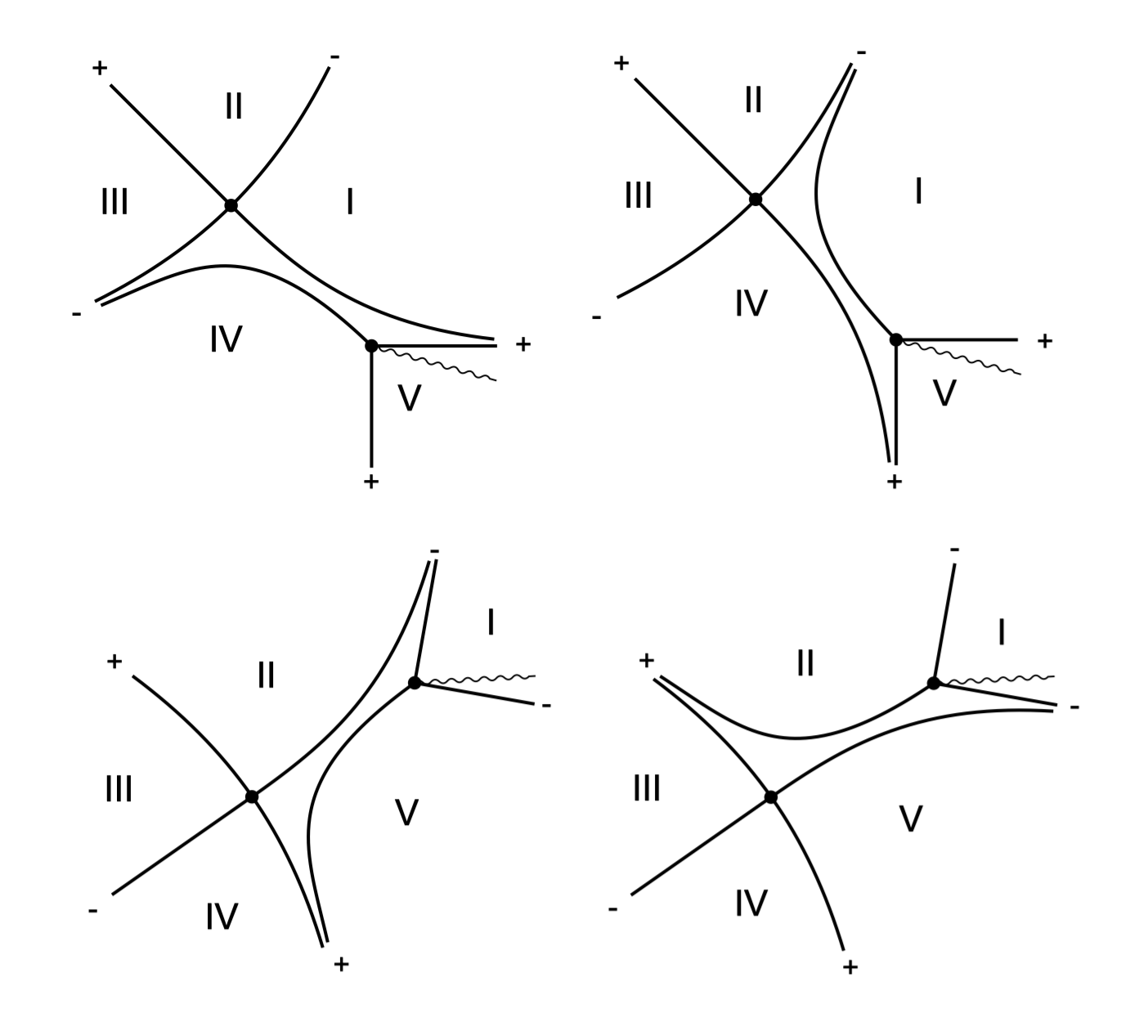}
     \caption{The Stokes graphs before (left) and after (right) the Stokes transitions at $\arg(t) = 3\pi/5$ (top) and $\arg(t) = 7\pi/5$ (bottom). The labels $+$ and $-$ denote which of the two $\psi_\pm$ solutions is dominant on that particular Stokes line. The curly line denotes a branch cut. The top left graph lies in the $\cS_1$ sector, the top right and bottom left graphs in the $\cS_2$ sector (hence they have the same topology) and the botom right graph in the $\cS_3$ sector.}
     \label{fig:4sgraphs}
\end{figure}

Having derived the Stokes transformations (\ref{eq:G2})-(\ref{eq:aij}) around the double turning point, we can now derive explicit expressions for the monodromy data (\ref{eq:G1}) of our entire linear problem, which allow us to establish connection formulae for the transseries expansions. In order to obtain the full monodromy data, we consider -- for every non-linear Stokes transition in the $\bC_t$-plane -- three adjacent linear Stokes regions in the $\bC_x$-plane. Following the original computation of Takei \cite{Tak1}, we start with the non-linear Stokes transition that occurs at $\arg(t) = 3\pi/5$ and consider the upper two graphs displayed in figure \ref{fig:4sgraphs}. The left and right Stokes graphs in that figure show the situation before and after the non-linear Stokes transition. We consider in this case the two consecutive connection formulae that occur in the counterclockwise direction between linear solutions of regions $\rnc{1}$, $\rnc{2}$ and $\rnc{3}$. The reason for considering only two connection formulae is that these provide two Stokes multipliers that, as discussed in section \ref{sec:mono}, fix all the other Stokes multipliers. We can express these two multipliers in terms of $\rho, \xi$ and $E$, which all depend on the Painlevé I solution $\lambda(t)$ (cf. equations (\ref{eq:ers}) and (\ref{eq:aij})). Therefore, once we have examined two consecutive transitions in the linear problem, we have in principle fixed its complete monodromy data in terms of the Painlevé I solution.

\sk
The transition that we are describing here occurs at  $\arg(t) = 3\pi/5$, which is equivalent to $\arg(z)= 8\pi/5$, or in terms of the 'inverse Écalle time'\footnote{We use the inverse of the Écalle time because in the true Écalle time, $z^{-5/4}$, we are rotating clockwise which is opposite to the direction of rotation in the original complex $t$-plane.} $\arg\left(z^{5/4}\right) = 2\pi$. We use a terminology related to the latter convention and therefore call this transition the \textit{$2\pi$-transition}. Subsequently, what we can do is keep rotating $t$ counterclockwise in the complex $\bC_t$-plane until we reach the next non-linear Stokes line at $\arg(t) = 7\pi/5$ or $\arg\left(z^{5/4}\right) = 3\pi$, at which point the \textit{$3\pi$-transition} occurs, which will be the second transition that we study. The corresponding Stokes graphs now look like the ones displayed at the bottom left and bottom right of figure \ref{fig:4sgraphs} (displaying the situation before and after crossing the non-linear Stokes line respectively). In this case we consider the connection formulae between solutions from regions $\rnc{2}$, $\rnc{3}$ and $\rnc{4}$. It will be convenient to introduce the (double) sectors $\cS_k$ where $k<\arg(z^{5/4})<k+1$ or equivalently $\frac45 k-\pi < \arg(t) < \frac45 (k+1)-\pi$. With these definitions we see that the $2\pi$-transition separates the $\cS_1$- and $\cS_2$-sectors, and that the $3\pi$-transition separates the $\cS_2$- and $\cS_3$-sectors, etc.

\sk
To fully construct the monodromy data of the linear problem, on top of the data for the double turning point that we have now extensively studied, we also need the mappings across Stokes lines connected to the simple turning point. These transitions are much simpler (see e.g. \cite{Iwa2}) and are known to be, in the basis $(\psi_+, \psi_-)^T$ and in the counterclockwise direction, the following:
\begin{equation}
    \begin{pmatrix}
         1 & i\\
         0 & 1
    \end{pmatrix}
    \hspace{1cm} \text{or} \hspace{1cm}
    \begin{pmatrix}
         1 & 0\\
         i & 1
    \end{pmatrix},
    \label{Airymat}
\end{equation}
depending on whether $\psi_+$ is the dominant solution (labeled $+$ in Figure \ref{fig:4sgraphs}) or $\psi_-$ is dominant (labeled $-$), respectively\footnote{Note our conventions: $\psi_+$ is the solution that starts out as the dominant one along the positive real axis, but we keep that same label all throughout the complex $x$-plane, so at certain anti-Stokes lines dominance switches and $\psi_-$ will become the dominant solution.}. 

\sk
To connect this result to the expressions for the double turning point, however, we need to deal with one more subtlety: following the conventions of \cite{Tak1}, we consider solutions \textit{normalized at infinity}:
\begin{equation}
    \psi_\pm(x) \sim \exp\left(\pm\eta \int_{-2\lambda_0}^x S_{-1}\diff x' \pm \int_\infty ^x (S_{\text{odd}}-\eta S_{-1})\diff x'\right),
    \label{eq:AiryStokes}
\end{equation}
where the normalization is encoded in the lower boundaries of the integrals. This normalization matches the one used in studying the double turning points before. However, the matrices  (\ref{Airymat}) are valid for solutions normalized at the turning point. Hence, whenever we cross a Stokes line that emanates from the simple turning point, we first adjust the normalization to that simple turning point, then cross the Stokes line using the transitions in (\ref{Airymat}), and finally return to the original normalization at infinity. When crossing the simple Stokes line between regions $\rnc{1}$ and $\rnc{2}$ in the upper right panel of Figure \ref{fig:4sgraphs}, for example, this amounts to the consecutive transformations
\begin{equation}
    \begin{pmatrix}
            \cV & 0 \\
            0 & \cV^{-1}
    \end{pmatrix}
     \begin{pmatrix}
            1 & i \\
            0 & 1
    \end{pmatrix}
     \begin{pmatrix}
            \cV^{-1} & 0 \\
            0 & \cV
    \end{pmatrix}
    = 
     \begin{pmatrix}
            1 & i\cV^2 \\
            0 & 1
    \end{pmatrix}
\end{equation}
where 
\begin{equation}
    \begin{split}
    \cV :=& \exp\left(\int_{-2\lambda_0}^\infty (S_{\text{odd}}-\eta S_{-1})\diff x\right)\\
    =& \exp\left(\frac{1}{2} \oint_{\lambda_0}(S_{\text{odd}}-\eta S_{-1})\diff x\right)\\
    =&\exp\left(i\pi E/4\right)
    \end{split}
\end{equation}
is also known as a \textit{Voros symbol}. The second equality comes from doubling the integration contour along both sides of the branch cut and then closing it at infinity and deforming, and the third equality from using (\ref{eq:Eres}). Plugging this in, we see that crossing a Stokes line of a simple turning point counterclockwise along a plus or minus direction amounts to a transformation
\begin{equation}
     \begin{pmatrix}
            1 & i\exp\Big(-i \pi E/2) \\
            0 & 1
    \end{pmatrix}
    \hspace{5mm} \text{or} \hspace{5mm} 
     \begin{pmatrix}
            1 & 0\\
            i\exp\Big(i \pi E/2) & 1
    \end{pmatrix},
    \label{eq:G3}
\end{equation}
respectively. Including these matrices we can now write down the exact expressions for the Stokes multipliers related to our two chosen non-linear Stokes phenomena. Combining (\ref{eq:G1}), (\ref{eq:G2}) and (\ref{eq:G3}) we have
\begin{equation}
    2\pi\text{-direction}\hspace{5mm} \cS_1:
    \begin{cases}
    m_{1} = a_{12}\\
    m_{2} = a_{23}
    \end{cases}
     \hspace{1cm} \cS_2:
     \begin{cases}
    m'_{1} = a_{12}+i\exp\Big(i\pi E/2\Big)\\
    m'_{2} = a_{23}
    \end{cases}
    \label{eq:M1}
\end{equation}
\begin{equation}
    \hspace{4mm}
    3\pi\text{-direction}\hspace{5mm} \cS_2:
    \begin{cases}
    m_{2} = a_{23}\\
    m_{3} = a_{34}
    \end{cases}
     \hspace{10mm}  \cS_3: 
     \begin{cases}
    m'_{2} = a_{23}+i\exp\Big(-i\pi E/2\Big)\\
    m'_{3} = a_{34}
    \end{cases}
    \label{eq:A2}
\end{equation}
with the $a_{jk}$ as given in (\ref{eq:aij}). Here, the $m_{k}$ are the Stokes multipliers corresponding to crossing from region $k$ to region $k+1$ {\em before} the non-linear Stokes transition occurs, and the $m_k'$ the multipliers for the same crossings {\em after} the non-linear Stokes transition has occured.

\section{Stokes phenomena for 2-parameter transseries}
\label{sec:Stok2par}
In section \ref{sec3.1}, using the multiple-scale analysis of \cite{AKT3}, we learned that we can construct a transseries expansion for the Painlevé I solution $\lambda(t)$ parametrized by two free parameters that were called $\alpha$ and $\beta$. In this section we want to show that for each of the two Stokes transitions that we studied in the previous section, we can express a pair of Stokes multipliers $(m_{i}, m_{i+1})$ purely in terms of transseries parameters $(\alpha, \beta)$. We can accomplish this on both sides of the non-linear Stokes line, and as a result we can construct a formula that connects the transseries on one side, defined by parameters $(\alpha, \beta)$, to the transseries on the other side, defined by $(\alpha', \beta')$. Such connection formulae were first constructed in \cite{Tak4, KT3} for two-parameter transseries and in \cite{Iwa1} for the 2-parameter tau functions. We will derive our formulae following \cite{Tak1} with a slightly different normalization that fits our 'stringy conventions' -- up to a rescaling of the transseries parameter -- and which allows for an easy and straightforward generalization to arbitrary Stokes transitions.
\sk
As formulated above, the result holds only for the specific two nonlinear Stokes transitions that occur in the $\bC_z$-plane\footnote{So far we have worked mostly with the original variables of \cite{Tak1}, i.e. $t$ and $\lambda_0 = \sqrt{-t/6}$. Since our main objective is studying transseries solutions to  (\ref{eq:P1}) we now switch to the more convenient variables $z$ and the inverse Écalle time $z^{5/4}$.} at $\arg(z^{5/4})= 2\pi$ and $\arg(z^{5/4})= 3\pi$, but we can easily generalise our result to arbitrary transitions, which is what we do afterwards. We show that all the different Stokes transitions are related to each other using a transformation $T$ that we introduce shortly, and that all transitions along different nonlinear Stokes lines are essentially described by only two distinct Stokes automorphisms.

\subsection{Connection formulae for two Stokes transitions}
\label{sec:connect}
Let us first establish connection formulae for our first Stokes transition, at $\arg(z^{5/4})= 2\pi$. We start from the situation shown in the upper left panel of figure \ref{fig:4sgraphs}. From equations (\ref{eq:AKTtrans}) and (\ref{eq:ers}) we see that we can write
\begin{equation}
    \left(\alpha \lambda_0^{5\alpha\beta}e^{\tau}+\beta \lambda_0^{-5\alpha\beta}e^{-\tau}\right) = (12\lambda_0)^{1/4}\Lambda_0 = \frac{1}{2\sqrt{2}}\bigg( (\rho_0+2\xi_0)-(\rho_0-2\xi_0)\bigg).
    \label{eq:G4}
\end{equation}
This rewriting is convenient: we can now take the first two relations of (\ref{eq:aij}) to express $\rho_0 \pm 2 \xi_0$ in terms of the connection coefficients $a_{ij}$, and then use (\ref{eq:M1}) to further rewrite that expression in terms of the Stokes multipliers $m_i$, leading to
\begin{equation}
    \begin{split}
        \alpha \lambda_0^{5\alpha \beta}e^\tau &= -\frac{m_{1}}{2\sqrt{\pi}}\frac{C_+}{C_-}e^{i\pi E_0/4}(2\eta)^{-E_0/2} \Gamma(1+E_0/4) \cdot \left(1+\cO(\eta^{-1/2})\right),\\
        \beta \lambda_0^{-5\alpha \beta}e^{-\tau}  &= -\frac{im_{2}}{2\sqrt{\pi}}\frac{C_-}{C_+}e^{-i\pi E_0/2}(2\eta)^{E_0/2} \Gamma(1-E_0/4) \cdot \left(1+\cO(\eta^{-1/2})\right).
    \end{split}
\end{equation}
The single equation(\ref{eq:G4}) has now split into  two separate equations because on both sides we find terms that either carry an instanton transmonomial $e^{\tau}$ or its inverse $e^{-\tau}$. On the right hand side, these factors are hidden in the prefactors $C_+/C_-$ and $C_-/C_+$ respectively, as is shown in appendix \ref{app:matching}. From (\ref{eq:Cpm}), we take the expressions for $C_\pm$ to leading order in $\eta^{-1/2}$ and plug them in to find
\begin{equation}
    \begin{split}
        \alpha \lambda_0^{5\alpha \beta}  &= -\frac{m_{1}}{2\sqrt{\pi}}(64\eta(3\lambda_0)^{5/2})^{-E_0/4} e^{-i\pi E_0/4}\Gamma(1+E_0/4) \cdot \left(1+\cO(\eta^{-1/2})\right),\\
        \beta \lambda_0^{-5\alpha \beta}  &= -\frac{im_{2}}{2\sqrt{\pi}}(64\eta(3\lambda_0)^{5/2})^{E_0/4}  \Gamma(1-E_0/4) \cdot \left(1+\cO(\eta^{-1/2})\right).
    \end{split}
    \label{eq:ab}
\end{equation}
Finally, by multiplying both equations ($\ref{eq:ab}$) with one another one finds the rather simple relation
\begin{equation}
    \alpha \beta = -\frac{E_0}{8}.
    \label{eq:G5}
\end{equation} 
Any potential $\cO(\eta^{-1/2})$ terms in this result vanish, by definition of the free parameters $\alpha$ and $\beta$: in solving the equations (\ref{Lsystem}) one finds that the product $\alpha \beta$ must be an $\eta$-independent constant \cite{AKT3} and so indeed the $\eta^{-1/2}$-corrections need to vanish. Plugging (\ref{eq:G5}) back into (\ref{eq:ab}) leads to 
\begin{equation}
    \begin{split}
        \alpha  &= -\frac{m_{1}}{2\sqrt{\pi}}(576\sqrt{3}\eta)^{-E_0/4} e^{-i\pi E_0/4}\Gamma(1+E_0/4),\\
        \beta &= -\frac{im_{2}}{2\sqrt{\pi}}(576\sqrt{3}\eta)^{E_0/4} \Gamma(1-E_0/4).
    \end{split}
    \label{eq:A1}
\end{equation}
Here, again the $\cO(\eta^{-1/2})$ corrections need to vanish in order for the transseries solution (\ref{eq:AKTtrans}) to be self-consistent: allowing for such corrections would break the ordering of the equations (\ref{Lsystem}) that allowed us to solve for the transseries in the first place, and would lead to a redefinition of the $\Lambda_{n/2}$. Therefore, we can require that just like the product $\ga \gb$, also the parameters $\alpha$ and $\beta$ themselves are free of $\cO(\eta^{-1/2})$ corrections. 

\sk
Note that there {\em is} still an overall $\eta$-dependence in the expressions (\ref{eq:A1}), but such a single factor does not change the structure of (\ref{Lsystem}) and therefore is not in contradiction with our previously made assumptions. As before, we now have the choice to set the $\eta$-parameter to any particular value we like, in particular the value $\eta=1/6$ that corresponds to the `stringy' conventions in e.g. \cite{ASV1}. When we plug this value into (\ref{eq:A1}), we finally obtain
\setlength{\fboxsep}{10pt}
\begin{equation}
 \boxed{\begin{aligned}
        \alpha  &= -\frac{m_{1}}{2\sqrt{\pi}}(96\sqrt{3})^{-E_0/4} e^{-i\pi E_0/4}\Gamma(1+E_0/4)\\
        \beta &= -\frac{im_{2}}{2\sqrt{\pi}}(96\sqrt{3})^{E_0/4} \Gamma(1-E_0/4).
    \end{aligned}
    \label{eq:A3}}
\end{equation}
We can invert this relation to express the Stokes multipliers $m_1$ and $m_2$ in terms of transseries parameters $\alpha$ and $\beta$, which in fact gives a slightly more elegant result as we can now re-express $E_0$ in terms of $\ga \gb$ using (\ref{eq:G5}):
\begin{equation}
 \boxed{
    \begin{aligned}
        m_{1} =& -\frac{2\alpha \sqrt{\pi}(96\sqrt{3})^{-2\alpha \beta}}{\Gamma(1-2\alpha \beta)}e^{-2\pi i \alpha \beta} \\
        m_{2} =& \frac{2i\beta \sqrt{\pi}(96\sqrt{3})^{+2\alpha \beta}}{\Gamma(1+2\alpha \beta)}.
    \end{aligned}
    \label{eq:mij}
 }
\end{equation}
Jumping slightly ahead of our story, a glance at these expressions already indicates that the relation between transseries parameters $\alpha$, $\beta$ and Stokes multipliers $m_1$, $m_2$ may not be one-to-one. Moreover, due to the gamma functions in the denominators, the Stokes multipliers $m_1$ and $m_2$, as functions of the product $\alpha \beta$, appear to have singularities evenly spaced along the real line at nonzero integer values of $2\alpha\beta$, meaning that at certain values of $\ga$ and $\gb$ one might expect the transseries to be ill-defined as an asymptotic description of a true Painlevé I solution. We discuss this and further peculiar properties of the above equations in section \ref{sec:results}.

\sk
Before doing that, to arrive at our goal of expressing the Stokes transition as an operation on the transseries parameters $\ga$ and $\gb$, we still need to repeat this computation for the Stokes multipliers $(m'_{1},m'_{2})$, defined {\em after} the transition at $\arg(t) = 3\pi/5$. (The upper right panel in figure \ref{fig:4sgraphs}.) The computation is essentially the same with some minor straightforward changes, and leads to 
\begin{equation}
    \begin{aligned}
        \alpha' =& \frac{-m'_{1}+i e^{i \pi  E'_0/2}}{2\sqrt{\pi}}\big(96\sqrt{3}\big)^{-E'_0/4} e^{ - i\pi E'_0/4} \Gamma(1+E'_0/4) \\
        \beta' =& -\frac{i m'_{2}}{2\sqrt{\pi}}\big(96\sqrt{3}\big)^{E'_0/4}  \Gamma(1-E'_0/4)
    \end{aligned}
    \label{eq:ab2}
\end{equation}
as the counterpart of (\ref{eq:A3}), where now  $E'_0= -8\alpha'\beta'$ analogous to (\ref{eq:G5}). We can again invert this expression and obtain
\begin{equation}
    \begin{aligned}
        m'_{1} =& -\frac{2\alpha' \sqrt{\pi}(96\sqrt{3})^{-2\alpha' \beta'}}{\Gamma(1-2\alpha' \beta')}e^{-2\pi i \alpha' \beta'} + i e^{-4\pi i \alpha' \beta'} \\
        m'_{2} =& \frac{2i\beta' \sqrt{\pi}(96\sqrt{3})^{+2\alpha' \beta'}}{\Gamma(1+2\alpha' \beta')}.
    \end{aligned}
\end{equation}
Note that these expressions for the multipliers (as well as (\ref{eq:mij})) are in agreement with those found in \cite{Tak4} when adjusted for the normalization\footnote{Our expressions for $(m_1, m_2)$ in the sectors $\cS_1$ and $\cS_2$ match those of $(S_2, S_3)$ in \cite{Tak4} when the function $\chi$ given in that paper is reformulated as $\chi(E) = 2\sqrt{\pi} (96\sqrt{3})^{E/4} e^{i\pi E/2}/\Gamma(1+E/4)$.}.

\sk
Finally, to connect the pair $(\alpha', \beta')$ to the pair $(\alpha, \beta)$, we fix the monodromy data, i.e.\ we impose that $m_{i} = m'_{i}$. As discussed before, according to the theory of isomonodromic deformations, this requirement forces the transseries parameters on both sides of the transition to describe the asymptotics of the same Painlevé I solution $\lambda(t)$.  The final missing link is then an expression for $E_0'$ in terms of the original variables $\alpha$ and $\beta$. Such an expression is easily found by multiplying equations (\ref{eq:ab2}) with one another, giving 
\begin{equation}
        -\frac{E'_0}{8}= \frac{im_{1} m_{2}+m_{2}e^{\pi i E'_0/4}}{4 \pi}e^{-\pi i E_0'/4}\Gamma(1+E_0'/4)\Gamma(1-E'_0/4). 
\end{equation}
With equations (\ref{eq:mij}) and some reshuffling we finally obtain
\bea
     e^{i \pi  E'_0/2} & = & \frac{e^{-4\pi i \alpha\beta}}{1-2\beta \sqrt{\pi}(96\sqrt{3})^{2\alpha\beta}/\Gamma(1+2\alpha\beta)} \nonumber \\[0.3cm]
     & = & \frac{1+m_{1}m_{2}}{1+i m_{2}}.
     \label{eq:E}
\eea
A similar relation was also found in \cite{KT3}. Putting everything together, the equations that we have derived now establish a connection formula\footnote{Here we give more of a recipe than a single explicit formula. Such a formula, though one could now in principle write it down without a problem, would be a dreadfully long expression due to all the logarithms and gamma functions. Rather, at the end of this section we formulate a general expression for global transseries parameters.} for the transseries parameters on either side of the Stokes line. First, equation (\ref{eq:mij}) computes the Stokes multipliers in terms of $(\ga,\gb)$. Subsequently, equation (\ref{eq:E}) allows one to obtain an expression for $E'_0$, which finally can be plugged into equations (\ref{eq:ab2}) to obtain the new transseries parameters $(\alpha', \beta')$. Evidently, in this procedure there is a branch choice ambiguïty in the last step, also noted in \cite{KT3}, since one needs to take a logarithm in (\ref{eq:E}) to get $E'_0$. We address this subtlety in section \ref{sec:multiplesolutions}.

\sk
As a very first check, we can easily see that the formulae we have found are consistent with the familiar Stokes phenomenon of a plain formal series solution (or \textit{zero-parameter transseries} solution), where $\alpha = \beta = 0$. In this case, we see from (\ref{eq:mij}) that both Stokes multipliers $m_1$ and $m_2$ are also zero, and we find the mapping
\begin{equation}
    (\alpha, \beta) = (0, 0) \qquad \longmapsto \qquad (\alpha', \beta')=\left(\frac{i}{2\sqrt{\pi}}, 0\right).
\end{equation}
which Takei also derived in his original calculation \cite{Tak1}. When we translate these transseries parameters back to the $(\sigma_1, \sigma_2) = -3^{1/4}(\alpha, \beta)$ used in \cite{ASV1}, then this is equivalent to 
\begin{equation}
    (\sigma_1, \sigma_2) = (0, 0) \longmapsto (\sigma_1', \sigma_2')=(S_1, 0),
\end{equation}
with the familiar Stokes constant 
\begin{equation*}
    S_1 = -\frac{3^{1/4}}{2\sqrt{\pi}}\;i
\end{equation*}
that has been computed via several different approaches in e.g.\ \cite{Kap1, Kap2, Dav1, Tak1, Cos4, Cos5}.

\sk
Taking this one step further, we can also consider the explicit form of the transition of a one-parameter transseries across the Stokes line. In the case that $\alpha$ is nonzero the transition is rather familiar:

\be
    (\alpha, 0) \qquad \longmapsto \qquad \left(\alpha+\frac{i}{2\sqrt{\pi}}, 0 \right),
\ee
and we have $m_1 = -2\alpha \sqrt{\pi}$ and $m_2 = 0$. The more non-trivial case is the one where $\beta$ is non-zero. If we choose the more convenient parametrization $\beta = -\frac{i m }{2\sqrt{\pi}}$, then we learn that

\be
 (0, -\frac{i m}{2  \sqrt{\pi}}) \mapsto \left( \frac{i(96\sqrt{3})^{\frac{\log(1+im)}{2\pi i}}\Gamma(1-\frac{\log(1+im)}{2\pi i})}{2\sqrt{\pi(1+im)}}, \frac{(96\sqrt{3})^{-\frac{\log(1+im)}{2\pi i}}\Gamma(1+\frac{\log(1+im)}{2\pi i})}{2\sqrt{\pi}} \right),
\ee
where the Stokes multipliers are $m_1 = 0$ and $m_2 =m$. Lastly, one can consider the transition of a general two-parameter transseries, but this connection is an overly complicated expression that we refrain from writing out completely. Rather, in section \ref{sec:GST} we formulate a general expression for the transseries parameters computed directly form the Stokes multipliers.

\sk
With the computations for the Stokes phenomenon at the $2\pi$-direction complete, let us now also establish the connection formulae for the Stokes phenomenon across the $3\pi$-direction. We are now dealing with multipliers as described by (\ref{eq:A2}), corresponding to the lower two panels of figure \ref{fig:4sgraphs}. The computation is of course very similar to that for the $2\pi$-transition, and leads to the following expressions for the transseries parameters\footnote{To not overly clutter notation, we again use $(\ga,\gb)$ before the transition and $(\ga',\gb')$ afterwards, rather than starting from $(\ga',\gb')$ and transitioning to e.g.\  $(\ga'',\gb'')$}:
\begin{equation}
    \begin{aligned}
        \beta &= -\frac{im_{2}}{2\sqrt{\pi}}(96\sqrt{3})^{E_0/4} \Gamma(1-E_0/4)\\
        \alpha  &= \frac{m_{3}}{2\sqrt{\pi}}(96\sqrt{3})^{-E_0/4} e^{i\pi E_0/4}\Gamma(1+E_0/4),
    \end{aligned}
    \label{eq:A4}
\end{equation}
where we chose to keep the order of $m_2, m_3$ and therefore switched the order of $\ga, \gb$ -- or when inverted
\begin{equation}
    \begin{aligned}
        m_{2} =& \frac{2i\beta \sqrt{\pi}(96\sqrt{3})^{+2\alpha \beta}}{\Gamma(1+2\alpha \beta)} \\
        m_{3} =& \frac{2\alpha \sqrt{\pi}(96\sqrt{3})^{-2\alpha \beta}}{\Gamma(1-2\alpha \beta)}e^{2\pi i \alpha \beta} .
    \end{aligned}
    \label{eq:mjk}
\end{equation}
 When we repeat the procedure of fixing the monodromy data and expressing the transseries parameters after the transition in terms of the parameters before, this leads to the following formulae:
\begin{equation}
    \begin{aligned}
        \beta' &= -\frac{im_{2}+e^{-i\pi E'_0/2}}{2\sqrt{\pi}}(96\sqrt{3})^{E'_0/4} \Gamma(1-E'_0/4) \\
        \alpha'  &= \frac{m_{3}}{2\sqrt{\pi}}(96\sqrt{3})^{-E'_0/4} e^{i\pi E'_0/4}\Gamma(1+E'_0/4)
    \end{aligned}
    \label{eq:ab3}
\end{equation}
where now $E'_0$ can be calculated via the relation
\begin{equation}
    e^{-i \pi E'_0/2} = \frac{e^{4\pi i \alpha\beta}}{1+2i\alpha \sqrt{\pi}(96\sqrt{3})^{-2\alpha\beta}e^{2\pi i \alpha \beta}/\Gamma(1-2\alpha\beta)} = \frac{1+m_{2}m_{3}}{1+i m_{3}}.
    \label{eq:E2}
\end{equation}
Of course, the differences between the results (\ref{eq:A3})-(\ref{eq:E}) for the $2\pi$ direction and (\ref{eq:A4})-(\ref{eq:E2}) for the $3\pi$ direction are small -- essentially only signs, factors of $i$ and integer powers of $\exp(i\pi E_0)$ -- but those subtle differences will become important once we start combining consecutive Stokes phenomena, which is the subject we now turn to.

\subsection{General Stokes transitions}
\label{sec:GST}
 In order to study the full monodromy of transseries solutions to the first Painlevé equation, we need connection formulae for Stokes transitions in all possible directions. Let us first introduce some notation: let $\mS_{2\pi}$ and $\mS_{3\pi}$ denote the Stokes automorphisms across the Stokes lines that we studied in the previous subsection, corresponding to the $2\pi$- and $3\pi$-transitions.  These and other automorphisms act on transseries solutions $\lambda_{\alpha, \beta}(t)$ described by particular parameters $(\alpha, \beta)^T$ as
 \begin{equation}
     \begin{pmatrix}
          \alpha' \\
          \beta'
     \end{pmatrix} = \mS_{n\pi}
     \begin{pmatrix}
           \alpha\\
          \beta
     \end{pmatrix},
 \end{equation}
for some $n\in \bZ$, where we stick to our convention that $\alpha'$ and $\beta'$ denote the transseries parameters after the transition and $\alpha$ and $\beta$ those before. From our computation in the previous subsection, we know what  $\mS_{2\pi}$ and $\mS_{3\pi}$ are, and we wish to generalise these to arbitrary $\mS_{n\pi}$. 
\begin{figure}
    \centering
    \includegraphics[width = 1 \textwidth]{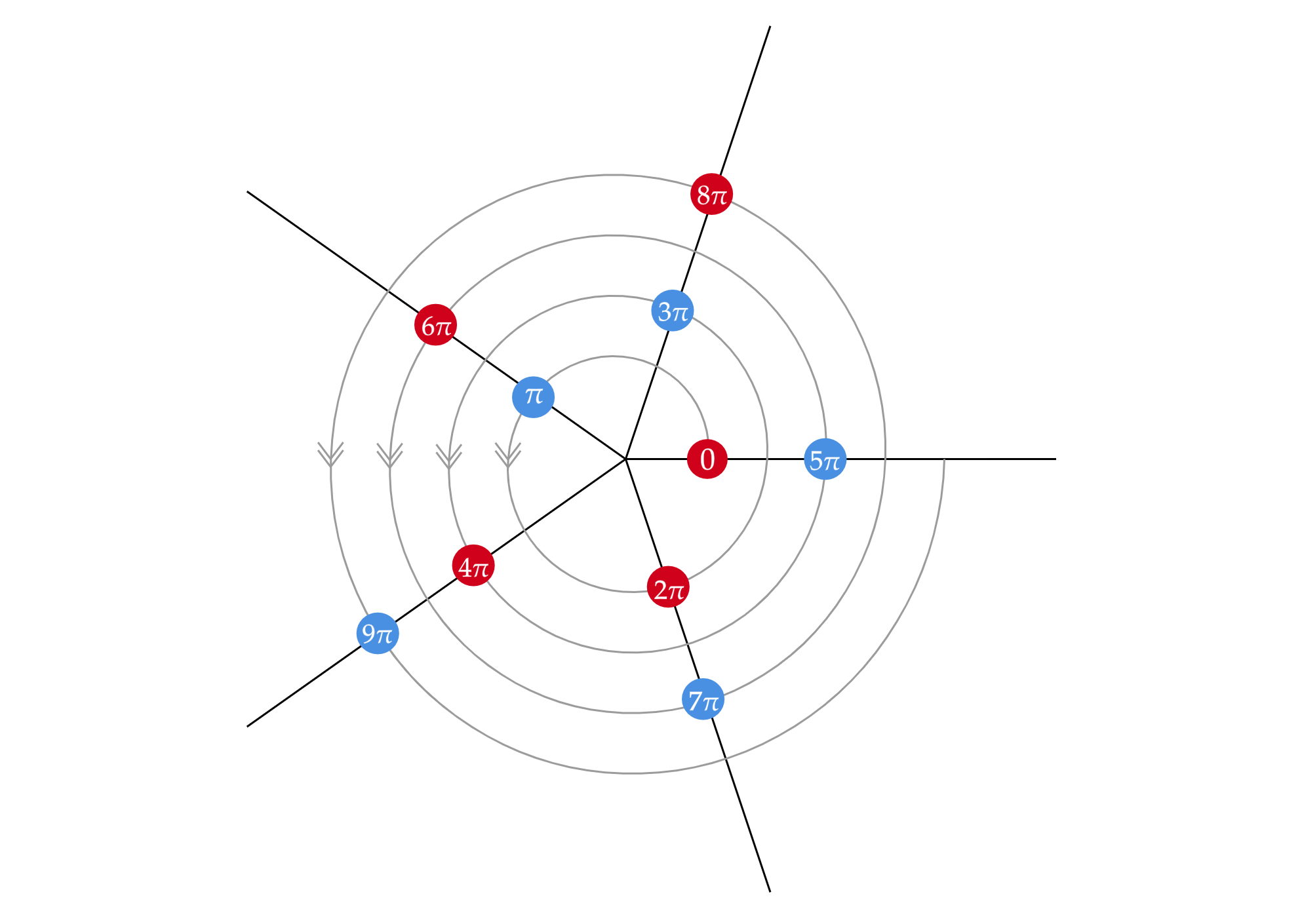}
    \caption{The monodromy as a composition of consecutive Stokes transitions $\mS_{n\pi}$ in the $\bC_z$-plane. In red and blue we indicate those transitions that are related to $\underline{\mS}_0$ and $\underline{\mS}_\pi$ respectively. After 10 transitions, the Painlevé I transseries solution is back to its original form up to an application of the map $T^5$ -- see equation (\ref{eq:B1}).}
    \label{fig:monodromy}
\end{figure}
\begin{figure}
     \centering
     \includegraphics[width=1 \textwidth]{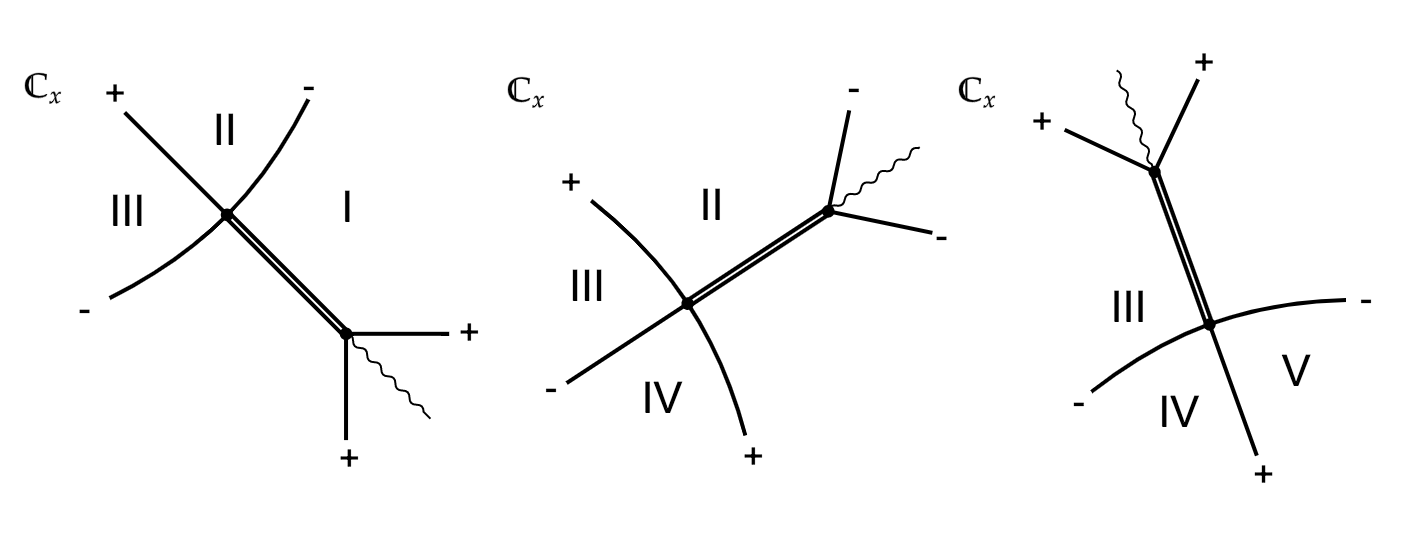}
     \caption{The degenerate Stokes graphs that are encountered when we cross the $2\pi$-, $3\pi$- and $4\pi$-transitions (from left to right) with for each transition the relevant asymptotic regions $\rnc{1}$ up to $\rnc{5}$ indicated. The wavy line denotes the branch cut emanating from the simple turning point. Note that the left and right diagram are identical up to a relabeling of the sectors, whereas the middle diagram moreover has the $+$ and $-$ signs indicating the dominant solutions reversed.}
     \label{fig:diff}
\end{figure}

\sk
Figure \ref{fig:monodromy} shows the various transitions that we encounter when continuing a transseries solution around the origin in the complex $z = -t/6$ plane. Note that in this process, the rays at angles $2\pi/5$ act as Stokes lines and as anti-Stokes lines in an alternating fashion. Continuing from our choice of transitions in the previous subsection, after we cross the $2\pi$ and $3\pi$ Stokes lines, we continue to the $4\pi$ transition, for which the degenerate Stokes graph is displayed in figure \ref{fig:diff} on the right. For comparison, we have also displayed the degenerate Stokes graphs of the two preceding transitions at $2\pi$ and $3\pi$ in this figure. If we look at the graphs closely, we see that the connection problems for the $2\pi$ and $4\pi$ transitions are identical up to a relabeling of the sectors, and therefore up to a relabeling of the Stokes multipliers $m_i$ and the connection coefficients $a_{jk}$. The connection coefficients we need are $a_{34}$ and $a_{45}$ instead of $a_{12}$ and $a_{23}$, and for the Stokes multipliers we use  $m_3$ and $m_4$ instead of $m_1$ and $m_2$. The new connection coefficients are therefore acquired by straightforward extension of our analysis of the Weber equation to subsequent regions $\rnc{3}$, $\rnc{4}$ and $\rnc{1}$\footnote{When going from $\rnc{4}$ to $\rnc{1}$ in figure \ref{fig:Weberrnc}, one should remain on the same sheet of the Riemann surface, and therefore not insert a branch cut but instead just naively extend the connections (\ref{eq:phiStokes}).}.  Following this approach, we learn that $a_{34} = -e^{-i\pi E/2} a_{12}$ and $a_{45} = -e^{i\pi E/2} a_{23}$, and so we find before and after the $4\pi$-transition:
\bea
    4\pi\text{-direction} \qquad
    \begin{aligned}
    \cS_3:&\begin{cases}
    m_{3} = a_{34} = -e^{-i\pi E/2} a_{12},\\
    m_{4} = a_{45} = -e^{i\pi E/2} a_{23},
    \end{cases} \\
     \cS_4:&\begin{cases}
    m'_{3} = a_{34}+ie^{-i\pi E/2} = -e^{-i\pi E/2} a_{12}
    +ie^{-i\pi E/2},\\
    m'_{4} = a_{45} = -e^{i \pi E/2}  a_{23},
    \end{cases}
    \end{aligned}
    \label{eq:M2}
\eea
analogous to (\ref{eq:M1}). As expected, the $2\pi$- and $4\pi$-expressions look very similar, and in fact when we look at formulas (\ref{eq:mij}) and (\ref{eq:ab2}) more closely, we see that there is a shortcut to computing the $4\pi$ transition. We can first shift the transseries parameters 
\begin{equation}
    (\alpha, \beta) \;\mapsto\; (-\alpha e^{-i\pi E_0/2}, -\beta e^{i\pi E_0/2}),
\end{equation} 
where as before only $E_0$ appears since $\ga$ and $\gb$ have no $\eta^{-1/2}$-corrections. Then we apply $\mS_{2\pi}$, and finally shift back the resulting transeries: 
\begin{equation}
    (\alpha', \beta') \;\mapsto\; (-\alpha' e^{i\pi E'_0/2}, -\beta e^{-i\pi E'_0/2}).
\end{equation}
Introducing the operation
\begin{equation}
    T = -
    \begin{pmatrix}
         e^{-i\pi E_0(\alpha, \beta)/2} &0\\
         0&e^{i\pi E_0(\alpha, \beta)/2} 
    \end{pmatrix},
\end{equation}
this shortcut can be summarized in the following simple formula\footnote{Note that $T$ depends on the transseries parameters that it alters. Therefore, since the automorphisms $\mS_{n\pi}$ affect these parameters as well, the matrices $T$ and $T^{-1}$ in this equation are not each other's inverse: $T$ on the right hand side depends on the transseries parameter before $\mS_{2\pi}$ is applied, and $T^{-1}$ on the left hand side depends on the parameters after that transition.}:
\begin{equation}
    \mS_{4\pi} = T^{-1} \mS_{2\pi}T.
\end{equation}
Further investigation of the Stokes graphs and the corresponding connection coefficients tells us that we can generalise this relation to arbitrary Stokes transitions:
\begin{equation}
    \mS_{(m+2n)\pi} = T^{-n} \mS_{m\pi}T^{n}
\end{equation}
for all $m, n\in \bZ$. This saves us the time of computing all further individual transitions, and shows that all automorphisms can be related to essentially only two distinct ones. Following the notation of \cite{ASV1} we shall denote these two distinct `basis' automorphisms by $\underline{\mS}_0$ and $\underline{\mS}_\pi$ where we use an underline to distinguish them from the plain automorphisms $\mS_{n\pi}$.

\begin{figure}
    \centering
    \includegraphics[width = 1 \textwidth]{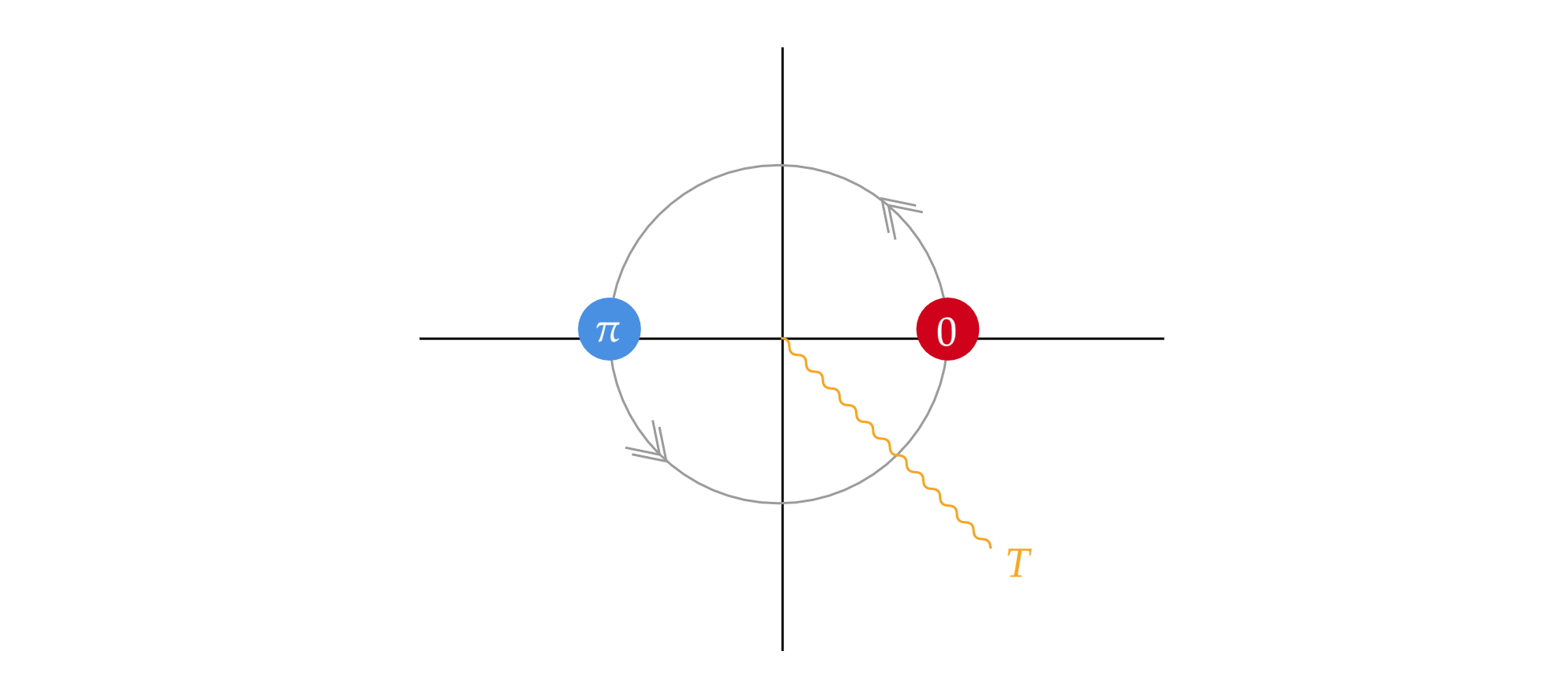}
    \caption{A single monodromy in the complex plane of the inverse Écalle time, $z^{5/4}$. It includes the two transitions $\underline{\mS}_0$ and $\underline{\mS}_\pi$ and the transformation $T$ that is interpreted as implementing a shift between branches.}
    \label{fig:Ecallemono}
\end{figure}

\sk
To wrap up our discussion of the general Stokes transitions, a few comments regarding the transformation $T$ are in order:
\begin{itemize}
    \item $T$ is an invertible transformation: as long as we don't change transseries parameters we can insert $TT^{-1}$ in our expressions and thereby `shift' the automorphisms to other branches. This means for example that the following two operations are equivalent:
    \begin{equation}
        T^5 \mathfrak{S}_{9\pi}  \mathfrak{S}_{8\pi} \mathfrak{S}_{7\pi}\mathfrak{S}_{6\pi}\mathfrak{S}_{5\pi} \mathfrak{S}_{4\pi}  \mathfrak{S}_{3\pi} \mathfrak{S}_{2\pi}\mathfrak{S}_{\pi} \mathfrak{S}_{0} = (T\mathfrak{S}_{\pi} \mathfrak{S}_{0})^5
        \label{eq:B1}
    \end{equation}
    We shall see in the next section that the above product of transformations acts as the identity on an arbitrary pair of transseries parameters $(\alpha, \beta)^T$.

    \item The operator $T^2$ can also be interpreted as implementing a change of branch choice on the constants of proportionality $C_\pm(t, \eta)$. When matching local solutions $\varphi_\pm(\zeta)$ to global solutions $\psi_\pm(x)$ around the double turning point, we had to compute these constants -- see equations (\ref{eq:psiphiconnection})-(\ref{eq:Cpm}) and Appendix \ref{app:matching} -- which required us to pick a logarithmic branch in evaluating expression (\ref{eq:D0}). We chose the branch that leads to a factor $\exp(\mp i \pi  E_0/4)$ in the expressions for $C_\pm$, but we could also have picked the other branch that yields a factor $\exp(\pm i \pi  E_0/4)$. Our connection coefficients $a_{ij}$ scale with $C_\pm/C_\mp$, and so this would lead to alternative connection coefficients $\tilde a_{ij}$ that are different from (\ref{eq:aij}) in the following way:
\begin{equation}
    \begin{pmatrix}
         \tilde a_{12}\\
         \tilde a_{23}
    \end{pmatrix} = \begin{pmatrix}
         e^{-i\pi E_0} &0\\
         0&e^{i\pi E_0} 
    \end{pmatrix} \begin{pmatrix}
         a_{12} \\
         a_{23}
    \end{pmatrix} =
    T^2\begin{pmatrix}
         a_{12} \\
         a_{23}
    \end{pmatrix}=
    \begin{pmatrix}
         a_{56} \\
         a_{67}
    \end{pmatrix} .
\end{equation}
Hence, we see that the ambiguity hidden in the computation of $C_\pm$ is equivalent to computing the connection coefficients on a different branch, in this case leading us to $\mS_{6\pi}$ instead of $\mS_{2\pi}$. Note that both of these two non-linear Stokes transitions require us to study the Stokes phenomena of the double turning point in the regions $\rnc{1}$, $\rnc{2}$ and $\rnc{3}$ in figure \ref{fig:Weberrnc}, only on a different sheet. Hence applying $T^2$ can also be interpreted as crossing a branch cut of the double turning point.
\item The transformation $T$ also has a straightforward interpretation when we consider the monodromy of transseries solutions in the coordinates of inverse Écalle time $z^{5/4}$. At the end of section \ref{sec3.1} we already noticed that (when we exclude the overall factor $\sqrt{z}$) all the terms in the transseries scale with $\alpha^n \beta^m z^{-\frac58 (n+m)} z^{\frac52 \alpha\beta (n-m)}$. When we complete one full rotation by $2\pi$ counterclockwise in the complex $z^{5/4}$ plane, which is equivalent to mapping $z\mapsto z e^{8\pi i /5}$, we observe that
\bea
 \alpha^n \beta^m z^{-\frac58 (n+m)+\frac52 \alpha\beta (n-m)} &\mapsto &(-1)^{n+m}e^{4\pi i  \alpha\beta(n-m)} \alpha^n \beta^m z^{-\frac58 (n+m)+\frac52 \alpha\beta (n-m)} \ret
 & =& \left(-\alpha e^{4\pi i  \alpha\beta}\right)^n\left(-\beta e^{-4\pi i  \alpha\beta}\right)^m  z^{-\frac58 (n+m)+\frac52 \alpha\beta (n-m)}.\ret
 & &
 \label{eq:Trot}
\eea
In the second line we absorbed the new factors in the transseries parameters and hence we learn that our rotation is equivalent to the applying a single $T$ transformation -- see figure \ref{fig:Ecallemono}. A single insertion of $T$ thus implements a logarithmic plus fractional power branch change, and when going five times around the origin this diagram schematically represents the right-hand side of expression (\ref{eq:B1}). This representation of the monodromy matches the picture that emerges from studying the first Painlevé equation using resurgence as in e.g.\ \cite{ASV1}: there are essentially only two unique automorphisms that in the terminology of \cite{ABS1} correspond to \textit{forward motion} and \textit{backward motion along the alien chain}.
\item A related transformation was found in \cite{Tak4} (see remark 3 in that paper) that we denote here by $\Tilde{T}$. It is a version of the transformation that we call $T$, but one that relates the expressions for all five Stokes multipliers across a single Stokes transition, whereas our $T$ relates two multipliers across a double transition. Essentially, we have that $ \tilde{T}^2 \simeq T$.
\end{itemize}

This T-transformation allows us to formulate a general expression for the transseries parameters of solutions in arbitrary sectors $\cS_k$. For a given set of Stokes multipliers $\{ m_j\}$, the transseries solution in the sector $\cS_k$ has the following transseries parameters
\setlength{\fboxsep}{10pt}
\begin{equation}
 \boxed{\begin{aligned}
        \alpha  &= (-1)^{(k_\alpha+1)/2} \ \frac{m_{k_\alpha}}{2\sqrt{\pi}}(96\sqrt{3})^{-E_0/4} e^{i\pi E_0(k_\alpha-2)/4}\Gamma(1+E_0/4)\\
        \beta &= (-1)^{k_\beta/2}\ \frac{im_{k_\beta}}{2\sqrt{\pi}}(96\sqrt{3})^{E_0/4}e^{-i\pi E_0 (k_\beta-2)/4} \Gamma(1-E_0/4)
    \end{aligned}
    }
    \label{eq:generaltrans}
\end{equation}
where $k_\alpha = 1+2\lfloor \frac{k}{2}\rfloor$ and $k_\beta = 2\lfloor \frac{k+1}{2}\rfloor$. The value of $E_0$ can be computed through
\be
    1+m_{k}m_{k+1} = \exp\left((-1)^{k+1} i \pi E_0\right),
    \label{eq:genE}
\ee
as we shall explain in the next section. One can check that this expression is consistent with the aforementioned properties of the $T$ transformation.

\sk
Equations (\ref{eq:generaltrans}) represent more of a \textit{top-down} approach to study the transseries solutions to the first Painlevé equation. Instead of going from sector to sector computing the next set of transseries parameters using $\mS_{n\pi}$, we can immediately determine the complete set of monodromy data $\{m_j\}$ and from there derive the transseries parameter values in all the double sectors $\cS_k$ in one go. This is obviously desirable from a computational point of view, but also has its perks once we start studying the monodromy of the transseries solutions in the next section.

\sk
Finally, we want to anticipate a key property of the Painlevé I transseries expansion that we briefly hinted at already: as we cross ten succesive Stokes transitions we end up with the same transseries expansion, provided that we correct for the branch using the appropriate $T$-transformation. This can already be seen from the $\bZ_5$ symmetry that resides in the Stokes multipliers and the $\bZ_2$ structure from the two essentially distinct Stokes transitions $\underline{\mS}_0$ and $\underline{\mS}_\pi$ seen in figure \ref{fig:Ecallemono}. We verify this in the next section.

\section{Results}
\label{sec:results}
With our explicit formulae for two-parameter transseries parameters in any double sector, we possess all the tools for a global study of the two-parameter transseries solutions. In subsection \ref{sec:symm} we prove that a rotation of $8\pi$ around the origin of the complex $z$-plane -- which corresponds to ten succesive Stokes transitions -- brings us back to the original transseries representation, provided that we apply the appropriate $T$-transformations. We also provide two examples were we explicitly write out the transseries parameter values in all succesive sectors before completing an $8\pi$ rotation . Then in subsection \ref{sec:multiplesolutions}, we show that a single Painlevé I transcendent does not have a single unique transseries representation, but that a whole class of transseries expansions exist that correspond to a single transcendent. Subsequently, we study the relation between the transseries representations and the different types of Painlevé I transcendents in subsection \ref{sec:TPS}. More specifically, we are interested in seeing which type of transseries -- zero, one or two parameter -- appears when we rotate a tronquée, tritronquée or elliptic type transcendent around the origin of the complex $z$-plane. In subsection \ref{sec:test} we perform some high-precision numerical tests of our claims and results. Finally, we show how the Stokes constants of the first Painlevé equation can be derived from our connection formulae and how these formulae confirm the results of \cite{BSSV}, most notably their connection formulae for two-parameter transseries solutions of the Painlevé I equation.

\subsection{Symmetry of the monodromy}
\label{sec:symm}

In the previous section we already formulated a general expression for the transseries parameters which required us to calculate the quantity $E_0$ from the Stokes multipliers. This was done through equation (\ref{eq:genE}), the form of which we would like to clarify. Sticking to notation of double sectors $\cS_k$ that we gave above, we see from the expressions for the Stokes multipliers $m_1$ and $m_2$ in (\ref{eq:mij}) that
\begin{equation}
    1+m_{1}m_{2} = \exp\left(i \pi E_0/2\right)
\end{equation}
in $\cS_1$. After crossing the $2\pi$ transition, we subsequently find 
\begin{equation}
    1+m_{2}m_{3} = \exp\big(-i \pi E_0/2\big)
\end{equation}
in $\cS_2$, which can be checked using either the expressions for $m_2$ and $m_3$ in (\ref{eq:mjk}) or using the transformation (\ref{eq:E}) and the properties of the Stokes multipliers (\ref{eq:mult}). This pattern continues indefinitely and can therefore be captured in the single equation (\ref{eq:genE}).

\sk
Using the $\bZ_5$ symmetry of the Stokes multipliers, it is easy to see that
\begin{equation}
    E_0 \Bigg|_{\cS_k} = -E_0 \Bigg|_{\cS_{k+5}}+4n
    \label{Esymm}
\end{equation}
for some integer $n \in \bZ$. If we ignore the choice of branch encoded in the $4n$ term by assuming that $n=0$ for now -- we discuss the appearance of a non-zero $n$ in the next subsection -- then this implies that the product of transseries parameters $\alpha\beta$ returns to {\em minus} itself after five transitions. Hence, after ten transitions we find ourselves back at the same value for $E_0$ and therefore with the same product $\alpha\beta$.

\sk
We would now like to explicitly demonstrate that the transseries parameters return to their original values as well, after ten succesive Stokes transitions. First of all, for ten transitions we need to rotate our solutions four times around the origin in the complex $z$-plane which corresponds to sending $z\mapsto ze^{8\pi i}$. Using the same logic that we presented in section \ref{sec:GST} -- in particular regarding equation (\ref{eq:Trot}) -- one can check that if we rotate $z$ by $8\pi$ counterclockwise, we find
 \begin{equation}
    \lambda(z e^{8\pi i}, \alpha, \beta) = \lambda(z, - e^{20\pi i \alpha \beta} \alpha, - e^{-20\pi i \alpha \beta} \beta) = T^{5} \lambda(z, \alpha, \beta).
\end{equation}
That is, the 'naive' rotation by $8\pi$ is equivalent to applying $T^{-5}$ to our transseries parameters.

\sk
Secondly, we can use our solution for the global transseries parameters (\ref{eq:generaltrans}) to see what the effect is of ten succesive Stokes transitions. This amounts to simply shifting $k \mapsto k+10$ from which we learn that 
\be \underbrace{\mS\mS.....\mS\mS}_{\text{10 times}} \lambda(z, \alpha, \beta) =  \lambda(z, - e^{-20\pi i \alpha \beta} \alpha, - e^{20\pi i \alpha \beta} \beta) = T^{-5} \lambda(z, \alpha, \beta).
\label{eq:B4}
\ee
Hence, when we put these two results together, we learn that
\be
\underbrace{\mS\mS.....\mS\mS}_{\text{10 times}} \lambda(ze^{8\pi i}, \alpha, \beta) =   \lambda(z, \alpha, \beta)
\label{eq:B5}
\ee
This equation tells us that a transseries expansion rotated four times around the origin, with the inclusion of all its Stokes jumps, yields the exact same expansion as we started with. Hence, this proves that a general two-parameter transseries expansion returns to its original transseries parameter values after ten succesive transitions.

\sk
Finally, we would like to demonstrate the aforementioned properties of the transseries solutions by presenting two examples. We take two pairs of initial transseries parameters $(\alpha, \beta)$ in $\cS_1$ and calculate transseries parameters in the subsequent sectors $\cS_{k>1}$. The two parameter pairs that we use for our example are $(\alpha, \beta) = (\frac{1}{4}, \frac{1}{5})$ and $(\frac{1}{5}, \frac{1}{2})$. The transseries parameter values after each transition are displayed in table \ref{T1}. We have computed exact values for all the parameters, but we only show numerical values (with a few exceptions) because the analytical expressions are in general awfully long and complicated. The table shows that the transseries parameters do indeed return to their original values, as long as we include the shift $T^5$ at the end. For the $(\frac{1}{4}, \frac{1}{5})$ transseries this shift is in fact unnecessary: in this case we find $T^5 = 1$. Also note that even though our starting pairs are rather similar (in both cases simple, real and rational numbers smaller than 1 are used), their behavior as we go along the sequence of transitions is already quite different.

\begin{figure}[ht]
\centering
\begin{minipage}{.5\textwidth}
  \centering
  \includegraphics[width=.95\linewidth]{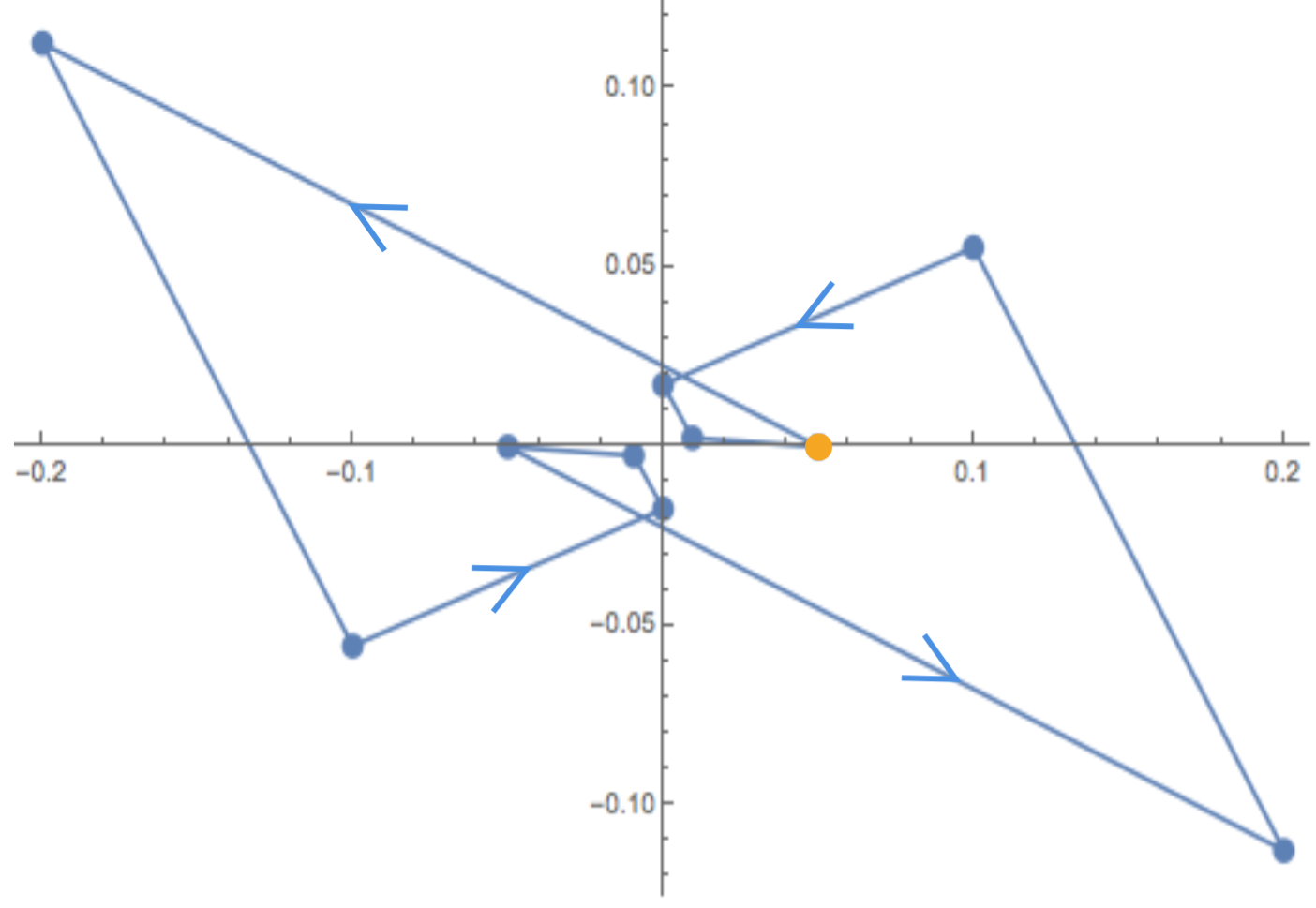}
\end{minipage}%
\begin{minipage}{.5\textwidth}
  \centering
  \includegraphics[width=.95\linewidth]{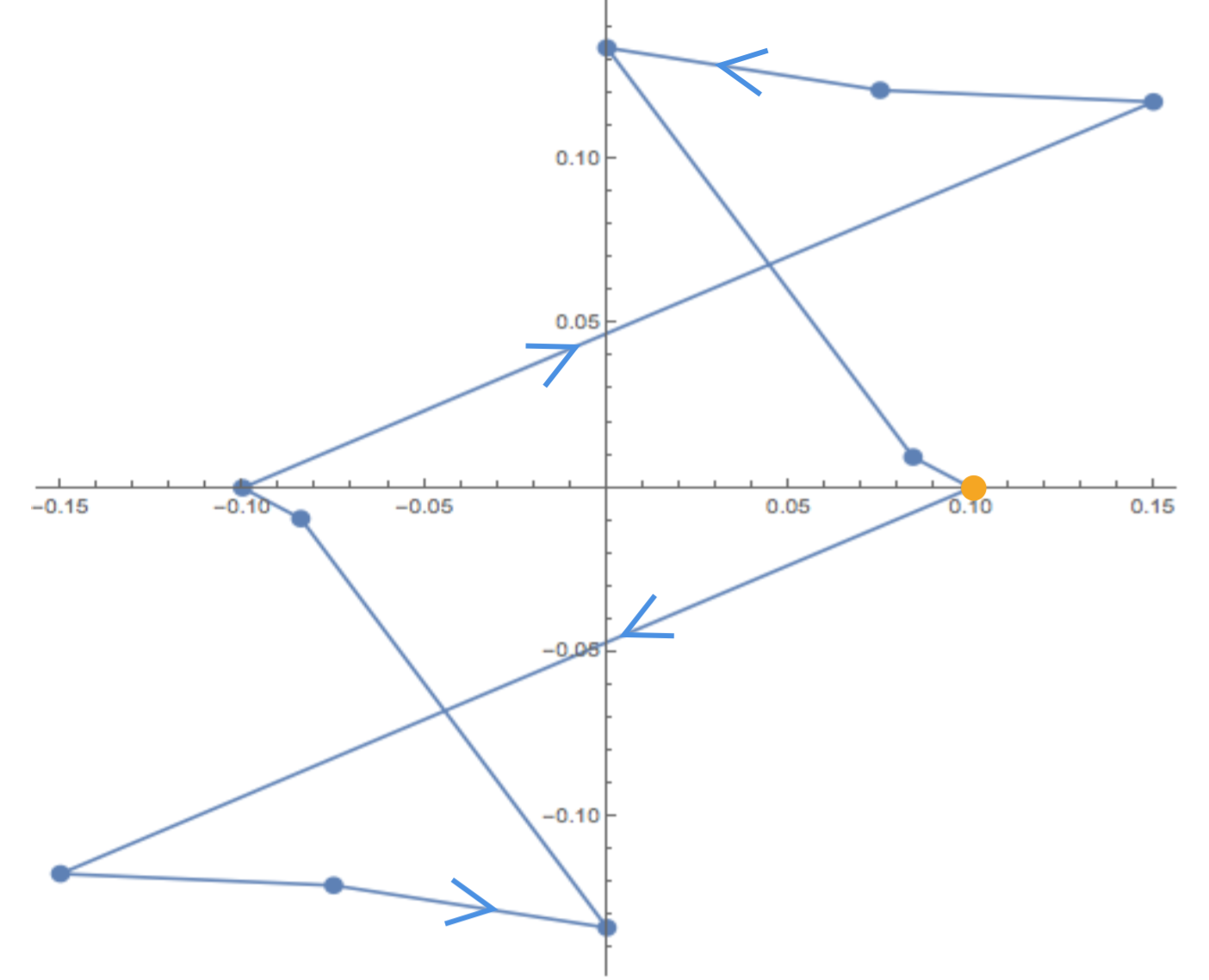}
\end{minipage}
\caption{The complex values of $\alpha\beta$ for the monodromy of the two transseries solutions described in table \ref{T1}. The transseries with the initial pair of parameters $(\frac{1}{4}, \frac{1}{5})$ is depicted on the left, starting at the orange dot where $\alpha\beta = 0.05$. The transseries starting with $(\ga,\gb) = (\frac{1}{5}, \frac{1}{2})$ is depicted on the right, starting at the orange dot at $\alpha\beta = 0.1$. The arrows indicate the direction in which the parameter values move.}
\label{F1}
\end{figure}

\sk
We can also study these transseries solutions graphically by plotting the change in the product $\alpha\beta$ in the complex plane as a result of the different Stokes transitions. This plot is shown in figure \ref{F1} for both examples in table \ref{T1}. As expected, the graphs are symmetric around the origin. Moreover, the loops in the graph are closed -- recall that the transformation $T$ leaves the product $\alpha\beta$ invariant, and therefore each path ends where it started, regardless of whether we include the $T^5$ operation at the end or not.

\begin{table}[t]
    \centering
    \begin{tabular}{|c||c|c||c|c|}
    \hline
         &$\alpha$ & $\beta$ &  $\alpha$ & $\beta$  \\
         \hline
         \hline
         $\cS_1$& $1/4$ & $1/5$ &  $1/5$ & $1/2$ \\
         \hline
         $\cS_2$ & $-0.035-0.051i$ & $0.329-3.684i$  &$-0.016+0.016i$ & $0.825+8.416i$\\
         \hline $\cS_3$ &$-0.009+0.063i$ & $-0.639+1.694i$&$-0.010+0.048i$ & $-2.098+2.013i$\\
         \hline $\cS_4$ &$-0.021+0.045i$ & $-0.318+0.146i$ &$0.024+0.090i$ & $-1.382-0.365i$\\
         \hline$\cS_5$ &$-0.041+0.042i$ &$0.083+0.144i$  &$-0.159-0.373$ &$0.103-0.182i$\\
         \hline$\cS_6$ &$-1/5$ & $1/4$  &$-i/2$ & $-i/5$\\
         \hline$\cS_7$ &$0.107+0.010i$ & $1.762-1.211i$  &$-32.890-335.620$ & $-0.0004+0.0004i$\\
         \hline$\cS_8$ &$9.713+3.663i$ & $0.011+0.002i$  &$-2.666+129.522$& $0.0009-0.0006i$\\
         \hline$\cS_9$ &$-0.252-0.548i$ & $-0.026-0.012i$  &$24.356-92.292i$& $-0.0014+0.0004i$\\
         \hline$\cS_{10}$ &$-0.123+0.131i$ & $-0.026-0.048i$  &$-0.148-0.238i$& $-0.187+0.238i$\\
         \hline $\cS_{11}$ &$1/4$ & $1/5$  &$-1/5$& $-1/2$\\
         \hline\hline
          $T^5$ &$1$ & $1$  &$-1$& $-1$\\
          \hline
    \end{tabular}
    \caption{The transseries parameter pairs that appear while rotating two different transseries solutions around the origin in the $z$-plane. Each row denotes the parameters in a single Stokes sector $\cS_k$. The final row shows the value of the $T^5$ transformation that we need to muliply with the transseries parameters in $\cS_{11}$ in order to return the starting parameter values. }
    \label{T1}
\end{table}

\subsection{Non-uniqueness of transseries representations}
\label{sec:multiplesolutions}
In using our equations relating transseries parameters $(\ga,\gb)$ to Stokes multipliers $m_i$, there is a subtle but important detail that we already touched upon but so far have not yet fully adressed. This is the fact that Stokes multipliers and transseries parameters are not related to one another in a one-to-one fashion. If we take for example the equations (\ref{eq:mij}) mapping $(\ga,\gb)$ to $m_i$, we can see that there are multiple pairs of transseries parameters that yield the same Stokes mulipliers $m_{1,2}$. This ambiguity is more manifest when we invert this relation back to (\ref{eq:A3}) -- or its generalization (\ref{eq:generaltrans}) for arbitrary double sectors. Knowing that $1+m_1m_2 = \exp(i\pi E_0/2)$, we learn that the ambiguity rests in the branch of the logarithm that we need to choose in order to compute $E_0$, which in turn determines the product $\ga \gb$. Once this choice is made, the individual $\ga$ and $\gb$ can be determined from (\ref{eq:A3}). As a result, given a specific choice of $m_{1,2}$, we can generate a whole sequence of pairs of transseries parameters that solve (\ref{eq:mij}).

\sk
Take for example our two-parameter transseries with $(\alpha, \beta) = (\frac{1}{5}, \half)$ in $\cS_1$, corresponding to $E_0=-\frac45$. From (\ref{eq:mij}) we can derive the first two Stokes multipliers 
\begin{equation}
    m_1 = \frac{e^{-6\pi i/5}\sqrt{\pi}}{5 \times 3^{3/10}\Gamma(\frac{4}{5})} \hspace{3cm}   m_2 = \frac{2i \times 3^{3/10}\sqrt{\pi}}{\Gamma(\frac{6}{5})} 
\end{equation}
and subsequently recompute $E_0$. In this last step we can pick whichever branch of the logarithm we like, so we do not necessarily find back the exact same value for $E_0$ that we started with. Plugging these different values of $E_0$ back into (\ref{eq:A3}) yields an infinite multitide of outcomes, of which we have displayed some in table \ref{tab:abpairs}. We find the original transseries pair $(\frac{1}{5}, \half)$ but also other pairs whose product $\alpha\beta$ is shifted by a multiple of $\half$ with respect to the original product of $\frac{1}{10}$. This shift corresponds to the $4n$ ambiguity that we encountered in relation (\ref{Esymm}). If we plot more graphs like figure \ref{F1} for other transseries parameters, the points in these graphs end up within the strip $-\frac{1}{4}<\Re(\alpha\beta)<\frac{1}{4}$ as long as in computing the monodromy we consistently take the principal branch of the logarithm -- i.e. we assume a branch cut along the negative axis for the logarithm.

\begin{figure}
    \centering
    \includegraphics[width= 1 \textwidth]{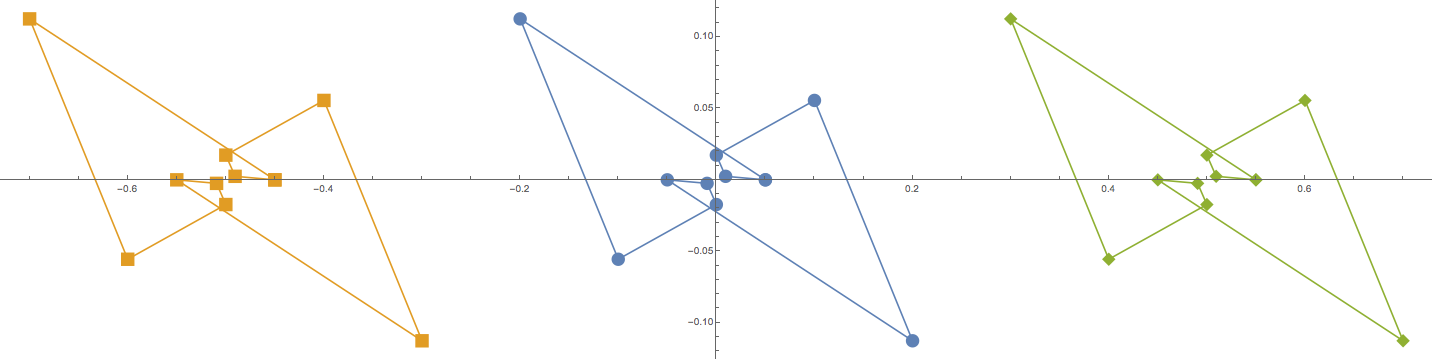}
    \includegraphics[width= 1 \textwidth]{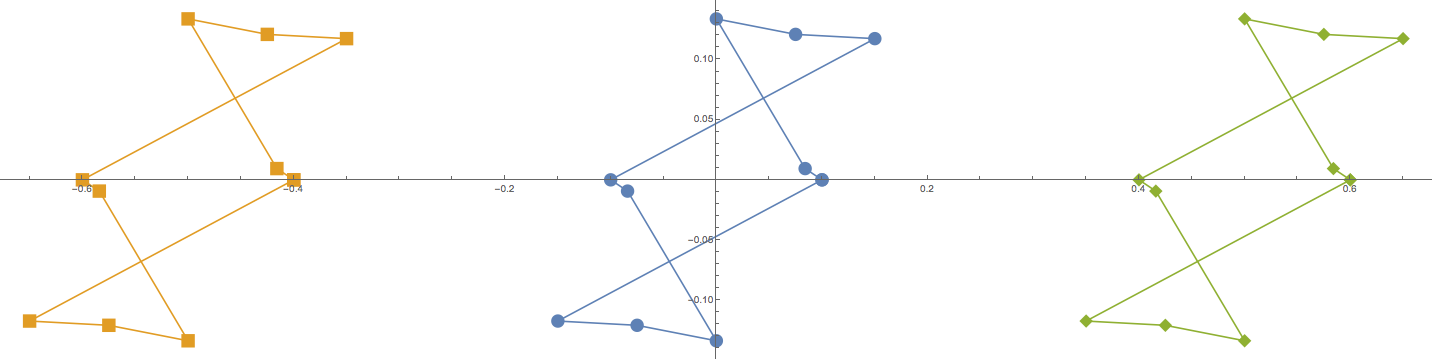}
    \caption{The monodromy of transseries solutions $(\frac{1}{4}, \frac{1}{5})$ (top) and $(\frac{1}{5}, \frac{1}{2})$ (bottom) and their `neighbouring' transseries pairs in the complex $\alpha\beta$ plane.}
    \label{F2}
\end{figure}

\sk
What happens if, as a starting point, we take one of the transseries pairs whose product lies outside this strip? Provided that we consistently pick the branch of the logarithm that corresponds to the strip on which our starting value $\alpha\beta$ lies, we expect to find a new path which remains in this alternative strip. For example, let us take the two pairs 'neighbouring' the fundamental one in table \ref{tab:abpairs}, $(-1/600\sqrt{3}, 240\sqrt{3})$ and $(96\sqrt{3}, 1/160\sqrt{3})$, and study their monodromy. We once again plot the product $\alpha\beta$ in the complex plane but now, while computing the monodromy, we also consistently take a neighbouring branch of the logarithm whenever we compute $E_0$. This leads to the three graphs shown in the top panel of figure \ref{F2}. For completeness we have also displayed the original graph centered at the origin, and repeated this procedure for the $(\frac{1}{4}, \frac{1}{5})$ transseries. We learn that we get copies of the original graph, simply translated by $\half$ to either the left or right. Clearly, these graphs in general are symmetric around a point in $\half \bZ$ which is in accordance with (\ref{Esymm}) when including the the $4n$ ambiguity. 

\sk
The upshot of this discussion is that the product $\alpha\beta$ of any transseries solution under the Stokes transitions traces out a point symmetric path on the cylinder $\bC \; / \; \frac12 \bZ$. The different `neighbouring' transseries pairs under the identification with $\frac12\bZ$ are related to one and the same Painlevé I solution as defined by the Stokes multipliers $m_i$. Therefore, the different transseries obtained in this way should be thought of as expansions of the exact same Painlevé transcendent. Thus, the two-parameter transseries representations are {\em not unique}: if we resum the different expansions related to given Stokes multipliers in an appropriate way, we expect to find the exact same function.

\begin{table}[]
    \centering
    \begin{tabular}{|c||c |c| c| c |c|}
    \hline 
    \rule{0pt}{3ex}$\alpha$ & $\frac{1}{96000}$ & $-\frac{1}{600\sqrt{3}}$ & $\frac{1}{5}$ &$96\sqrt{3}$ &  $ 23040$\\[1.5ex]
    \hline
    \rule{0pt}{3ex}$\beta$ & $-86400$ & $240\sqrt{3}$ & $\frac{1}{2}$ & $\frac{1}{160\sqrt{3}}$ &$\frac{11}{230400}$\\[1.5ex]
     \hline
    \rule{0pt}{3ex}$\alpha\beta$ & $-\frac{9}{10}$& $-\frac{4}{10}$ & $\frac{1}{10}$ & $\frac{6}{10}$& $\frac{11}{10}$\\[1.5ex]
    \hline
\end{tabular}
    \caption{Pairs of transseries parameters $\alpha$ and $\beta$ that all correspond to the same Stokes multipliers $m_1$ and $m_2$.}
    \label{tab:abpairs}
\end{table}

\subsection{Special solutions}
\label{sec:TPS}
So far, in studying the monodromy of asympototic solutions, we have considered the full two-parameter transseries expansions. In the Boutroux classification that we discussed in section \ref{sec:P1BC}, these expansions generally correspond to Painlevé transcendents that have poles in all five sectors of the $z$-plane. As mentioned in that section, however, there are two special types of solutions, the tronquée and tritronquée solutions. We would now like to understand how our connection formulas apply to these special situations and what we can learn about the monodromy of the special solutions. We shall see shortly that for both the tronquée and tritronquée solutions, different transseries representations appear in the same sector that are consistent with either the presence or absence of poles in that region. For the sake of simplicity, we only consider transseries expansions on the 'fundamental strip' where $|\Re(\alpha\beta)|<\frac14$. We start with a brief discussion of the three classes of Painlevé I solutions.

\sk
To understand the distinction between the three types of solutions we need to look at the Stokes multipliers $m_i$ that parametrize the Painlevé transcendents once more. We already encountered the fact that the relations (\ref{eq:mult}) between the $m_i$ allow us to express the monodromy data in terms of two independent variables. Let us make the discussion explicit by choosing $m_1$ and $m_2$ as our parameters. We then find that, under the condition $m_1 m_2 \neq -1$, the five different $m_i$ are 
\begin{equation}
    \{m_1, m_2, \frac{i-m_1}{1+m_1m_2}, i(1+m_1m_2), \frac{i-m_2}{1+m_1m_2}\}.
    \label{eq:Z1}
\end{equation}
On the other hand, when $m_1 m_2 = -1$ we get
\begin{equation}
    \{i, i, m_3 , 0, i-m_3\}
    \label{eq:Z2}
\end{equation}
as the only solutions to (\ref{eq:mult}). The full manifold of complex dimension two that the Stokes multipliers describe is called the \textit{monodromy manifold} -- see e.g.\ \cite{KK, YN} -- and is in fact a cubic surface \cite{SvdP}:
\be
    m_{j} m_{j+1} m_{j+2}+m_j+m_{j+2} = i
\ee

\sk
Now we can classify three types of Painlevé transcendents. These are:
\begin{itemize}
    \item \textbf{Elliptic type solutions}. These correspond to points on the monodromy manifold where $m_j \neq 0$ for all $j$.
    \item \textbf{Tronquée solutions}. These correspond to points on the monodromy manifold where $m_j = 0$ for a single $j$. From (\ref{eq:Z1}) and (\ref{eq:Z2}) one can see that this leads to five one-parameter families of tronquee solutions.
    \item \textbf{Tritronquée solutions}. These correspond to points on the monodromy manifold where $m_j = m_{j+1} = 0$ for a single $j$, where as usual we treat $j$ as a cyclic index. This leads to five possible tritronquée solutions that can be interpreted as the intersection points of two loci of tronquée solutions. Note that this fixes the other three Stokes multipliers to be $m_{j+2}=m_{j+3}=m_{j+4}=i$.
\end{itemize}
This classification exhausts all possibillities. In summary, we have a complex 2-dimensional manifold parametrizing all solutions. Embedded in that manifold are five loci of complex dimension one (i.e.\ five surfaces) that parametrize five families of tronquée solutions. Each of these five surfaces at two different points intersects another one of these surfaces, thus defining five distinct tritronquée solutions. Notice the $\bZ_5$ symmetry that is clearly manifest in this description.

\sk
What do the transseries expansions of the special solutions and the monodromy of their parameters look like? First of all, note that whenever we encounter a transseries where one or both of the parameters ($\alpha$,$\beta$) vanish, we are dealing with an asympotic expansion of a tritronquée or tronquée solution. This can be seen from equations like (\ref{eq:mij}) and (\ref{eq:mjk}) that show that one or two vanishing transseries parameters imply that one or two Stokes multipliers vanish\footnote{\label{note1}Note that the reverse statement is false. There are infinitely many pairs of non-zero $(\alpha, \beta)$ that force either $m_i$ or $m_{i+1}$ to vanish. This can be achieved by picking $\alpha\beta \in \half \bZ \backslash \{0\}$ -- recall the discussion of section \ref{sec:multiplesolutions} regarding the different branches. In this section, however, we stick to the fundamental strip, i.e.\ $-\frac14 \leq \Re(\alpha\beta)<\frac{1}{4}$.}. Hence, a zero-parameter expansion -- i.e. a formal power series without any instanton transmonomials -- must be the expansion of a tritronquée solution. Similarly, a one-parameter transseries can be the expansion of either a tronquée or tritronquée solution. 

\sk
In the fundamental strip, two-parameter transseries appear for both elliptic type solutions and tronquée solutions, but not for tritronquées. One would naively think that any two subsequent non-zero Stokes multipliers give rise to a two-parameter transseries expansion in some double sector, but this is not necessarily the case. The exception is the situation in which (at least) two subsequent Stokes multipliers are $m_i = m_{i+1} = i$\footnote{This `glitch' does not occur for general elliptic type solutions. Relations (\ref{eq:mult}) show that any pair of subsequent Stokes multipliers $m_i = m_{i+1} = i$ forces $m_{i+3} = 0$ rendering the solution a (tri)tronquée one.}. This becomes a problem once we reach the Stokes sector in which we would have to solve
\begin{equation}
    \exp\big(\pm 4\pi i \alpha \beta\big)=\exp\big(\mp i \pi E_0 /2\big) = 1+m_im_{i+1} = 0,
\end{equation}
where the sign depends on the double sector in which the solution lies -- see the discussion of section \ref{sec:symm}. The expression above implies that in order to find the product $\alpha\beta$, we must compute the logarithm of either zero or infinity, thus leading to an ill-defined transseries solution. When we actually apply our connection formulae to the individual transseries parameters we find that one of these jumps to zero, whereas the other diverges. 

\sk
At this moment we do not have a satisfying reformulation of these expansions with diverging parameters. Clearly the parameters $(\alpha,\beta)$ obtained using the Stokes transformations fail to describe the transseries globally, and so we probably require some sort of `gauge transformation' to describe this corner of the transseries parameter space in terms of regular transseries. We call the would-be expansions where $(\ga,\gb)$ are ill-defined \textit{singular transseries expansions}. Hence, on the fundamental strip we never expect to find a two-parameter transseries expansion for a tritronquée solution, since all its non-zero Stokes multipliers have the value $i$. This is also consistent with the fact that tritonquées only have a single sector filled with poles. There is a loophole however: outside of the fundamental strip, it is possible to encounter a two-parameter transseries that corresponds to a one-parameter transseries on the fundamental strip through the $\frac12 \bZ$-symmetry -- see also footnote \ref{note1}.

\begin{table}[]
    \centering
    \begin{tabular}{|c|c|c|c|c|}
    \hline 
    & Two-parameter & One-parameter & Zero-parameter & Singular\\
    \hline
    \hline
     Elliptic type & \checkmark &&&\\
    \hline
    Tronquée& \checkmark&\checkmark&&\checkmark\\
    \hline
    Tritronquée &\checkmark* &\checkmark&\checkmark &\checkmark\\
    \hline
\end{tabular}
    \caption{The type of transseries expansion (column) that can appear when considering a specific class of solutions (row). The * indicates that for a tritronquee solution, a two-parameter transseries can appear, but only outside the fundamental strip where $\alpha\beta \in \frac12 \bZ \backslash \{0\}$ .}
    \label{tab:solutions}
\end{table}

\sk
If we now put together all the pieces, table \ref{tab:solutions} emerges. To summarise, we have Painlevé transcendents that fall into one of three Boutroux classes as described above, indicated in the rows in table \ref{tab:solutions}. They can be described by 0, 1, 2 or singular transseries parameters $(\ga,\gb)$ as indicated in the columns. We would like to stress once more that a transcendent always belongs to a specific Boutroux class, but that the non-elliptic ones can be described by different types of transseries, depending on what Stokes sector we are in. 

\sk
Let us now by computation study the monodromy of these tritronquée and tronquée solutions. We pick the following initial transseries parameter pairs; $(0, 0)$ for the tritronquée solution and $(1,0)$ for the tronquée solution. These parameter choices fix the Stokes multipliers $(m_1, m_2, m_3, m_4, m_5)$, which are $(0, 0, i, i, i)$ and $(-2\sqrt{\pi}, 0, i+2\sqrt{\pi}, i , i)$ respectively. From the Stokes multipliers we can compute transseries parameters in each subsequent Stokes sector using (\ref{eq:generaltrans}). The results are shown in table \ref{tab:tronquee}. The table confirms our expectations regarding the types of transseries that appear for the various types of Painlevé transcendents. All other tronquée or tritronquée solutions show the same qualitative behaviour: the same transseries types appear in exactly the same order, possibly cyclically permuted. For instance, one could consider the alternative tronquée solution starting with a transseries expansion $(\alpha, \beta)$ = $(0, 1)$. This solution is defined by the multipliers $(0, 2i\sqrt{\pi}, i, i, i-2i\sqrt{\pi})$ and therefore jumps to a two-parameter transseries description after the first Stokes transition. Subsequently, it follows the same pattern: singular-two-one-one-two-singular-two-one and so forth.

\sk
The behaviour of the tronquée and tritronquée solutions allows us to predict in which sectors the poles of the Painlevé transcendent lie. This prediction requires us to make two mild assumptions. First of all, whenever one of the transmonomial terms $e^{\pm\tau}$ is `on' in a given sector -- i.e. the corresponding transsseries parameter is non-zero -- and it is exponentially {\em growing} in the large $z$ limit, then we interpret this transseries as an analytic continuation of the Fourier expansion along an anti-Stokes line of a series whose coefficients are built from Weierstrass elliptic functions\footnote{This is an avenue that Takei has more recently explored in \cite{Tak2, Tak3}. }. Boutroux already showed that to leading order the asymptotic behaviour of the Painlevé transcendents is described by these elliptic functions in the pole sectors. Hence the presence of exponentially large transmonomials should tell us that the corresponding Stokes sector is filled with poles. The second assumption is that the singular transseries expansions appear in \textit{algebraic}\footnote{This terminology comes from \cite{KK} where sectors filled with poles were described using Weierstrass elliptic functions (`elliptic asymptotics') and the pole-free sectors were described using so-called `algebraic asymptotics'.} Stokes sectors: those sectors in which no poles appear.

\begin{figure}
    \centering
    \includegraphics[width = 1 \textwidth]{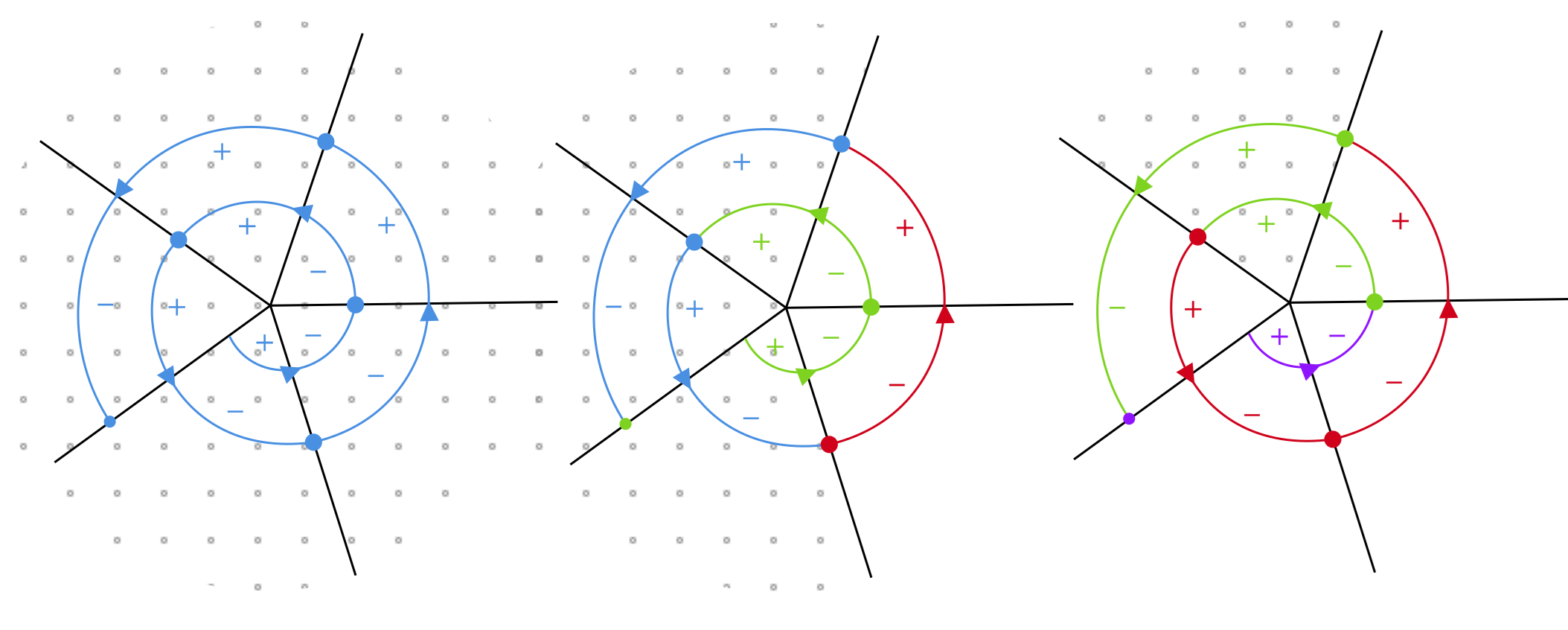}
    \caption{Three diagrams displaying the monodromy of transseries expansions for solutions of the elliptic type (left), tronquée solutions (middle) or tritronquée solutions (right). The gray dots in the background mark sectors in which the Painlevé I transcendent asymptotically contains poles. Starting with a two- (left), one- (middle) or zero-parameter transseries (right), we continue the expansions in the  counterclockwise direction around the origin and observe that after crossing a Stokes line, the transseries parameters change. The colors blue, green and purple denote two-, one- and zero-parameters transseries respectively. The color red denotes \textit{singular transseries expansions}. The plus or minus signs indicate whether $e^{+\tau}$ or $e^{-\tau}$ is the dominant instanton transmonomial. This dominance changes whenever the solutions cross an anti-Stokes line, which always bisects the angle between two consecutive Stokes lines.}
    \label{fig:special}
\end{figure}

\sk
Applying these two assumptions, figure \ref{fig:special} emerges. It shows the monodromy (with a total rotation of $4\pi$ in the variable $z$ this time, since the pattern of poles simply repeats after that) of the three classes of Painlevé transcendents. We see that elliptic type solutions, which have 2-parameter transseries expansions everywhere, have only sectors filled with poles. For the tronquée and tritronquée solutions we see the appearance of algebraic sectors where the poles are absent. In these sectors we find either a singular or a zero-parameter transseries, or a one-parameter transseries where the non-zero parameter corresponds to the subdominant transmonial $e^{\pm\tau} \ll 1$ as $z \xrightarrow[]{} \infty$.

\begin{table}[]
    \centering
    \begin{tabular}{|c|c|c||c|c|c|}
    \hline
         \multicolumn{3}{|c||}{Tritronquée} & \multicolumn{3}{|c|}{Tronquée}\\ 
         \hline
         $\alpha$ & $\beta$ & \begin{tabular}{@{}c@{}}Transseries \\ type\end{tabular} &$\alpha$ &$\beta$ & \begin{tabular}{@{}c@{}}Transseries \\ type\end{tabular} \\
         \hline
         \hline
         $0$ & 0 & zero& 1 & 0 & one \\
         \hline
         $i/2\sqrt{\pi}$ & 0 & one &$1+i/2\sqrt{\pi}$& 0&one \\
         \hline
         0 & $\infty$ & singular & \small{$-0.257+0.407i$} & \small{$0.315+0.108i$} & two \\
         \hline
         $\infty$ &0& singular & $\infty$&0 & singular\\
         \hline
         $-i/2\sqrt{\pi}$ & 0 & one&\small{$-4.044-4.941i$} &\small{$-0.023+0.002i$} & two \\
         \hline
         0 & 0 & zero & 0 & $i$ & one\\
         \hline
         0& $-1/2\sqrt{\pi}$ & one & 0 & $i - 1/2\sqrt{\pi}$ & one\\
         \hline
         $\infty$ & 0 & singular &\small{$-0.006-0.013i$}&\small{$2.510+11.125i$}  &two\\
         \hline
         0 & $\infty$ & singular& 0 & $\infty$ & singular\\
         \hline
         0 & $1/2\sqrt{\pi}$ & one & \small{$0.001i$} & \small{$-117.729+117.387i$}& two \\
         \hline
    \end{tabular}
    \caption{Two sequences of transseries parameters $(\ga,\gb)$ starting from a given pair and applying successive Stokes transitions. The left hand side of the table gives the parameters for the tritronquée solution that one obtains starting from $(\ga,\gb)=(0,0)$; the right hand side does the same for the tronquée solution obtained from starting with $(\ga,\gb)=(1,0)$. Along the way, different types of transseries occur, in agreement with table \ref{tab:solutions}. All parameter values can be computed exactly, but where these expressions are too complicated we have written down numerical approximations.}
    \label{tab:tronquee}
\end{table}

\subsection{Tests: numerical resummation}
\label{sec:test}
We would now like to check some of our results numerically. To this end, we resum formal transseries solutions $u(z)$ to the Painlevé I equation into actual (numerical) transcendents. To achieve this, we use the following procedure:
\begin{itemize}
    \item We pick a point $z_0$ in the complex $z$-plane where the transseries will be summed, and pick values $(\ga,\gb)$ -- or equivalently $(\gs_1,\gs_2)$ -- for the transseries parameters.
    \item We Borel-Écalle sum the perturbative $z$-expansions in the individual sectors to obtain values for $u^{(n|m)}(z_0)$ and ${u'}^{(n|m)}(z_0)$. In practice, we use a diagonal Padé approximant after computing the Borel transform and before doing the inverse (Laplace) transform -- see e.g.\ \cite{ASV1} for details on this procedure. For most tests, we compute instanton sectors up to $n=7$ and $m=7$, with the perturbative expansions of the $(0|0)$, $(1|0)$ and $(0|1)$ sectors containing about 500 terms, the other $(n|0)$ and $(0|m)$ sectors about 230 terms, and all further $(n|m)$ sectors about 150 terms.
    \item We sum over $n$ and $m$, with the appropriate $(\gs_1,\gs_2)$-dependent prefactors, to get numerical values for $u(z_0)$ and $u'(z_0)$. In this step, it is important to check that $\gs_1$ and $\gs_2$ are small enough for the $n$ and $m$ sums to reasonably converge. Depending on the value of the instanton transmonomials $e^{\pm A {z_0}^{5/4}}$, `small enough' can be of order 1, but sometimes one of the parameters also needs to be of order $10^{-6}$.
    \item The boundary conditions $u(z_0)$ and $u'(z_0)$ are then fed into a numerical solver that solves the Painlevé I equation. Because of the presence of a large number of poles in the solution, existing solvers do not do a very good job, so we wrote our own Mathematica code based on the strategy outlined in \cite{FW}. The main idea in this strategy is to use Padé approximants (rather than Taylor approximations) for $u(z)$ near a lattice point, which is very beneficial in dealing with the poles.
    \item The output of the numerical solver is a contour plot of the transcendent in the complex $z$ plane. Because we have Padé approximants at all lattice points that the solver uses, we can also very precisely read off the pole locations\footnote{For us, the pole locations only serve as a precision test on our connection formulae, but there is interesting work on other methods to compute these locations and how to interpret them -- see e.\ g.\ \cite{Cos1, Cos2, Cos3, ASV2}.}. These pole locations can then be used for high-precision comparison between computations.
\end{itemize}
All code was written in Mathematica. Of the above steps, the second one is the most time consuming: performing the Borel-Padé-Écalle sums to obtain values for $u^{(n|m)}(z_0)$ and ${u'}^{(n|m)}(z_0)$ takes ~10 hours on a laptop. Fortunately, the answer only depends on $z_0$, so the results can be saved and then used for different values of $\gs_1$ and $\gs_2$. The other parts of the computation -- of which running the numerical solver is then the most time consuming -- take a few minutes on a laptop.

\sk
With the above procedure, we reach 30-45 decimal places of precision in the boundary conditions $u(z_0)$ and $u'(z_0)$. Further precision is of course lost in running the numerical solver, so in the best cases the pole locations we find are correct up to about 25 decimal places. In the worst cases, we find pole locations that are still correct up to 5-10 decimal places. The precision of the results is always easy to test: for example by running the boundary value computation using a few less coefficients, and comparing the resulting values. The best precision test for the numerical solver is to simply run it again: on each run, a pseudo-random path through the lattice points is chosen, and so the output slightly differs from run to run.

\sk
{\bf Tritronquée solution}

\begin{figure}[th]
 \centering
 \includegraphics[width=0.3\textwidth]{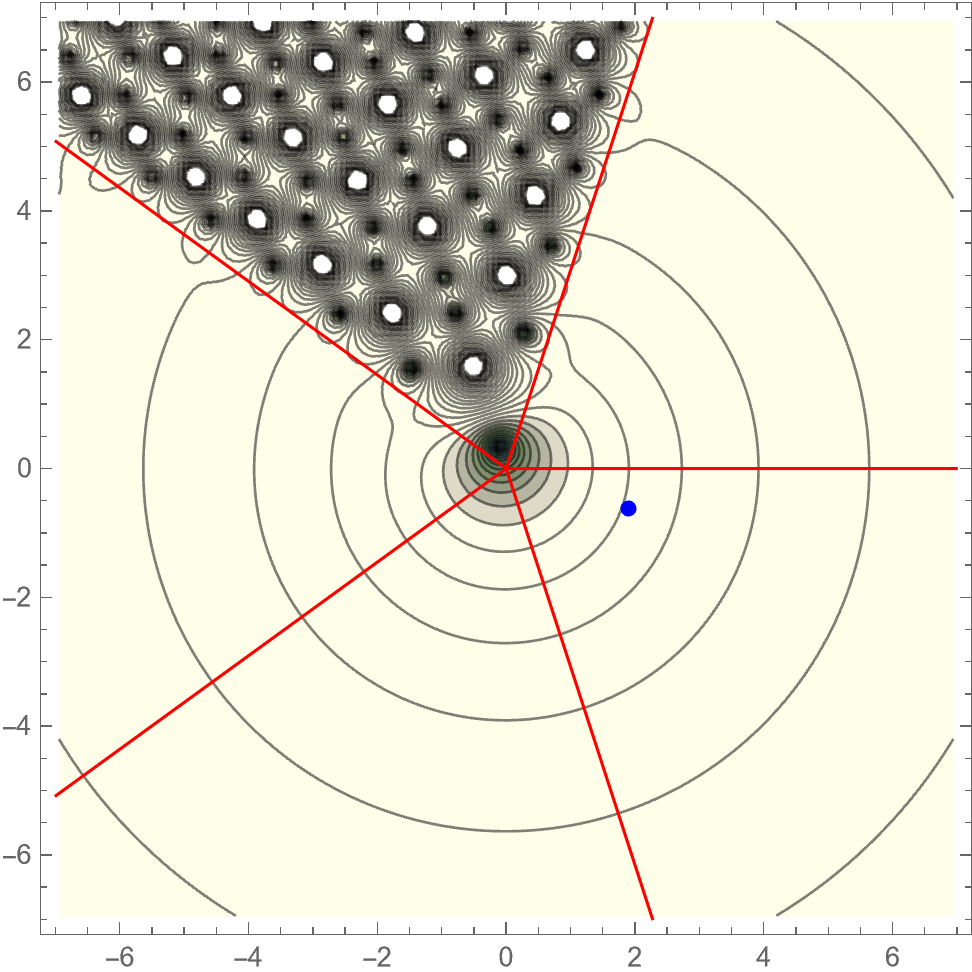}\hspace{1em}
 \includegraphics[width=0.3\textwidth]{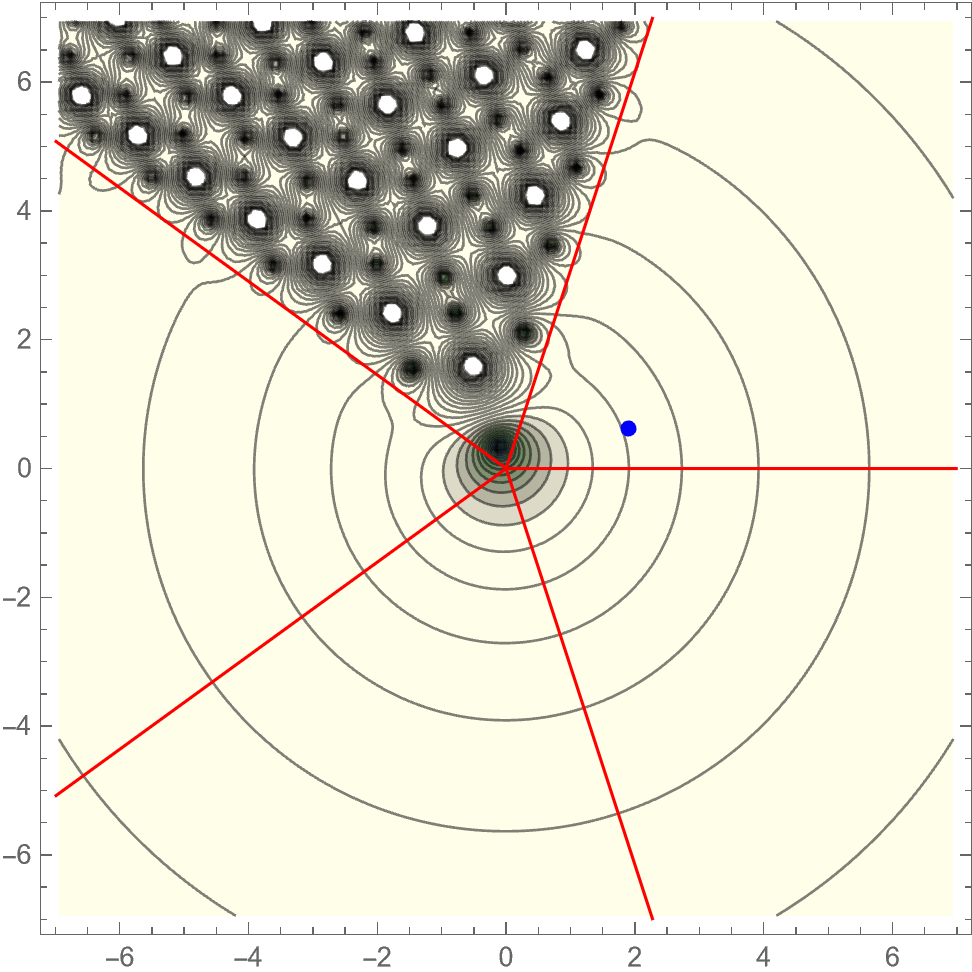}\hspace{1em}
 \includegraphics[width=0.3\textwidth]{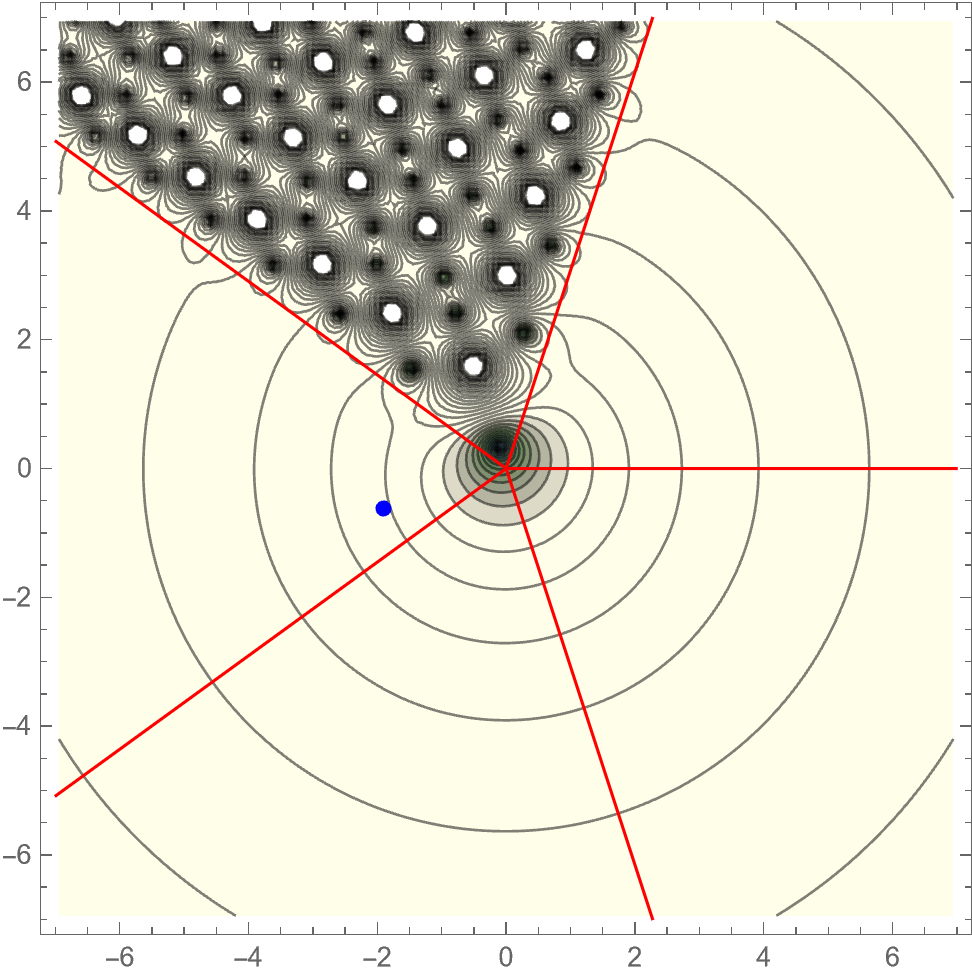}
 \caption{Left: the tritronquée transcendent obtained by setting $\gs_1 = \gs_2 = 0$ with boundary conditions at $z_0 = 2 e^{-i\pi/10}$ (indicated by the blue dot). Middle: the same transcendent after crossing the Stokes line with $\arg(z)=0$, with boundary conditions at $z_0 = 2 e^{+i\pi/10}$. Right: the same transcendent after crossing the Stokes line with $\arg(z)=-4\pi/5$, with boundary conditions at $z_0 = 2 e^{-9i\pi/10}$.}
 \label{fig:test008}
\end{figure}

\sk
As a first test, let us compute a tritronquée solution. We start at $z_0=2 e^{-i\pi/10}$ with trivial transseries parameters
\be
 \gs_1=0, \qquad \gs_2 = 0.
\ee
In this case, we only need to sum the perturbative sector; this results in boundary conditions
\bea
 u(z_0)  & \approx & 1.3926891004 - 0.2247957342 \; i \ret
 u'(z_0) & \approx & 0.3515399634 + 0.0604002386 \; i.
 \label{eq:bdytri1}
\eea
where for readability we display $\sim$10 decimal places, but our code provides around 30 correct decimals. The resulting transcendent can be seen in the left plot of figure \ref{fig:test008}. Clearly, the solver finds a tritronquée solution: four of the five sectors are free of poles.

\begin{figure}[th]
 \centering
 \includegraphics[width=0.25\textwidth]{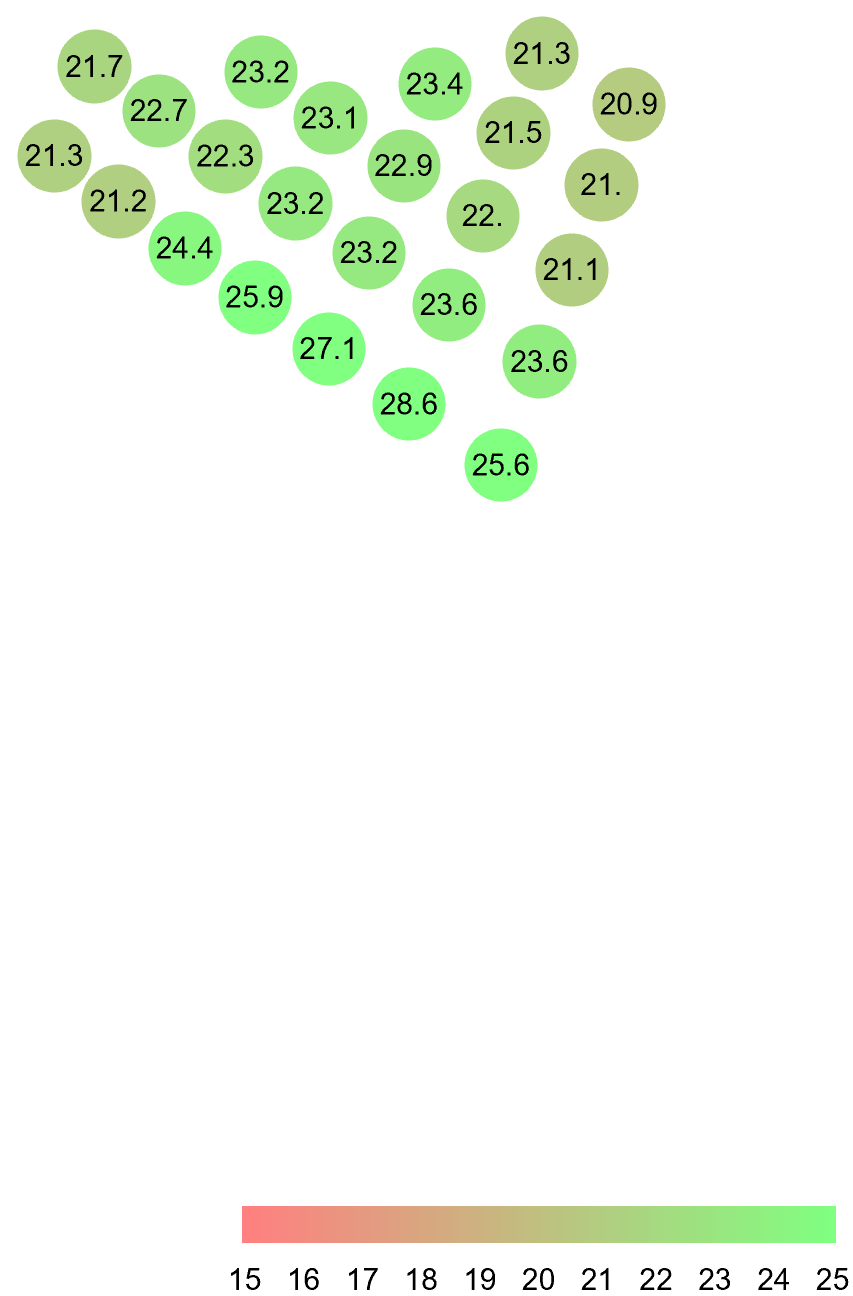}\hspace{8em}
 \includegraphics[width=0.25\textwidth]{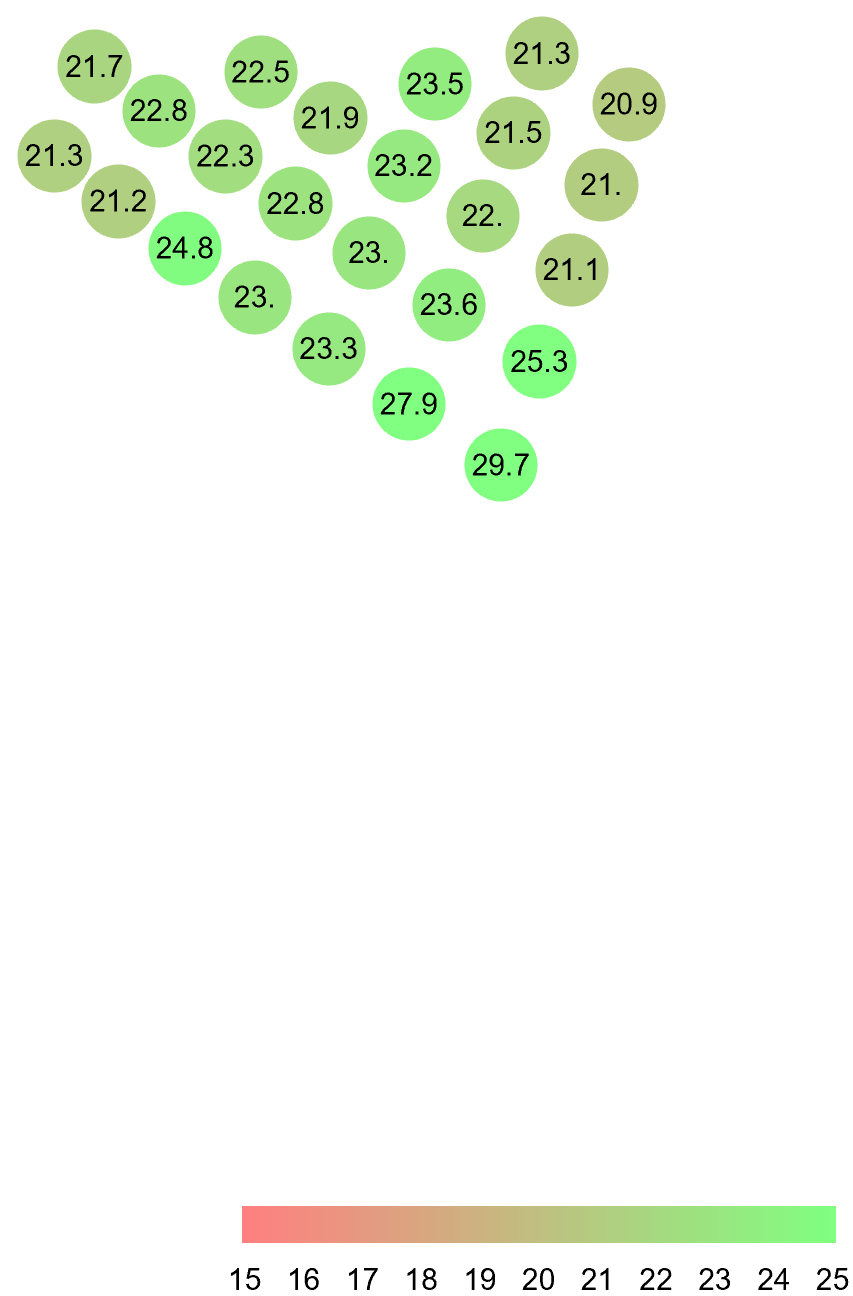}
 \caption{Left: Comparison of the pole locations of the tritronquée transcendent before and after the $\mS_0$ Stokes transition. The pole locations agree to 20-25 decimal places. Right: Comparison of the pole locations of the same transcendent before and after the $\mS_{-\pi}$ transition, with similar precision.}
 \label{fig:comp008}
\end{figure}

\sk
Now, we cross the Stokes line at $\arg(z)=0$ and move the point where we compute our boundary conditions to $z_0 = 2 e^{+i\pi/10}$. We apply the Stokes transition $\mS_0$, which gives
\be
 \gs_1=S_1, \qquad \gs_2 = 0.
\ee
These values are of course not at all surprising: for a one-parameter transseries (here actually a zero-parameter transseries) the Stokes automorphism is simply the classical map $\gs_1 \to \gs_1 + S_1$, where we add the first Stokes constant. Nevertheless, it is a good check for our code, as well as a nice way to show that a tritronquée solution can be obtained using {\em nonzero} parameters (compare table \ref{tab:tronquee}), to compute the transcendent using the resulting new boundary conditions. This now involves summing the $(n|0)$ sectors, and we obtain
\bea
 u(z_0) & \approx  & 1.3922689644 + 0.2254276382 \; i \ret
 u'(z_0) & \approx & 0.3534835301 - 0.0628969021 \; i.
\eea
Note that these numbers are close (but not exactly equal) to the complex conjugates of (\ref{eq:bdytri1}), a consequence of the fact that in the ``empty sectors'' the transcendent to first order behaves as $\sqrt{z}$. Inserting the boundary conditions in the solver, we obtain the middle plot of figure \ref{fig:test008}. Clearly, we find the same tritronquée transcendent; the difference between the solutions is not visible to the naked eye. To check the precision of our method, we therefore compare the pole locations, that turn out to agree to 20-25 decimals of precision. We show the comparison in the left plot in figure \ref{fig:comp008}.

\sk
Of course, we can now similarly perform the $\mS_{-\pi}$ transition. We cross the Stokes line at $\arg(z)=-4\pi/5$ and compute with $z_0 = 2 e^{-9i\pi/10}$. The new transseries parameters are
\be
 \gs_1=0, \qquad \gs_2 = i S_1,
\ee
and the boundary conditions are
\bea
 u(z_0) & \approx  & 0.2158403465 - 1.3937874422 \; i \ret
 u'(z_0) & \approx & 0.0494139094 + 0.3556190264 \; i.
\eea
The resulting transcendent is shown in the right plot of figure \ref{fig:test008}, which again to the naked eye is exactly the same as the original transcendent. We compare the pole locations before and after the $\mS_{-\pi}$ transition in the right plot in figure \ref{fig:comp008}, and again find 20-25 decimals of precision.

\sk
{\bf Tronquée solution}

\begin{figure}[th]
 \centering
 \includegraphics[height=0.33\textwidth]{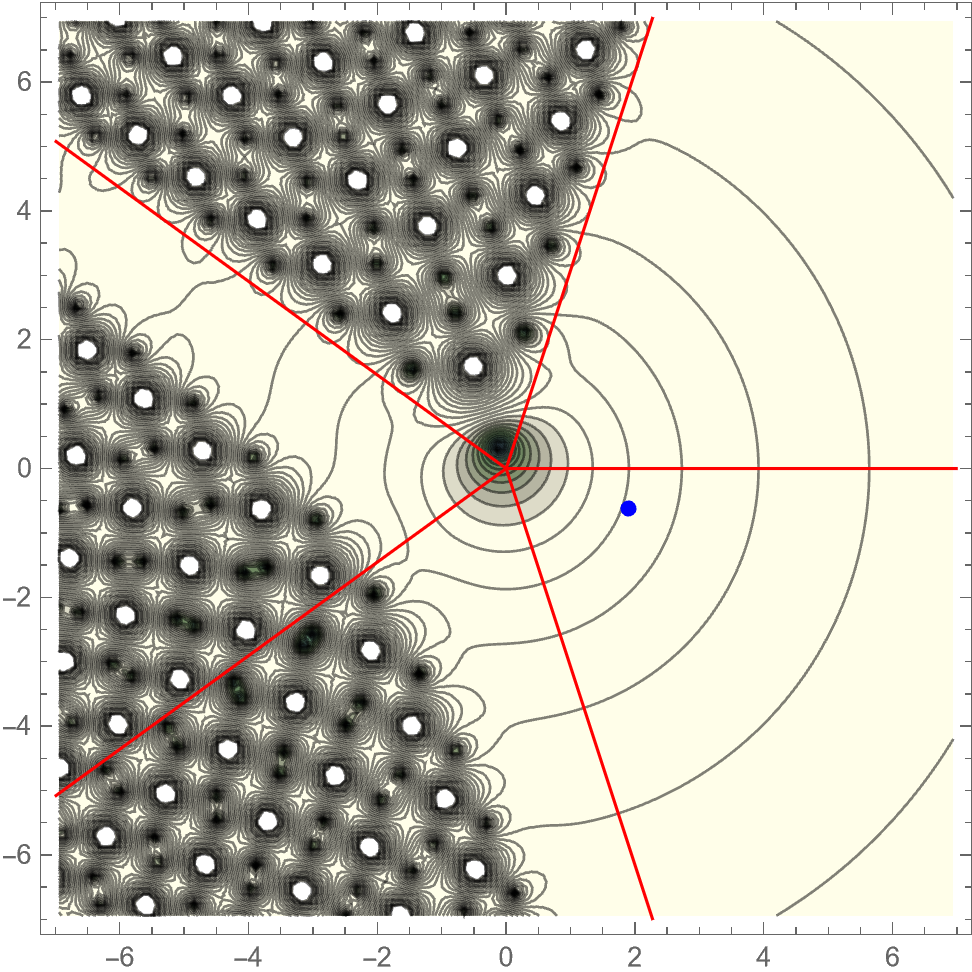}\hspace{1em}
 \includegraphics[height=0.33\textwidth]{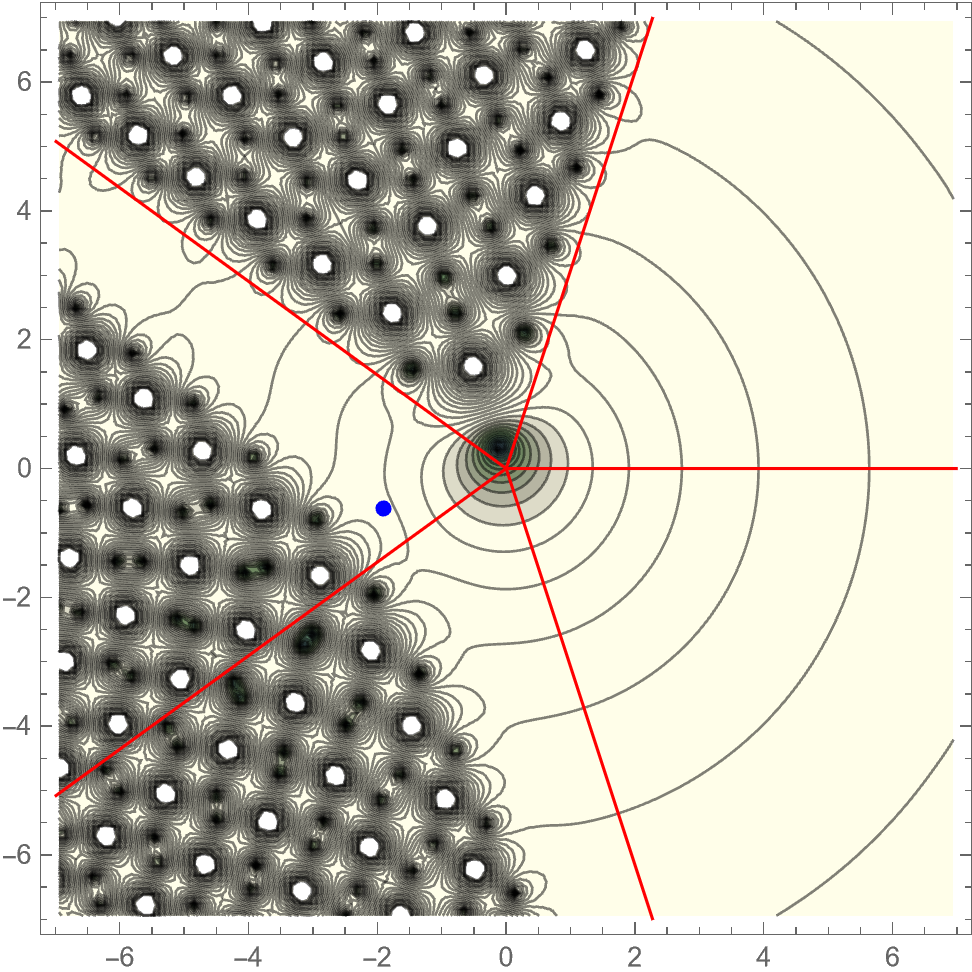}\hspace{1em}
 \includegraphics[height=0.33\textwidth]{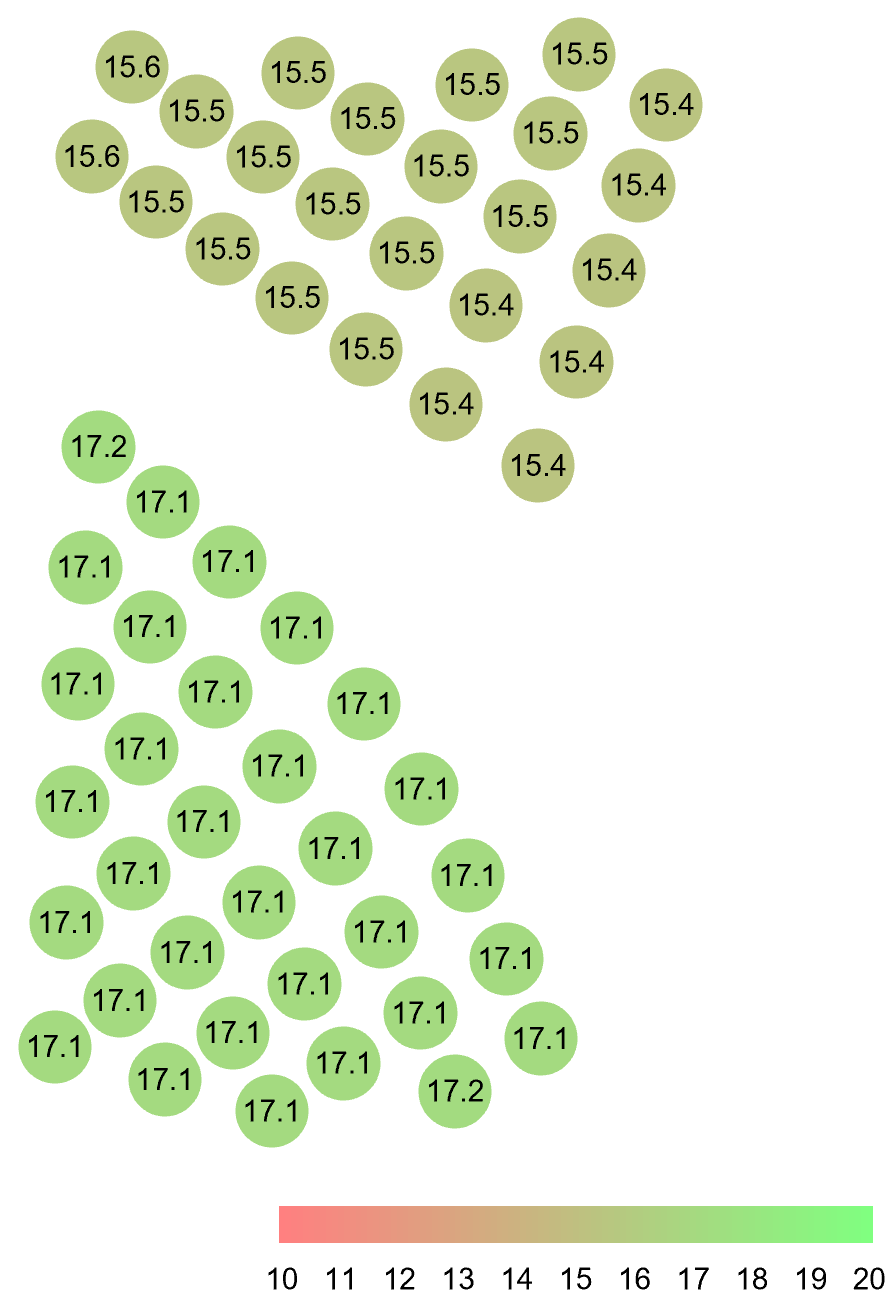}
 \caption{Left: the tronquée transcendent obtained by setting $\gs_1 = 10^{-4}$ and $\gs_2 = 0$ with boundary conditions at $z_0 = 2 e^{-i\pi/10}$. Middle: the same transcendent after crossing the Stokes line with $\arg(z)=-4\pi/5$, with boundary conditions at $z_0 = 2 e^{-9i\pi/10}$. Right: comparison between the pole locations.}
 \label{fig:test009}
\end{figure}

\nopagebreak
\sk
For a much more nontrivial test, let us now move on to a tronquée solution. To this end, we again start at $z_0 = 2 e^{-i\pi/10}$, but this time with transseries parameters
\be
 \gs_1=10^{-4}, \qquad \gs_2 = 0,
\ee
The value of $\gs_1$ may seem very small, but in fact one only needs very small parameters to see the effect of them being nonzero: whereas $\gs_1=\gs_2=0$ gives us a tritronquée solution, the above values make two sectors of poles `come in from infinity' leading to the left plot in figure \ref{fig:test009}, which we computed with boundary values
\bea
 u(z_0) & \approx  & 1.3926889302 - 0.2247956210 \; i \ret
 u'(z_0) & \approx & 0.3515406360 + 0.0603997152 \; i.
\eea
Note the small difference in boundary conditions with (\ref{eq:bdytri1}) (of course, we computed $\sim$20 more relevant decimal places), but the large resulting difference in the lower left sectors of the resulting plot.

\sk
We could now cross the Stokes line at $\arg(z)=0$, but of course the result will not be too interesting: since we start with a one-parameter transseries with only $\gs_1 \neq 0$, the result after $\mS_0$ will simply be another transseries with $\gs_1 \to \gs_1 + S_1$, as was the case for the tritronquée solution. Much more interesting is the $\mS_{-\pi}$ transition at $\arg(z) = -4\pi/5$, as now we really need the full Stokes automorphism to obtain two nontrivial parameters: after $\mS_{-\pi}$ (or rather, $(\mS_{-\pi})^{-1}$ since we cross the Stokes line in the clockwise direction) we obtain
\bea
 \gs_1 & \approx & 0.0001000244 + 0.0000000135 \; i \ret
 \gs_2 & \approx & 0.3711670363 - 0.0000999634 \; i,
\eea
which is close to $\gs_2 \to \gs_2 + i S_1$, but certainly not exactly equal to it\footnote{We also compared pole locations after performing the incorrect automorphism $\gs_2 \to \gs_2 + i S_1$, and obtained less than 5 correct decimal places, compared to the 15-17 correct decimal places after the true automorphism.}. To turn these values into boundary conditions, we now must Borel-Padé sum all $(n|m)$-sectors; we obtain
\bea
 u(z_0) & \approx  & 0.1927306646 - 1.4281545677 \; i \ret
 u'(z_0) & \approx & 0.2122633198 + 0.3988221351 \; i.
\eea
The resulting plot, as usual indistinguishable from the original one, can be seen in the middle part of figure \ref{fig:test009}. In the right part of that figure we compare the pole locations; the result is of course less precise than in the tritronquée case (the main cause of this seems to be that we sum at a value of $z_0$ which is already quite close to a pole field, so that many instanton sectors contribute), but summing up to $n=m=7$ still leads to 15-17 decimal places of precision.

\sk
{\bf Elliptic type solution}

\begin{figure}[th]
 \centering
 \includegraphics[width=0.3\textwidth]{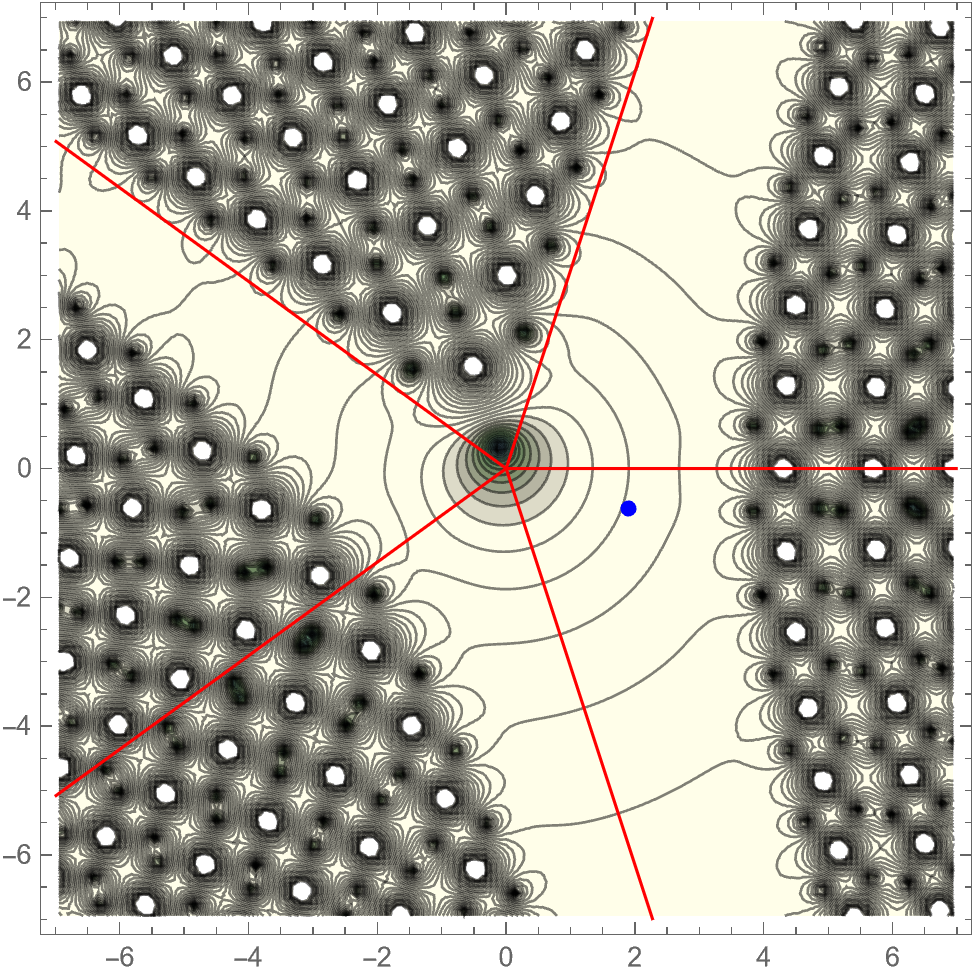}\hspace{1em}
 \includegraphics[width=0.3\textwidth]{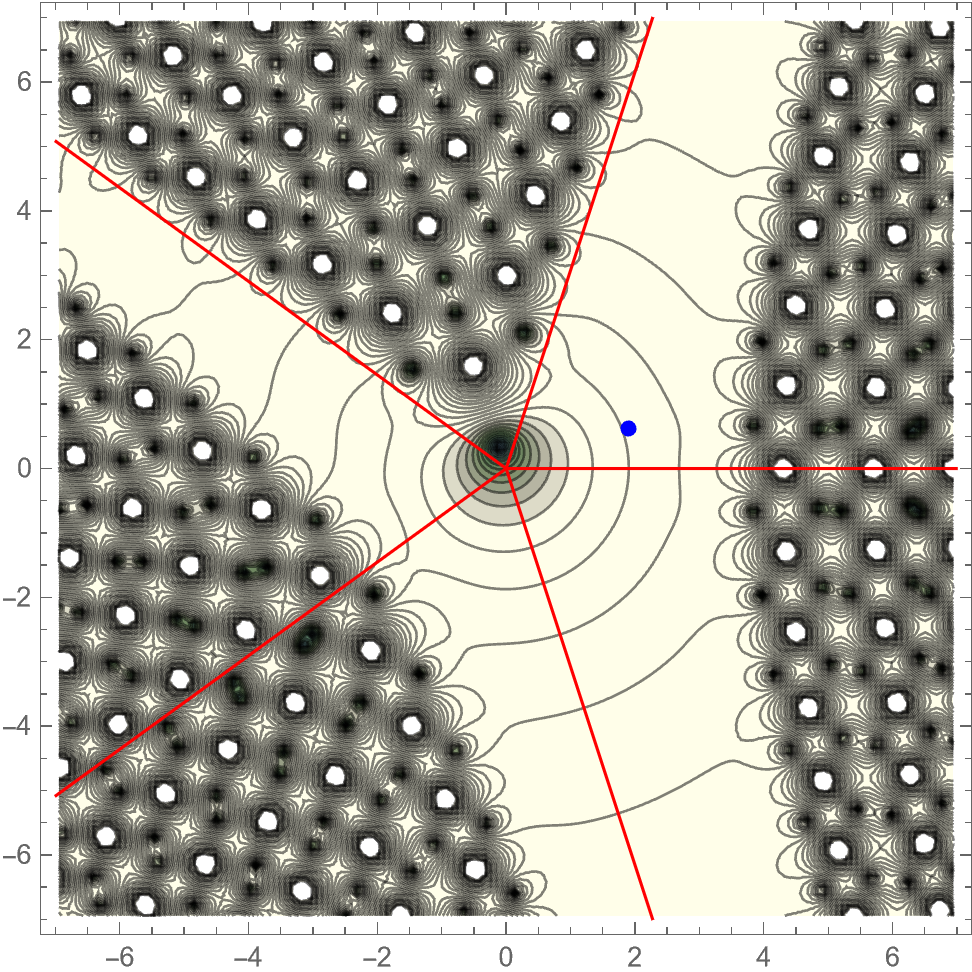}\hspace{1em}
 \includegraphics[width=0.3\textwidth]{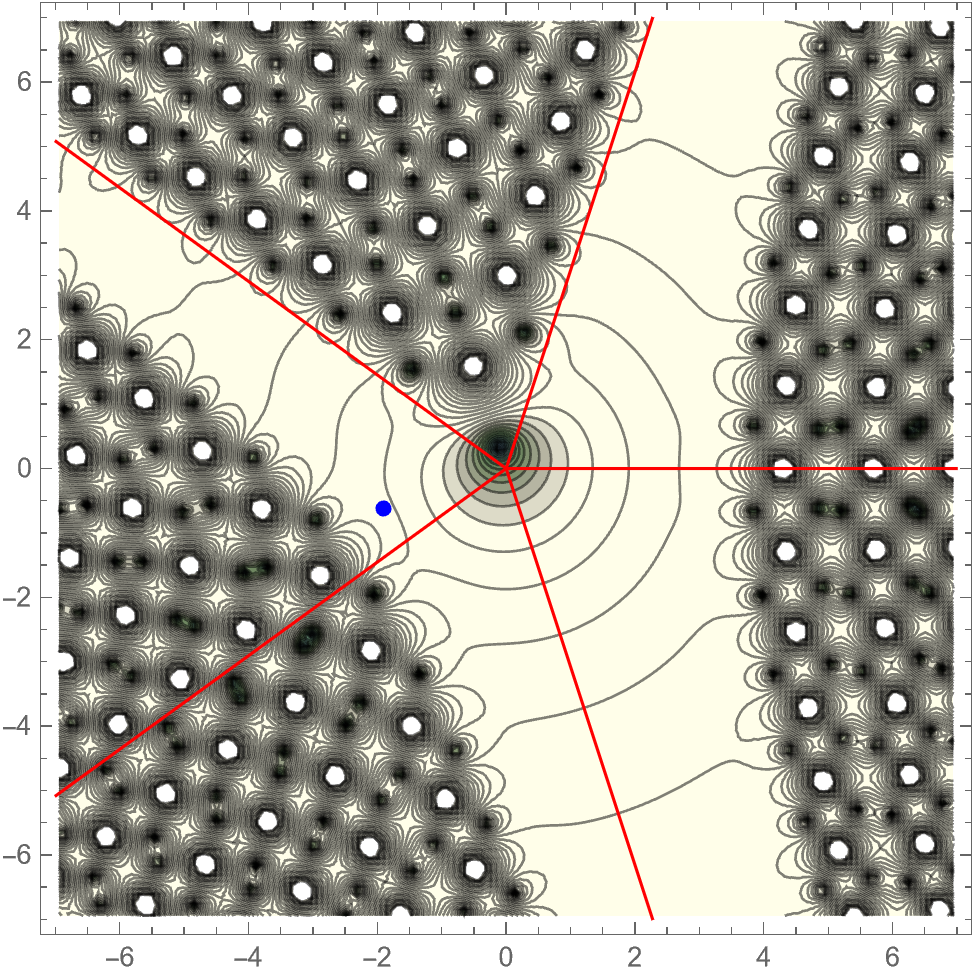}
 \caption{Left: the elliptic transcendent obtained by setting $\gs_1 = 10^{-4}$ and $\gs_2 = 10^{-6}$ with boundary conditions at $z_0 = 2 e^{-i\pi/10}$. Middle: the same transcendent after crossing the Stokes line with $\arg(z)=0$, with boundary conditions at $z_0 = 2 e^{+i\pi/10}$. Right: the same transcendent after crossing the Stokes line with $\arg(z)=-4\pi/5$, with boundary conditions at $z_0 = 2 e^{-9i\pi/10}$.}
 \label{fig:test011}
\end{figure}

\nopagebreak
\sk
Finally, let us test a full, two-parameter, elliptic type solution. We again start at $z_0 = 2 e^{-i\pi/10}$, and now turn on both transseries parameters: we take
\be
 \gs_1=10^{-4}, \qquad \gs_2 = 10^{-6},
\ee
leading to boundary conditions
\bea
 u(z_0) & \approx  & 1.3923652161 - 0.2250507705 \; i \ret
 u'(z_0) & \approx & 0.3501433116 + 0.0594838267 \; i.
\eea
The resulting transcendent can be seen in the left plot of figure \ref{fig:test011}. Clearly, now in all four sectors that were empty in the tritronquée case, poles have `come in from infinity'.

\sk
In the elliptic case, both the $\mS_0$ and $\mS_{-\pi}$ Stokes transitions are nonlinear, and so we check them both. After $\mS_0$ we work with $z_0 = 2 e^{+i\pi/10}$ and
\bea
 \gs_1 & \approx & 0.00009909427234707374 - 0.37125712513061910911 \; i \ret
 \gs_2 & \approx & 0.00000099999999999687 + 0.00000000000243962362 \; i,
\eea
where now we wrote 20 decimal places to show some further nontrivial digits. (Of course, these numbers -- though the expressions are rather ugly and not very informative -- can be computed exactly using our connection formulae.) After $(\mS_{-\pi})^{-1}$ we work with $z_0 = 2 e^{-9i\pi/10}$ and
\bea
 \gs_1 & \approx & 0.00010002440030337084 + 0.00000001346773630981 \; i \ret
 \gs_2 & \approx & 0.37116803583075350501 - 0.00009996367369018158 \; i.
\eea
For $\mS_0$ this leads to boundary conditions
\bea
 u(z_0) & \approx  & 1.3919451392 + 0.2256828906 \; i \ret
 u'(z_0) & \approx & 0.3520868133 - 0.0619805657 \; i,
\eea
and for $(\mS_{-\pi})^{-1}$ to boundary conditions
\bea
 u(z_0) & \approx  & 0.1927306627 - 1.4281545672 \; i \ret
 u'(z_0) & \approx & 0.2122633151 + 0.3988221421 \; i.
\eea
The resulting transcendents can be seen in the middle and right plots of figure \ref{fig:test011}, where as usual any small differences with the first plot are invisible.

\begin{figure}[th]
 \centering
 \includegraphics[width=0.35\textwidth]{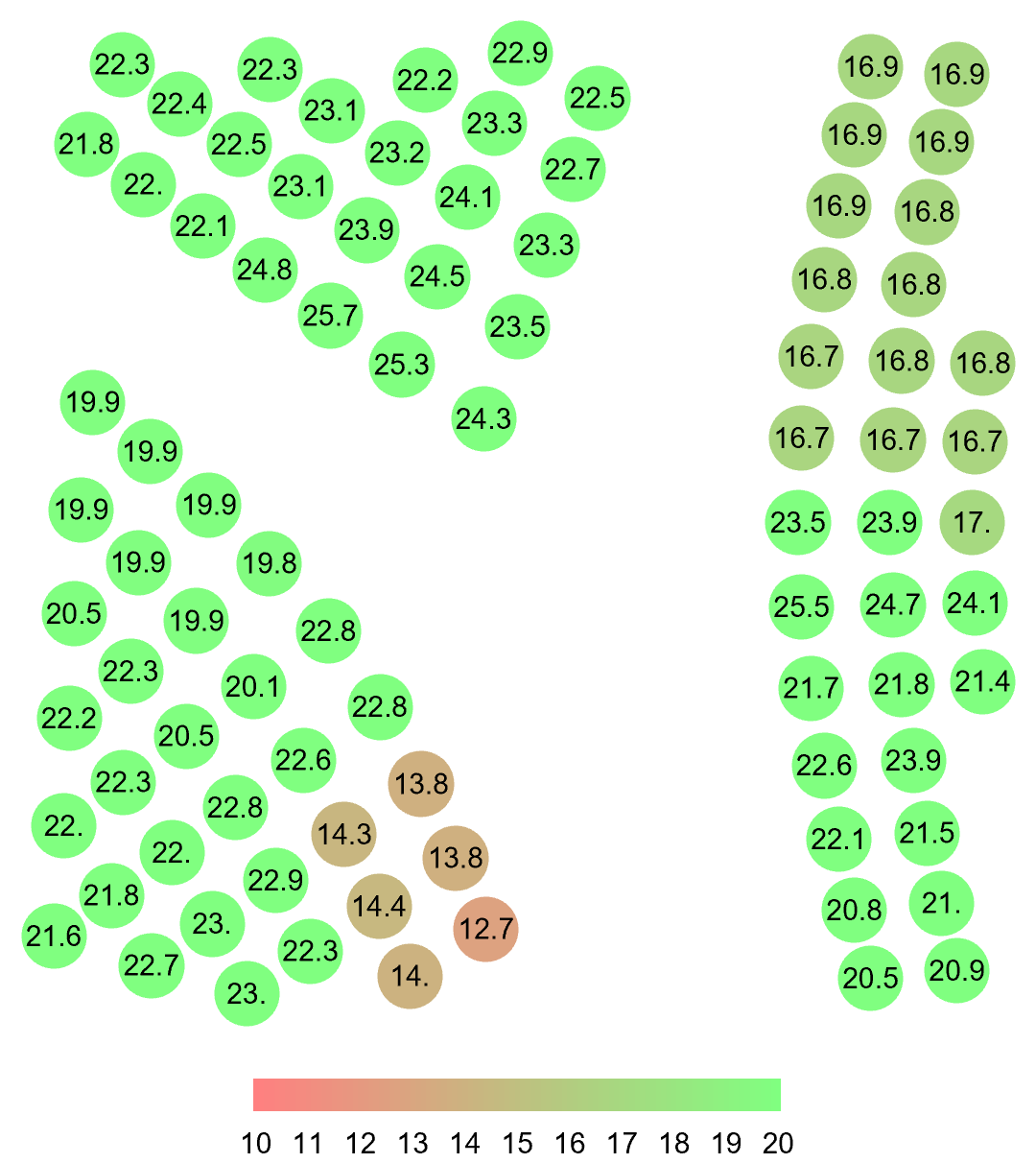}\hspace{8em}
 \includegraphics[width=0.35\textwidth]{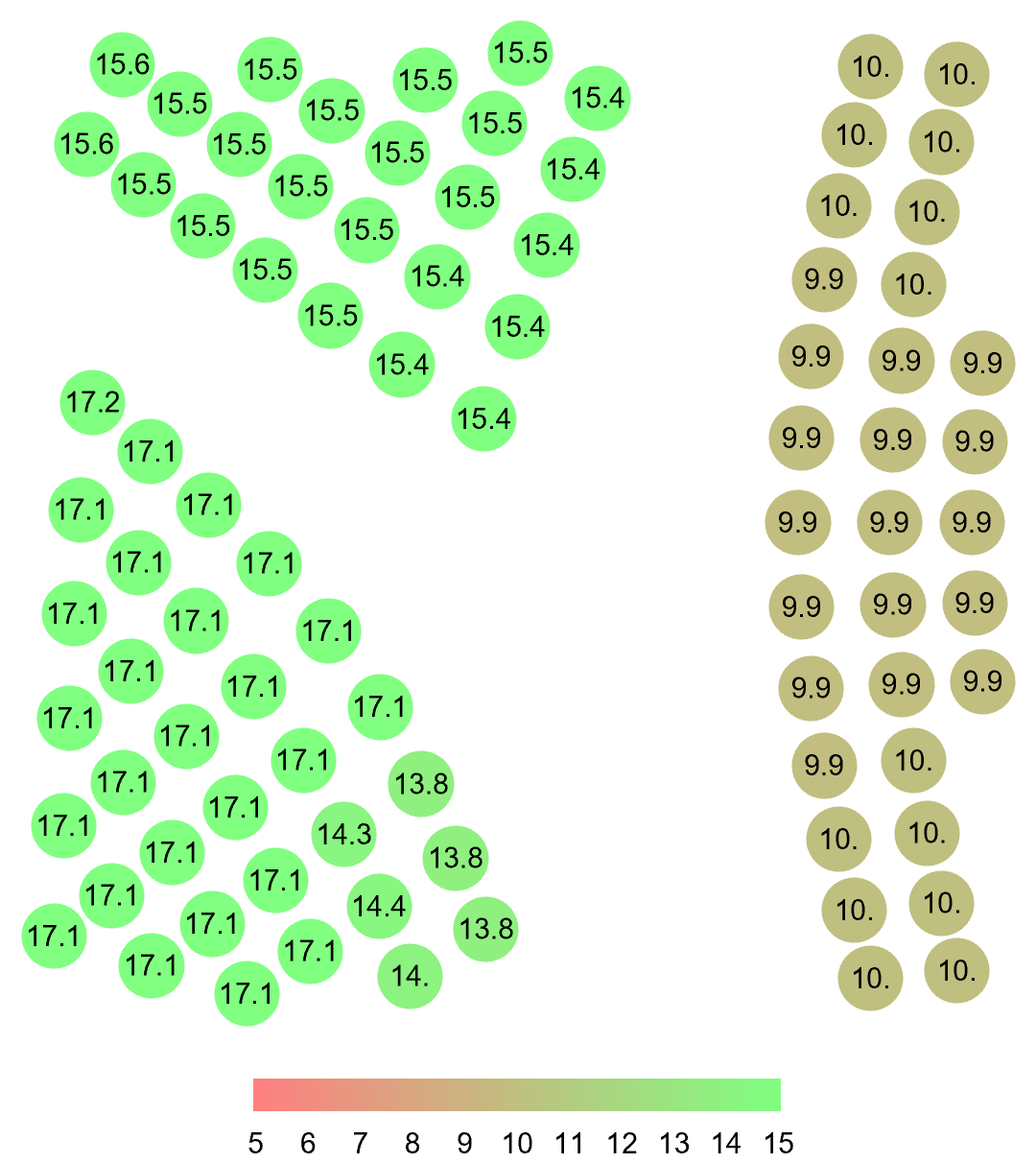}
 \caption{Left: Comparison of the pole locations of the elliptic transcendent before and after the $\mS_0$ Stokes transition. Most pole locations agree to 15-25 decimal places. Right: Comparison of the pole locations of the same transcendent before and after the $\mS_{-\pi}$ transition. The pole locations on the left agree to 15-17 decimal places; the `$\gs_2$-dependent' pole locations on the right to $\sim$10 decimal places.}
 \label{fig:comp011}
\end{figure}

\sk
The differences in pole locations before and after the $\mS_0$ and $\mS_{-\pi}$ transitions are plotted in figure \ref{fig:comp011}. The poles before and after the $\mS_0$ transition (left plot) agree generally to 15-25 decimal places, with a few outliers that agree to 12-15 decimal places. The lower precision is a random effect caused by the numerical solver: when lattice points are approached along a path that comes relatively close to one of the poles, some precision is lost. The paths are chosen randomly, so when we run the solver again, precision for these poles increases, but often a few other poles then get worse precision. By running the solver several times, we can in fact check all pole positions to a precision of over 20 decimal places.

\sk
The poles before and after the $\mS_{-\pi}$ transition (right plot) agree to 10-17 decimal places, where the upper limit of around 17 decimal places is again caused by the fact that $z_0$ is relatively close to the pole field. The lower precision poles are the ones in the two sectors on the right, whose location is mostly determined by $\gs_2$. Since in the original Stokes sector, $\gs_2=10^{-6}$ was taken smaller than $\gs_1=10^{-4}$, and we sum up to $\gs_1^7 \sim 10^{-28}$, this means that any contributions that we compute beyond order $\gs_2^5 \sim 10^{-30}$ are essentially irrelevant. Most likely, this explains the somewhat lower precision in the locations of the strongly $\gs_2$-dependent poles.

\sk
{\bf Towards even better numerics}

\nopagebreak
\sk
Unfortunately, the $\frac12 \bZ$-symmetry on the parameter space that we discussed in section \ref{sec:multiplesolutions} is much more difficult to test numerically. The main problem is that the symmetry shifts the parameter product $\ga \gb \to \ga \gb \pm \frac12$, which means that we have to resum at least one solution for which $|\ga \gb| \geq \frac14$. The main difficulty then lies in summing the diagonal $(n|n)$ sectors of the transseries, since those sectors are not suppressed by an instanton-type transmonomial, and are now multiplied by the $n$th power of the relatively large number $\ga \gb$ -- resulting in very slow convergence. One might attempt to compensate this by taking a large value of $z$, since these sectors also have a leading coefficient $z^{-\frac58 \gb_{nn}} = z^{-5/4}$, but this in turn means that one of the instanton transmonomials $e^{\pm A z^{5/4}}$ becomes very large (or both are of order 1), giving the sum over either the $(n|0)$ sectors or the $(0|m)$ sectors bad convergence properties.

\sk
One might attempt to improve our numerics further by making the individual Borel-Padé sums for the $(n|m)$ sectors more precise using the clever summation techniques of \cite{Cos6, Cos7}, though the results of \cite{BSSV} seem to suggest that for the particular case of Painlevé I these methods give only a slight improvement. Another promising approach, that could in particular resolve the issues for the $\frac12 \bZ$-symmetry discussed above, would be to sum over $n$ and $m$ exactly (following the idea of ``transasymptotic summation'' of \cite{Cos3, Cos5}) and only afterwards doing the asymptotic Borel-Padé sum. In this regard, the techniques of \cite{ASV2} are promising, and we hope to come back to this issue in the future.

\subsection{Exact expressions for the Stokes data}
\label{sec:StokesConstants}
From the resurgence point of view taken in e.g.\ \cite{ASV1, BSSV}, we can describe the Stokes automorphism not just as an operation acting on the transseries parameters, but as one acting on the different instanton sectors of the full transseries. This point of view is implemented by using the \textit{alien derivatives} which build up the automorphisms -- here with an underscore as they represent the two fundamental types of transitions -- for the first Painlevé equation (\ref{eq:P1}) in the following manner:
\begin{equation}
    \begin{split}
        \underline{\mS}_{0} &= \exp\left(e^{- A z^{5/4}}\Delta_{A}\right)\\
        \underline{\mS}_{\pi} &= \exp\left(\sum_{l=1}^\infty e^{l A z^{5/4}}\Delta_{-lA}\right),
    \end{split}
\end{equation}
where $A = \frac{8\sqrt{3}}{5}$ is the instanton action. The alien derivatives $\gD_{kA}$, when acting on the transseries sectors $u^{(n|m)}(z)$ of (\ref{eq:ufull}), yield linear combinations of sectors weighted by integer multiples of the \textit{Stokes constants}, indicating a deep (resurgent) relation between these different sectors. The collection of all these Stokes constants is sometimes called the \textit{Stokes data}\footnote{`Stokes data' can also  represent the collection of Stokes multipliers. We stick to the use of {\em monodromy data} for the Stokes multipliers of the linear problem and {\em Stokes data} for the Stokes constants of Painlevé I.} and it should contain the same information as our connection formulae. In this subsection we check that this is indeed the case by expanding our connection formulae in powers of the transseries parameters, which allows us to compute the Stokes constants up to arbitrary order. We can compare these results to the numerical values that were obtained in \cite{ASV1}, and more importantly, to the recent work of \cite{BSSV} where all of the Stokes data were obtained using a mix of asympotic analysis, alien calculus and numerical methods. The results of this latter work can be summarised in the connection formulae for $\underline{\mS}_0$ and $\underline{\mS}_\pi$ transition that they reconstructed from the Stokes data. Hence, in order to fully verify the agreement of our results with those of \cite{BSSV}, we show how their connection formulae relate to the ones derived here.

\sk
For definiteness, and to connect to the existing literature, let us compute the Stokes constants with respect to the forward motion along the alien chain \cite{ABS1}, i.e.\ those encoded in $\underline{\mS}_0$. For this purpose it is important which branch $\mS_{2n\pi}$ we pick to represent $\underline{\mS}_0$. A different choice leads to different Stokes constants, although they are related to the original ones by relatively simple translations. The right choice (i.e.\ the one matching the results of \cite{ASV1, BSSV}) turns out to be $\underline{\mS}_0 = \mS_{2\pi}$. In order to match our results to the literature we also need to rescale our parameters $(\alpha, \beta) = -3^{-1/4}(\sigma_1, \sigma_2)$, as we did in section \ref{sec3.1}. Obtaining the Stokes constants is then achieved in two steps.

\sk
First of all we take the $u^{(1|0)}$ sector which carries a single factor of $\sigma_1$, and study all the contributions it receives when the Stokes automorphism $\underline{\mS}_{0}$ acts on the complete transseries. Since we are moving forward along the alien chain, all contributions will come from lower degree sectors (see e.g.\ \cite{ABS1} and section 4 of \cite{ASV1}), i.e.\ sectors for which $n-m <1$. One then finds that after the Stokes transition the new transseries parameter $\sigma_1'$ can be expressed in terms of the old values and the Stokes constants in the following way:
\begin{equation}
     \sigma_1' = S_1^{(0)}+\sigma_1 +\bigg(\Big(S_2^{(0)}+\frac{1}{2}S_1^{(0)}S_1^{(1)}\Big)+S_1^{(1)}\sigma_1\bigg)\sigma_2 +\cO(\sigma_2^2),
\end{equation}
where $S_1^{(0)}=S_1$ denotes the already familiar leading Stokes constant. Now, we should be able to reproduce this expression using our connection formulae (\ref{eq:mij})--(\ref{eq:ab2}) and then read off the values of the Stokes constants. To this end, we express the new transseries parameter $\sigma_1' = -3^{1/4} \alpha'$ as a perturbative expansion in the old parameters $\sigma_1$ and $\sigma_2$. This yields
\begin{equation}
    \sigma_1' = \frac{-i3^{1/4}}{2\sqrt{\pi}}+\sigma_1 +\bigg(\frac{ i\pi -\gamma_E-\log(96\sqrt{3})}{2\pi}-\frac{2i(\gamma_E+\log(96\sqrt{3}))}{3^{1/4}\sqrt{\pi}}\sigma_1\bigg)\sigma_2 +\cO(\sigma_2^2).
\end{equation}
Equating this expression to the previous one we can now straightforwardly compute the Stokes constants one by one. When we generalise this procedure to higher order, we always find a linear system of equations that contains exactly as many equations as Stokes constants to solve for. The exact values of the first few constants are given in table \ref{T2}. The results exactly match the values of the {\em forward} Stokes vectors found in \cite{BSSV}, and in particular confirm the analytical value of the additional constant $\tilde A =  96\sqrt{3}$ that was found in that work.

\sk
Besides analytical expressions for the Stokes data, in \cite{BSSV} connection formulae for the $\underline{\mS}_0$ and $\underline{\mS}_{\pi}$ transitions were formulated. These formulae were reconstructed from the Stokes data and read
\be
    \begin{pmatrix}
         \sigma_1' \\
         \sigma_2'
    \end{pmatrix}
    =
    \begin{pmatrix}
         \frac{\left(P_0(\sigma_2 N^{(1)}\left(\sigma_1\sigma_2)\right)+\sigma_1\sigma_2\right) N^{(1)}\left(P_0(\sigma_2 N^{(1)}\left(\sigma_1\sigma_2)\right)+\sigma_1\sigma_2\right)}{\sigma_2 N^{(1)}(\sigma_1\sigma_2)} \\
         \frac{\sigma_2 N^{(1)}(\sigma_1\sigma_2)}{N^{(1)}\left(P_0(\sigma_2 N^{(1)}\left(\sigma_1\sigma_2)\right)+\sigma_1\sigma_2\right)},
         
    \end{pmatrix}
    \label{eq:BSSV0}
\ee
for the $\underline{\mS}_0$ transition and 
\be
    \begin{pmatrix}
         \sigma_1' \\
         \sigma_2'
    \end{pmatrix}
    =
    \begin{pmatrix}
     \frac{\sigma_1 N^{(-1)}(\sigma_1\sigma_2)}{N^{(-1)}\left(P_\pi(\sigma_1 N^{(-1)}\left(\sigma_1\sigma_2)\right)+\sigma_1\sigma_2\right)} \\
    \frac{\left(P_\pi(\sigma_1 N^{(-1)}\left(\sigma_1\sigma_2)\right)+\sigma_1\sigma_2\right) N^{(-1)}\left(P_\pi(\sigma_2 N^{(-1)}\left(\sigma_1\sigma_2)\right)+\sigma_1\sigma_2\right)}{\sigma_1 N^{(-1)}(\sigma_1\sigma_2)} 
    \end{pmatrix}
    \label{eq:BSSVpi}
\ee
for the $\underline{\mS}_\pi$ transition. Here the $N^{(\pm 1)}(x)$ are \textit{generating functions} for the Stokes data\footnote{The second of these equations differs by a minus sign from the equations in earlier versions of \cite{BSSV}, which was a consequence of earlier branch-cut choices in the literature.}:
\bea
    N^{(1)}(x) & = & N_1^{(1)}\frac{(96\sqrt{3})^{\frac{2 x}{\sqrt{3}}}}{\Gamma(1+\frac{2x}{\sqrt{3}})} \ret
    N^{(-1)}(x) & = &  i e^{\frac{2\pi i x}{\sqrt{3}}} N^{(1)}(-x).
\eea
Moreover, there are the functions
\be
    P_0(x) = i (N_1^{(1)})^2 \log\left(1-i \frac{x}{\left(N_1^{(1)}\right)^2}\right),
\ee
and $P_\pi(x) = -P_0(-x)$ and finally the constant $N_1^{(1)} =-\frac{3^{1/4}}{2\sqrt{\pi}}i$ which is in fact the leading Stokes constant $S_1$ that by now we have encountered several times.

\sk
The authors of \cite{BSSV} performed several numerical checks to test these expressions, and a way to analytically derive their formulae was described in the final subsection of that paper. Let us verify this derivation and fill in some details.

\sk 
We start by noting that
\bea
    P_0\left(\sigma_2 N^{(1)}(\sigma_1\sigma_2)\right)+\sigma_1\sigma_2 & = & \frac{i \sqrt{3}}{4\pi} \left(-\log\left(1-\frac{2\beta \sqrt{\pi} (96\sqrt{3})^{2\alpha \beta}}{\Gamma(1+2\alpha \beta)}\right) -4\pi i \alpha \beta\right) \ret
    &=&  \frac{i \sqrt{3}}{4\pi} \log\left(\frac{1+m_1m_2}{1+im_2}\right) \ret
    &=& -\frac{E_0'\sqrt{3}}{8}
    \label{eq:BSSV00}
\eea
where we switched to $(\alpha, \beta) = -3^{-1/4}(\sigma_1, \sigma_2)$ in the first line and in the second and third line used the Stokes multipliers and our own connection formula to rewrite the expression. Then we can write out the expression for $\beta' = -3^{-1/4}\sigma_2'$ in (\ref{eq:BSSV0}) to find that
\bea
    \beta' &=& \frac{ (96\sqrt{3})^{E_0'/4+2\alpha\beta}\Gamma(1-E_0'/4)}{\Gamma(1+2\alpha\beta)} \ \beta \ret
    &=& -\frac{im_2}{2\sqrt{\pi}}(96\sqrt{3})^{E_0'/4}\Gamma(1-E_0'/4)
\eea
which is precisely our expression (\ref{eq:ab2}) for $\beta'$ after the $2\pi$-transition in $\cS_2$. Hence, the connection formula for $\sigma_2'$ checks out. Verifying the connection for $\sigma_1'$ in (\ref{eq:BSSV0}) then becomes almost trivial: we first note that 
\be
    \sigma_1'  \sigma_2' = P_0\left(\sigma_2 N^{(1)}(\sigma_1\sigma_2)\right)+\sigma_1\sigma_2 = -\frac{E_0'\sqrt{3}}{8}.
\ee
Since this product is equivalent to the product $\alpha'\beta'\sqrt{3}$ from our own analysis, this also verifies the expression for $\sigma_1'$ and hence the connection (\ref{eq:BSSV0}) completely matches our $\mS_{2\pi}$ transition.

\sk
For connection (\ref{eq:BSSVpi}), the derivation is practically the same. Analogous to (\ref{eq:BSSV00}) we compute 
\bea
    P_\pi\left(\sigma_1 N^{(-1)}(\sigma_1\sigma_2)\right)+\sigma_1\sigma_2 & = & \frac{i \sqrt{3}}{4\pi} \left(\log\left(1+i \frac{2\alpha   \sqrt{\pi} e^{2\pi i \alpha \beta }(96\sqrt{3})^{-2\alpha \beta}}{\Gamma(1-2\alpha \beta)}\right) -4\pi i \alpha \beta\right). \ret
    &=&  \frac{i \sqrt{3}}{4\pi} \log\left(\frac{1+i m_3}{1+m_2m_3}\right) \ret
    &=& -\frac{E_0'\sqrt{3}}{8}.
    \label{eq:BSSVpipi}
\eea

Then we can write the expression for $\alpha' = -3^{-1/4}\sigma'_1$  in (\ref{eq:BSSVpi}) as

\bea
    \alpha'&  = &\frac{(96\sqrt{3})^{-E_0'-2\alpha \beta}e^{i \pi E'_0/4+2\pi i \alpha \beta}\Gamma(1+E'_0/4)}{\Gamma(1-2\alpha \beta)} \alpha \ret
    & = & \frac{m_3}{2\sqrt{\pi}}(96\sqrt{3})^{-E_0'/4}e^{i\pi E_0/4} \Gamma(1+E'_0/4)
\eea
which matches (\ref{eq:ab3}) and therefore represents one of the transseries parameters in $\cS_3$ after the $3\pi$-transition. Subsequently, using the same logic as before, we deduce that the expression for $\sigma_2'$ in (\ref{eq:BSSVpi}) also checks out and thereby confirm a match between the connection formula for the backwards motion in \cite{BSSV} and our $\mS_{3\pi}$ transition.

\begin{table}
    \centering
    \begin{tabular}{|c|r|c|c|}
    \hline
     & Numerical value & Analytical value& \begin{tabular}{@{}c@{}}In terms of \textit{proportionality} \\ \textit{ factors} of \cite{BSSV}\end{tabular}  \\
     \hline \hline
     $S_1^{(0)}$& $-0.371257...i$ & $-\frac{i 3^{1/4}}{2\sqrt{\pi}}$ & $N_1^{(1)}$ \\
     $S_2^{(0)}$& $0.500000...i$ &   $\frac{i}{2}$ & $2 N_1^{(2)}$\\ 
     $S_3^{(0)}$& $-0.897849...i$ &   $-\frac{2i\sqrt{\pi}}{3^{4/3}}$ & $3 N_1^{(3)}$\\
     $S_1^{(1)}$& $-4.879253...i$ & $-\frac{2i}{3^{1/4}\sqrt{\pi}}\Big(\gamma_E+\log(96\sqrt{3})\Big)$ & $2 N_0^{(1)}$\\
     $S_2^{(1)}$& $9.856875...i$ &  $i\sqrt{3}\Big(\gamma_E+\log(96\sqrt{3})\Big)$ & $3 N_2^{(2)}$ \\
     $S_1^{(2)}$& $-22.825711...i$ &  $-\frac{i 3^{1/4}}{\sqrt{\pi}}\left(-\frac{\pi^2}{6}+\Big(\gamma_E+\log(96\sqrt{3})\Big)^2\right)$ & $3 N_{-1}^{(1)}$ \\
     \hline
\end{tabular}
    \caption{Stokes constants for the forward motion along the alien chain of the Painlevé I transseries solution. Numerical values were found in \cite{ASV1}. Our analytical values exactly match those obtained by using the Stokes data generating function of \cite{BSSV}.}
    \label{T2}
\end{table}

\section{Conclusion}
In this paper, we have seen that two different perspectives on the study of the Painlevé I equation fit together nicely. On the one hand there is the long-established method of isomonodromic deformation that led to many insightful results, including the computation of the first Stokes constant \cite{Kap1, Tak1} and connection formulae for elliptic type asymptotics \cite{KK}. On the other hand, there is the more recent endeavor, starting with \cite{GIKM}, to study two-parameter transseries solutions to the Painlevé I equation by using resurgence, culminating in the results of \cite{BSSV}. Our hope is that this paper has bridged the gap between both approaches. To this end, we extended the work of Aoki, Kawai and Takei to show that methods from the former school of thought can provide a complete description of the Stokes phenomenon of two-parameter transseries expansions that the latter school has been studying recently. 

\sk
From our connection formulae we were able to extract Stokes constants up to arbitrary order, confirming the results of \cite{BSSV} which were obtained using alien calculus. The explicit relations between the transseries parameters and the Stokes multipliers led to some interesting new insights on how the  the different transseries representations relate to the three classes of Painlevé I transcendents. The relation between these two is clearly many-to-one. At several levels did we find a degeneracy of different transseries expansions that represent the same Painlevé transcendent. Of course, there is an ambiguïty in the choice of branches for $z^{-5/8}$ and $\log z^{-5/4}$ that leads to different transeries representations, and we have seen that those choices are all interrelated through the $T$ transformations, but we have also found other more non-trivial degeneracies. First of all, there is a $\frac12 \bZ$ symmetry in the transseries parameter product $\alpha\beta$, which allows for different transseries expansions of the same transcendent. Secondly, for the tronquée and tritronquée solutions in particular, we find that after a single $2\pi$ rotation around the origin in the $z$-plane, not only do we end up with a different pair of transseries parameters, but also with different {\em types} of transseries representations, as summarized in table \ref{tab:solutions}.

\sk
There are some clear follow-up questions to the work presented in this paper. First of all, the methods presented here can be applied to other Painlevé equations as well, and it would be interesting to see what the results are in those cases -- and in particular whether there is any universality in the connection formulae and Stokes constants that one would find\footnote{Note that for Painlevé I and a particular version of Painlevé II, there is certainly such a universality, which was exploited in \cite{BSSV} to treat these two equations using the same framework. Moreover, in \cite{KT4} a local reduction theorem is formulated that relates two-parameter solutions of the higher Painlevé equations to those of Painlevé I. Hence, there is substantial evidence to believe that there is such a universality.}. On the numerical side, it would be interesting to find methods to also compute with arbitrarily large values of the transseries parameters $\ga$ and $\gb$, in the domain where the instanton sums do not converge and where, therefore, some form of analytic continuation is needed. A third interesting open question would be to geometrically understand the space of transseries parameters $(\ga,\gb)$, including all identifications that we have mentioned, and show that it is equal to the well-known space of Stokes multipliers, or perhaps some covering thereof.

\sk
In our minds, the most interesting question left open by our results is how the connection formulae for two-parameter transseries relate to the results of Kapaev and Kitaev in \cite{KK}. In that paper the authors managed to give a complete description of the elliptic asymptotics,
\be
u(z) = \sqrt{z} \ \wp\left(\frac{4}{5}z^{5/4}-s\ ;\ g_2\ ,\ g_3 \right)+\cO(z^{-3/4}),
\label{eq:EllAs}
\ee
of elliptic type Painlevé I transcendents in the whole complex $z$-plane. They explained how the parameters $s$, $g_2$ and $g_3$ depend on the Stokes multipliers and the argument of $z$, thereby providing a description of the non-linear Stokes phenomenon for elliptic type solutions. Since these are the same solutions that can be expanded in a two-parameter transseries, this begs the question how our connection formulae fit with the description of \cite{KK}. In order to answer such a question, one first needs to understand how the elliptic asymptotics (\ref{eq:EllAs}) of these solutions can be translated into a two-parameters transseries and vice versa. We believe that Takei has already made some compelling progress towards answering this question, and further interesting insights may come from \cite{ASV2}. The numerical checks that we have performed in this project, combined with Takei's recent work \cite{Tak2, Tak3}, strongly suggest that the two-parameter transseries are indeed the analytical continuation of the Fourier expansion of (Weierstrass) elliptic functions along the anti-Stokes line. Confirming this conjecture would not only provide an interpretation for the exponentially growing instanton factors of two-parameter transseries, but would also be a first step towards connecting our description of the Stokes automorphism to the results of Kapaev and Kitaev. We hope to return to this issue in the future.

\acknowledgments
We thank Ricardo Schiappa for discussing the results in \cite{BSSV} with us before publication, and Inês Aniceto, Salvatore Baldino,  Ricardo Schiappa, Maximilian Schwick and Roberto Vega for further discussions on topics closely related to those reported in this paper as well as comments on an earlier version of this work. We are grateful to Yoshitsugu Takei for sharing some of the original work that we were not able to find online or in libraries, and for useful discussions that provided us further insight in his work and that of his collaborators. We thank the organizers of the workshop `Applicable resurgent asymptotics: towards a universal theory' (ARA) at the Isaac Newton Institute in Cambridge for the opportunity to present and discuss an early version of the results of this research. A preliminary version of the results in this paper was in fact presented there \cite{talk-Alexander}. The research of AvS was supported by the grant OCENW.KLEIN.128, `A new approach to nonperturbative physics', from the Dutch Research Council (NWO).


\appendix

\section{The exact WKB method}
\label{app:exactWKB}
In this appendix we briefly recall the most important tools of the exact WKB method. We do so under the conventions used by Aoki, Kawai and Takei in \cite{Tak1, AKT2, KT,  AKT3}. Our description is mostly based on chapters 2 and 3 of \cite{Iwa2}, which also uses these conventions. 

\sk
We start with a Schrödinger type equation with a large parameter $\eta$, playing the role of $\hbar^{-1}$:
\begin{equation}
    \bigg(\frac{\diff^2}{\diff x^2} -\eta^2Q(x, \eta)\bigg) \psi(x, \eta) = 0.
    \label{eq:SVap}
\end{equation}
We want to think of the solutions $\psi(x, \eta)$ to this equation as living on the complex $x$-plane $\mathbb{C}_x$. The `potential' $Q(x, \eta)$ can itself have quantum corrections and is assumed to be a polynomial in $\eta^{-1}$:
\begin{equation}
    Q(x,\eta) = Q_0(x)+Q_1(x)\eta^{-1}+Q_2(x)\eta^{-2} +...+Q_N(x)\eta^{-N}.
\end{equation}
The coefficients $Q_n(x)$ are meromorphic functions on $\mathbb{C}_x$ and the `classical' part $Q_0(x)$ is called the \textit{principal term}. In the $\eta$-independent case where $Q(x,\eta) = Q_0(x) = 2(V(x)-E)$ we get the familiar time-independent Schrödinger equation from quantum mechanics, where $V(x)$ is the potential and $E$ the energy of a specific state. With a slight abuse of terminology, $Q(x,\eta)$ is also referred to as the potential, and we shall adopt this convention. 

\sk
The following ansatz \cite{Dun1} can be used to solve the equation (\ref{eq:SVap}):
\begin{equation}
    \psi(x,\eta) = \exp\left( \int^x S(x', \eta) \diff x'\right),
    \label{eq:ansatzWKB}
\end{equation}
where we omit the lower integration bound which only serves as an overal multiplicative constant that normalizes the solution. After pluging this ansatz back into the ODE, we obtain a Riccati equation for $S(x, \eta)$:
\begin{equation}
    S(x, \eta)^2+\frac{\diff S}{\diff x}(x, \eta) = \eta^2 Q(x, \eta).
    \label{Riccatieq}
\end{equation}
This equation can then be solved for $S(x, \eta)$ in terms of a formal power series in $\eta^{-1}$,
\begin{equation}
    S(x, \eta) = \sum_{n=-1}^\infty \eta^{-n} S_{n}(x),
\end{equation}
which lets us recursively determine the coefficients $S_n(x)$. The first of these is $S_{-1}(x) = \pm \sqrt{Q_0(x)}$, with a choice of sign that leads to two branches of solutions, $\{S_n^{\pm}(x)\}$. We do not always write the $\pm$ index in our notation; any result with the index omitted will apply to both $S^{+}$ and $S^{-}$, and similarly for the corresponding wave functions that we denote $\psi_+$ and $\psi_-$.  

\sk
A nice property of the solutions $S(x, \eta)$ is that we can split the formal series into an \textit{odd} and an \textit{even} part,
\begin{equation}
    S(x, \eta) = S_{\text{odd}}(x, \eta)+S_{\text{even}}(x, \eta),
\end{equation}
where $S_{\text{odd}} = (S^+ -S^-)/2$ and $S_{\text{even}} = (S^+ +S^-)/2$. Using these definitions and the Riccati equation it is a straightforward exercise to show that
\begin{equation}
    S_{\text{even}}(x, \eta) = -\frac{1}{2}\frac{\diff }{\diff x}  \log\big(S_{\text{odd}}(x,\eta)\big) .
\end{equation}
This allows us to rewrite the solution as 
\begin{equation}
    \psi_\pm(x) = \Big(S_{\text{odd}}(x, \eta)\Big)^{-1/2}\exp\bigg(\int^x S_{\text{odd}}(x', \eta)\diff x'\bigg).
    \label{eq:WKBsol}
\end{equation}
We can then, up to an intanton-like exponential prefactor, expand the full solution in powers of $\eta^{-1}$ and obtain a formal series:
\begin{equation}
    \psi_\pm(x) = \eta^{-1/2} \exp\left(\pm\eta\int^x \sqrt{Q_0(x')} \diff x'\right)\sum_{k=0}^\infty\eta^{-k}\psi_{\pm, k}(x).
    \label{eq:WKBsol2}
\end{equation}
Note that in the case of $Q_0(x) = 2(V(x)-E)$, we retrieve the original WKB approximation
\begin{equation}
    \psi_\pm(x) \sim \frac{1}{\sqrt{p(x)}} \exp\bigg(\pm \frac{i}{\hbar} \int^x p(x') \diff x'\bigg)(1+\mathcal{O}(\hbar))
    \label{eq:WKBsol3}
\end{equation}
where we have replaced $\eta= \hbar^{-1}$ and  $p(x) :=  \sqrt{2(E-V(x))}$ is the classical momentum. Usually, we normalise solutions (\ref{eq:WKBsol}) at one of the zeroes of $Q_0(x)$ -- that is, we start the integration at such a point -- but sometimes we want to normalise a solution {\em at infinity}. The only problem is that the leading part of the riccati solution $S_{-1}(x) = \sqrt{Q_0(x)}$ diverges at infinity. Hence, we write such solutions normalised at infinity as
\begin{equation}
    \psi_\pm(x) = \Big(S_{\text{odd}}(x, \eta)\Big)^{-1/2}\exp\bigg(\eta \int_a^x S_{-1} \diff x' +\int^x_\infty \left(S_{\text{odd}}(x', \eta)-\eta S_{-1}\right)\diff x'\bigg),
    \label{eq:WKBpole}
\end{equation}
where $a$ is a zero of $Q_0(x)$. One can check that the higher order terms of the Riccati solution are well-defined at infinity.

\sk
The leading part of the potential, i.e.\ the principal term $Q_0(x)$, determines a Riemann surface $\cS$ called the \textit{WKB curve} defined by
\begin{equation}
    y^2 = Q_0(x).
\end{equation}
In terms of this Riemann surface, many constructions become geometric in nature; for example, one can think of the factor in the exponential of (\ref{eq:WKBsol2}) as an integral of the one-form $y\diff x$. The surface $\cS$ is a double cover of the Riemann sphere where the two sheets branch at zeroes of $Q_0(x)$. Although $S_{odd}$ and hence $\psi$ are multivalued on the Riemann sphere, they are single valued on the WKB curve $\cS$. This allows us to integrate the formal power series $S_{odd}(x)$ along a closed cycle $\gamma$ encircling two zeroes and exponentiate to obtain
\begin{equation}
    \cV_\gamma = \exp\left(\oint_{\gamma}S_{odd}(x)\diff x\right),
\end{equation}
which is called a \textit{Voros symbol}. When $Q_0(x)$ has poles (including a possible pole at infinity), we can also consider a path $\beta$ starting at such a pole, encircling a zero and returning to the pole. In such a situation we can consider the alternative Voros symbol
\begin{equation}
    \cV_\beta = \exp\left(\int_{\beta}\Big(S_{odd}(x)-\eta S_{-1}(x)\Big)\diff x\right).
    \label{eq:VorSym}
\end{equation}
Here, the leading part $\eta S_{-1}$ of the Riccati solution is removed since it blows up at the pole. Writing the latter symbol as $\cV= e^{W_{\beta}}$, the object $W_{\beta}$ is often called the \textit{Voros coefficient}. There are many interesting properties and applications of the Voros symbols, the most straightforward one being that they alter the normalization of the WKB solutions -- which is what we use them for in this paper. For example, the path $\beta$ might start at infinity, go around the point $x=a$ for which $Q_0(a)=0$ and back to its starting point. Then we see that the Voros symbol $\cV_\beta^{1/2}$ from (\ref{eq:VorSym}) turns a solution normalised at the turning point $a$ into a solution normalised at infinity, (\ref{eq:WKBpole}).

\begin{figure}
    \centering
    \includegraphics[width = 0.8 \textwidth]{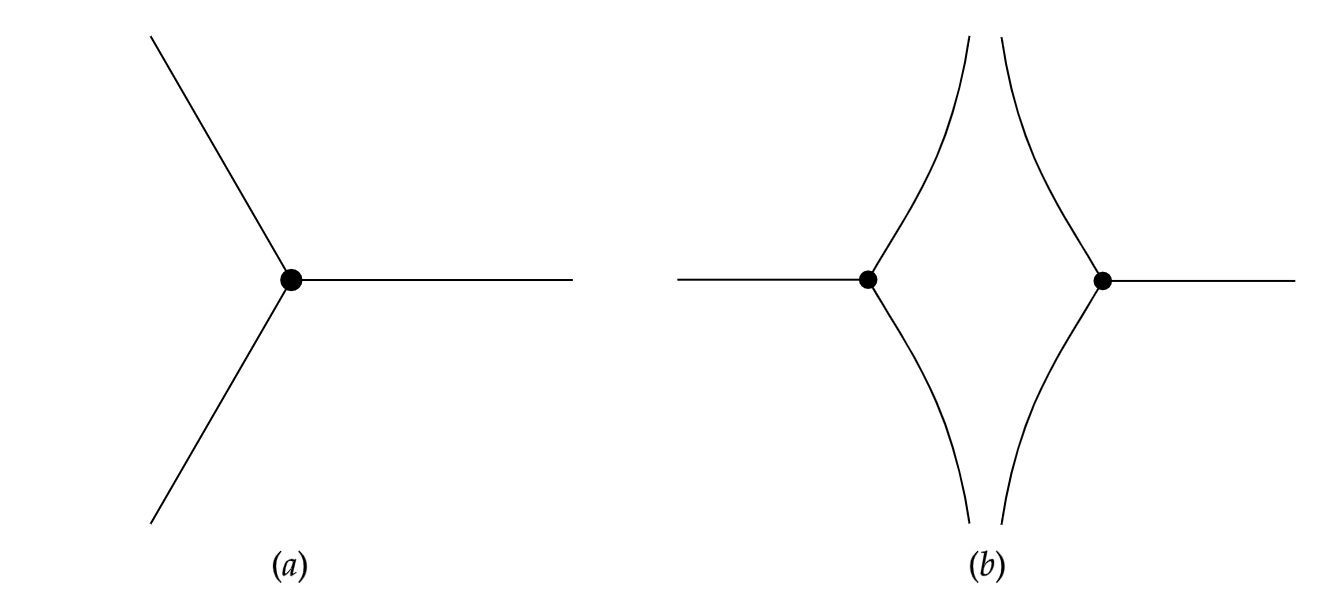}
    \caption{Two elementary examples of Stokes graphs: (a) for the Airy equation where $Q_0(x) = x$ and (b) for the quadratic potential $Q_0(x) = x^2-1$.}
    \label{fig:stokesexample}
\end{figure}

\sk
Finally, it will be useful to draw the zeroes of $Q_0(x)$ and the {\em Stokes lines} connected to these points in the complex $x$-plane $\bC_x$. These Stokes lines consist of the points $x$ for which
\be
\Im\left(\int_{a}^x\sqrt{Q_0(x')}\diff x'\right) = 0,
\ee
where $a$ is one of the zeroes of $Q_0(x)$. Figure \ref{fig:stokesexample} shows two such examples. At the Stokes lines, $\psi_+$ and $\psi_-$ are maximally growing/decaying, and as is the case for Painlevé I, on such a line the asymptotic behavior of a solution can jump. Therefore, one needs a basis transformation to relate the basis of growing and decaying solutions defined on one side of a Stokes line to that defined on the other side -- the {\em Stokes phenomenon}. These basis transformations will play a crucial role in our story. The diagram of Stokes lines is called the {\em Stokes graph} of our problem.

\section{Theory of isomonodromic deformation}
\label{app:isomon}
In this appendix we derive the Schrödinger differential equation for which the first Painlevé equation acts as a condition of isomonodromic deformation. The starting point is the following system of linear partial differential equations:
\begin{equation}
    \begin{split}
        \eta^{-1}\frac{\partial}{\partial x} Y&= A(x, t)Y,\\
        \eta^{-1}\frac{\partial}{\partial t} Y &= B(x, t)Y,\\
    \end{split}
    \label{Asystem}
\end{equation}
where $Y(x,t)$ is a two-component column vector of functions and $A$ and $B$ are the matrices
\begin{equation}
    \begin{split}
        A(x, t) & =\begin{pmatrix}
            \nu & 4(x-\lambda) \\
            x^2+x\lambda +\lambda^2+t/2 & -\nu
        \end{pmatrix},
        \\
        &\\
         B(x, t) & =\begin{pmatrix}
           0 & 2 \\
            x/2+\lambda & 0
        \end{pmatrix}.
    \end{split}
\end{equation}
The quantities $\nu(t)$, $\lambda(t)$ as well as $t$ itself can be regarded as external parameters of this linear problem. If we demand that variations of the solution $Y$ with respect to $t$ and $x$ commute, then we obtain the \textit{compatibility condition}
\begin{equation}
    \frac{\partial A}{\partial t} -\frac{\partial B}{\partial x} + \eta[A, B] = 0.
\end{equation}
A short computation shows that, as a condition on $\gl(t)$ and $\nu(t)$, this is equivalent to the Hamiltonian system (\ref{eq:ham}). The system has an irregular singularity at infinity around which the solution $Y$ has a non-trivial monodromy. In order to preserve the monodromy of the linear problem around this irregular singularity, the external parameters need to satisfy the compatibility condition and hence $\gl(t)$ and $\nu(t)$ produce solutions to the Painlevé I equation. For details, see e.g.\ \cite{JMU1} or chapter 4 of \cite{FKIN1}.

\sk
The linear system (\ref{Asystem}) with solution $Y = (Y_1, Y_2)^T$ can be reduced \cite{Iwa1} to the following set of scalar equations:
\begin{equation}
    \begin{split}
        &\left[\eta^{-2}\frac{\partial^2}{\partial x^2}- \frac{\eta^{-1}}{x-\lambda}\left( \eta^{-1} \frac{\partial}{\partial x}-\nu\right)-(4x^3+2tx+2H)\right] Y_1 =0\\[0.3cm]
        &\left[\eta^{-1}\frac{\partial}{\partial t}-\frac{1}{2(x-\lambda)}\left(\eta^{-1}\frac{\partial}{\partial x} -\nu\right)\right]Y_1 = 0
    \end{split}
\end{equation}
by identifying $Y_2 = (4(x-\lambda))^{-1}(\eta^{-1}\frac{\partial }{\partial x}-\nu)Y_1$. Of the first equation we can remove the first order derivative by introducing a new solution defined via  $Y_1 = 2\sqrt{x-\lambda}\Psi$, which yields the Schrödinger equation

\begin{equation}
    \left[\eta^{-2}\frac{\partial^2}{\partial x^2} -\big(4x^3+2tx+2H)+\eta^{-1} \frac{\nu}{x-\lambda}-\eta^{-2} \frac{3}{4(x-\lambda)^2}\right] \Psi =0.
\end{equation}


\section{Weber analysis}
\label{app:weber}
In this appendix, following \cite{Tak1} but with some minor adjustments, we review how one obtains the connection formulae for the local solutions $\varphi(\zeta)$ to (\ref{eq:S1}). The trick to achieve this is to map the local problem (\ref{eq:S1}) to a Weber differential equation. The solutions of that equation are known to be parabolic cylinder functions. The asymptotics of these solutions are well known and allow us to establish connection formulae for the Weber equation. By mapping the asymptotics back to the original local differential equation, we can obtain connection formulae for asymptotic solutions $\varphi(\zeta)$.

\sk
Let us start with the fact that, as is the case for any other Schrödinger type problem, the WKB ansatz $\varphi_\pm = \exp\bigg( \pm\int^{\zeta}T(\zeta', \eta)\diff \zeta'\bigg)$ for (\ref{eq:S1}) is recursively constructed by solving the associated Riccati equation
\begin{equation}
    T(\zeta, \eta)^2+\frac{\partial T}{\partial \zeta}(\zeta, \eta) = \eta^2 \tilde Q(\zeta, \eta),
\end{equation}
order by order in $\eta^{-1/2}$, analogous to (\ref{Riccatieq}). In the case of (\ref{eq:S1}), we find that the first three terms of the solution $T(\zeta, \eta) = \sum_{n=-1}^{\infty} T_{n/2}(\zeta) \eta^{-n/2}$ of the Riccati equation are
\begin{equation}
    \begin{split}
        T^{(\pm)}_{-1} &= \pm 2\zeta \\
        T^{(\pm)}_{-1/2} &= 0\\
        T^{(\pm)}_0 &= \pm \frac{E}{4\zeta}-\frac{1}{2}\frac{\diff}{\diff \zeta}\log(T^\pm_{-1}).
    \end{split}
\end{equation}
One can check that any higher order odd terms scale with negative powers of $\zeta$. Our solutions $\varphi_\pm$ are normalised at infinity -- just like the asymptotic expansions of the parabolic cylinder functions that we shall meet shortly. Analogous to equation (\ref{eq:WKBpole}) in appendix \ref{app:exactWKB} we write
\begin{equation}
    \begin{split}
        \varphi_\pm(\zeta) &= \frac{\sqrt{2\eta}}{\sqrt{ T_{\text{odd}}}} \zeta^{\pm E/4} \exp\left(\pm\eta \int_0^\zeta T_{\text{odd},-1}\diff \zeta' \pm \int_\infty^\zeta\left(T_\text{odd}-\eta T_{\text{odd},-1}-\frac{E}{4\zeta'}\right)\diff \zeta'\right) \\[0.3cm]
        &= \zeta^{-1/2\pm E/4} \exp\big(\eta\zeta^2\big)\Big(1+\mathcal{O}(\eta^{-1/2})\Big)
    \end{split}
    \label{eq:phiexp}
\end{equation}
In constructing these solutions, the prefactor $\sqrt{2\eta}$ has been added to cancel the same factor apearing in $\sqrt{\eta T_{\text{odd},-1}}$ in the denominator. Also, the factor $\zeta^{\pm E/4}$ comes from integrating $\pm\int_\infty^\zeta  E/(4\zeta')\diff \zeta'$ where we ignore the lower bound, which gives an infinite overall constant that we can remove after regularisation.

\sk
Next, we map this solution to a new function $w(z)$ that satisfies the Weber equation. This is done using the series of transformations described in \cite{Tak1}. We start by defining
\begin{equation}
    \begin{split}
        \tilde v &= \sqrt{\zeta-\xi/\sqrt{\eta}} \hspace{2mm}\varphi\\
        \tilde w &= \frac{1}{\zeta-\xi/\sqrt{\eta}} \bigg(\frac{\diff}{\diff\zeta}\tilde v +\sqrt{\eta}\rho \tilde v\bigg),
    \end{split}
\end{equation}
where $\xi$ is the parameter that was introduced in (\ref{eq:Qnew}), and subsequently perform a linear transformation
\begin{equation}
    \begin{pmatrix}
         w \\
         v
    \end{pmatrix} = 
    \begin{pmatrix}
         1/\sqrt{\eta} & \; 2\sqrt{\eta} \\
         -1/\sqrt{\eta} & \; 2\sqrt{\eta}
    \end{pmatrix}
    \begin{pmatrix}
         \tilde w \\
         \tilde v
    \end{pmatrix}.
\end{equation}
Then one can show that the following system of first order ODEs for $(v, w)$ holds:
\begin{equation}
    \frac{d}{d\zeta} \begin{pmatrix}
         w \\
         v
    \end{pmatrix}
    = \begin{pmatrix}
         2\eta\zeta & -\sqrt{\eta }(\rho-2\xi) \\
         -\sqrt{\eta} (\rho+2\xi) & -2\eta \zeta
    \end{pmatrix}
    \begin{pmatrix}
         w \\
         v
    \end{pmatrix}.
\end{equation}
If we substitute $y = 2\sqrt{\eta} \; \zeta$ and express the system as a single ODE for $w$, we obtain the \textit{Weber equation}:
\begin{equation}
    \frac{\diff^2w}{\diff y^2}+\bigg(\kappa+\half-\frac{y^2}{4}\bigg)w =0,
    \label{eq:weber}
\end{equation}
where $\kappa = -E/4-1$. Now let $\cG$ denote the mapping between solutions $\varphi(\zeta)$ and solutions of the Weber equation, i.e.\  $\cG \varphi(\zeta) = w(y)$. Then we have
\begin{equation}
    w(y) = \cG\varphi\left(\frac{y}{2\sqrt{\eta}}\right):= \frac{\sqrt{2}\eta^{1/4}}{\sqrt{y-2\xi}}\left(2\frac{\diff}{\diff y}\left(\varphi\left(\frac{y}{2\sqrt{\eta}}\right)\right)+\left(y+\rho-2\xi+\frac{1}{y-2\xi}\right)\varphi\left(\frac{y}{2\sqrt{\eta}}\right)\right).
    \label{eq:wtrafo}
\end{equation}
At this point we have not yet picked any specific solution, but it is known that (\ref{eq:weber}) is solved by parabolic cylinder functions $D_\kappa(y)$. In fact, two of these functions $D_\kappa(y)$ and $D_{-\kappa-1}(iy)$ form a basis of solutions, which allows us to study the Stokes phenomenon of the Weber equation. The asymptotics of these functions are given by 
\begin{equation}
    \Dtilde_\kappa (y) = e^{-y^2/4} y^\kappa \sum_{n=0}^\infty (-1)^n \frac{(k)_{2n}}{2^n y^{2n}n!} \hspace{1cm} \text{for}\hspace{5mm}  |\arg(y)| <\frac{3}{4}\pi, 
    \label{eq:expweb}
\end{equation}
where $(k)_{2n}$ is a falling factorial and $\Dtilde_\kappa(y)$ is a useful notation introduced to denote the large $y$ asymptotics of the functions $D_\kappa(y)$. Moreover, we have the following useful identity for the parabolic cylinder functions:
\begin{equation}
    D_\kappa(y) = e^{i \pi \kappa} D_\kappa(-y)+\frac{\sqrt{2\pi}}{\Gamma(-\kappa)}e^{i\pi(\kappa+1)/2}D_{-\kappa-1}(-iy).
    \label{eq:conweb}
\end{equation}
We can now consider the basis of solutions $D_{-\kappa-1}(-iy) $ and $D_\kappa(y)$ located in the first quadrant, whose asymptotic expansions are given by (\ref{eq:expweb}). When we move to the next quadrant (counterclockwise), we need to change the basis of solutions such that the asymptotic expansion (\ref{eq:expweb}) remains in its region of validity. We can use (\ref{eq:conweb}) to analytically continue solutions to the different quadrants and then expand them to compare the asymptotics in each quadrant; this leads to
 \begin{equation}
     \begin{split}
     &\begin{cases}
          D_{-\kappa-1}^{\rnc{1}}(-iy) &= D_{-\kappa-1}^\rnc{2}(-iy)\\
    D_{\kappa}^\rnc{1}(y) &=  e^{i\pi \kappa } D_{\kappa}^\rnc{2}(-y) +\frac{\sqrt{2\pi}}{\Gamma(-\kappa)}e^{i\pi(\kappa+1)/2} D_{-\kappa-1}^\rnc{2}(-iy)
    \end{cases}\\
    &\begin{cases}
    D_{-\kappa-1}^{\rnc{2}}(-iy) &= e^{-i\pi(\kappa+1)}D_{-\kappa-1}^\rnc{3}(iy) +\frac{\sqrt{2\pi}}{\Gamma(\kappa +1)}e^{-i\pi\kappa}  D_{\kappa}^\rnc{3}(-y)\\
     D_{\kappa}^\rnc{2}(-y) &=  D_{\kappa}^\rnc{3}(-y)
    \end{cases}\\
    &\begin{cases}
    D_{-\kappa-1}^{\rnc{3}}(iy) &= D_{-\kappa-1}^\rnc{4}(iy)\\
     D_{\kappa}^\rnc{3}(-y) &=- e^{i\pi \kappa }  D_{\kappa}^\rnc{4}(y) +\frac{\sqrt{2\pi}}{\Gamma(-\kappa)}e^{i\pi(\kappa+1)/2}D_{-\kappa-1}^\rnc{4}(iy)
    \end{cases}
     \end{split}
     \label{eq:B2}
 \end{equation}
where the Roman numeral in the superscript denotes the quadrant in which we use a solution. 

\sk
Having arrived at these connection formulas, we now want to map back to the $\varphi$-solutions. If we invert relation (\ref{eq:wtrafo}), we get 
\begin{equation}
    \varphi(\zeta) = \frac{1}{4\eta^{1/2}\sqrt{\zeta-\eta^{-1/2}\xi}}\left(\frac{1}{\rho-2\xi}\left(-2\frac{\diff w}{\diff y}+y w \right)+w \right)\Bigg|_{y=2\sqrt{\eta }\zeta}.
    \label{Gmap}
\end{equation}
This expression allows us to map the asymptotics of solutions $w(y)$ back to the asymptotics of solutions $\varphi(\zeta)$ to (\ref{eq:S1}). These are not necessarily the $\varphi_\pm(\zeta)$ solutions that we considered in (\ref{eq:localWKBsol}), but we should be able to express the resulting $\varphi(\zeta)$ in this basis. For example, if we take $w(y) \simeq \tilde D_\kappa(y)$ in the first quadrant and map it back to the local system (\ref{eq:S1}), then using (\ref{Gmap}) we find
\begin{equation}
    \begin{split}
        \cG^{-1} \tilde D_\kappa(y) &= \frac{1}{\rho-2\xi}(2\sqrt{\eta})^{-1-E/4}\exp\left(-\eta\zeta^2\right)\zeta^{-1/2-E/4} \; (1+\cO(\eta^{-1/2})) \\
        &= \frac{1}{\rho-2\xi}(2\sqrt{\eta})^{-1-E/4}\varphi_-(\zeta)
    \end{split}
\end{equation}
and similarly for $w(y) \simeq \tilde D_{-\kappa-1}(-iy)$ we have
\begin{equation}
    \begin{split}
        \cG^{-1} \tilde D_{-\kappa-1}(-iy) &= \frac{1}{2} (2\sqrt{\eta})^{-1+E/4}\exp\left(\eta\zeta^2\right)\zeta^{-1/2+E/4}e^{-i\pi E/8} \; (1+\cO(\eta^{-1/2})) \\
        &= \frac{1}{2}(2\sqrt{\eta})^{-1+E/4}e^{-i\pi E/8}\varphi_+(\zeta)
    \end{split}
\end{equation}
where in both cases the second line follows from (\ref{eq:phiexp}). Extending this procedure to the basis of solutions in each of the four quadrants yields:
\begin{equation}
     \begin{split}
     \rnc{1} &\begin{dcases}
          \cG^{-1} \tilde D_\kappa(y) &= \frac{1}{\rho-2\xi}(2\sqrt{\eta})^{-1-E/4}\varphi_-(\zeta)\\
           \cG^{-1} \tilde D_{-\kappa-1}(-iy)&= \frac{1}{2}(2\sqrt{\eta})^{-1+E/4}e^{-i\pi E/8}\varphi_+(\zeta)
    \end{dcases}\\
    \rnc{2} &\begin{dcases}
   \cG^{-1} \tilde D_\kappa(-y) &= -\frac{1}{\rho-2\xi}e^{i\pi E/4}(2\sqrt{\eta})^{-1-E/4}\varphi_-(\zeta)\\
           \cG^{-1} \tilde D_{-\kappa-1}(-iy)&= \frac{1}{2}(2\sqrt{\eta})^{-1+E/4}e^{-i\pi E/8}\varphi_+(\zeta)
    \end{dcases}\\
    \rnc{3} &\begin{dcases}
    \cG^{-1} \tilde D_\kappa(-y) &= -\frac{1}{\rho-2\xi}e^{i\pi E/4}(2\sqrt{\eta})^{-1-E/4}\varphi_-(\zeta)\\
           \cG^{-1} \tilde D_{-\kappa-1}(iy)&= \frac{1}{2}(2\sqrt{\eta})^{-1+E/4}e^{-3i\pi E/8}\varphi_+(\zeta)
    \end{dcases}\\
    \rnc{4} &\begin{dcases}
   \cG^{-1} \tilde D_\kappa(y) &= \frac{1}{\rho-2\xi}e^{i\pi E/4}(2\sqrt{\eta})^{-1-E/4}\varphi_-(\zeta)\\
           \cG^{-1} \tilde D_{-\kappa-1}(iy)&= \frac{1}{2}(2\sqrt{\eta})^{-1+E/4}e^{i\pi E/8}\varphi_+(\zeta).
           \end{dcases}
     \end{split}
     \label{eq:B3}
\end{equation}
Then, finally, we can use (\ref{eq:B2}) and (\ref{eq:B3}) to obtain the connection formula (\ref{eq:phiStokes}) for the local solutions $\varphi_\pm(\zeta)$. For example:
\begin{equation}
    \begin{split}
        \varphi^{\rnc{1}}_-(\zeta) &= (\rho-2\xi)(2\sqrt{\eta})^{1+E/4} \cG^{-1} \tilde D^{\rnc{1}}_{-E/4-1}(y)\\
        &= (\rho-2\xi)(2\sqrt{\eta})^{1+E/4} \bigg(-e^{-i\pi E/4}\cG^{-1}\tilde D^{\rnc{2}}_{-E/4-1}(-y) \\
        & \hspace{4.5cm} +\frac{\sqrt{2\pi}}{\Gamma(E/4+1)}e^{-i\pi E/8} \cG^{-1} \tilde D^\rnc{2}_{E/4}(-iy) \bigg) \\
        &= \varphi_-^{\rnc{2}}(\zeta)+\frac{\rho-2\xi}{2}\frac{\sqrt{2\pi}}{\Gamma(E/4+1)}e^{-i\pi E/4}(2\sqrt{\eta})^{E/2} \varphi^\rnc{2}_+(\zeta),
    \end{split}
\end{equation}
and analogously for the other quadrants.

\section{Matching local and global solutions}
\label{app:matching}
Let us recall that the global solutions $\psi_\pm$ of (\ref{eq:SL1}) and the local solutions $\varphi_\pm$ of the description (\ref{eq:S1}) near the double turning point, are related via equation (\ref{eq:psiphiconnection}). We can work out both sides of that last equation in order to construct an expression for $C_\pm(t, \eta)$. We start with the global solution on the left-hand side, for which the odd part of the solution to the associated Riccati equation is
\begin{equation}
    S_{\text{odd}}(x) = \eta\sqrt{Q_0(x)}+\frac{\cN_0^2-12\lambda_0\Lambda_0^2}{2\sqrt{Q_0(x)}}+\cO(\eta^{-1}).
    \label{eq:Sodd}
\end{equation}
Here, $Q_0(x) = 4(x-\lambda_0)^2(x+2\lambda_0)$ is the principal term of the potential $Q(x)$. In order to write out the WKB solution -- see (\ref{eq:WKBsol}) -- we need to integrate these leading terms, normalized at $-2\lambda_0$. Integrating the second term on the right-hand side of the above equation, where we use (\ref{eq:ers}), yields
\begin{equation}
\begin{split}
    \int_{-2\lambda_0}^x S_{\text{odd}, 0}(x')\diff x' &=\frac{E_0 }{4}\int_{-2\lambda_0}^x  \frac{(3\lambda_0)^{1/2}\diff x'}{(x'-\lambda_0)(x'+2\lambda_0)^{1/2}}\\
    &=\frac{E_0}{4}\log \frac{(x'+2\lambda_0)^{1/2}-(3\lambda_0)^{1/2}}{(x'+2\lambda_0)^{1/2}+(3\lambda_0)^{1/2}}\Bigg|_{-2\lambda_0}^x.
\end{split}
\label{eq:D0}
\end{equation}
The evaluation of this integral at $-2\lambda_0$ yields an ambiguous integration constant $\pm i\pi  E_0/4$, which was not written in \cite{Tak1}. The correct branch to pick for our purposes is $ -i\pi E_0/4$: in that case we find the connection formula as in section \ref{sec:connect}, which yields exactly the Stokes constants found by \cite{ASV1, BSSV}. Picking a different branch would lead to connection coefficients that relate to a different Stokes automorphism, as we explain in section \ref{sec:GST}.

\sk
When we plug (\ref{eq:Sodd}) and (\ref{eq:D0}) back into the WKB ansatz, we find that the global solutions take the form
\begin{equation}
\begin{split}
    \psi_\pm(x) = \left((x-\lambda_0)\sqrt{x+2\lambda_0}\right)^{-1/2} \exp\left(\pm \eta \int_{-2\lambda_0} ^x S_{-1} \diff x' \right) (x-\lambda_0)^{\pm E_0/4} \\
    \times\left(\sqrt{x+2\lambda_0}+\sqrt{3\lambda_0}\right)^{\mp E_0/2}e^{\mp i \pi E_0/4} \; (1+\mathcal{O}(\eta^{-1/2})).
\end{split}
\label{eq:D1}
\end{equation}
Next, we can compute the right-hand side of (\ref{eq:psiphiconnection}), where the local solutions appear. First of all, note that from the leading part of (\ref{eq:Qdef}) we find that
\begin{equation}
    \zeta_0(x) = \left(2\int_{\lambda_0}^x(x'-\lambda_0)(x'+2\lambda_0)^{1/2} \diff x' \right)^{1/2},
\end{equation}
allowing us to compute $\left(\frac{\partial\zeta_0}{\partial x}\right)^{-1/2}$. Keeping in mind the WKB solution (\ref{eq:localWKBsol}), we then find that
\begin{equation}
    \begin{split}
        \left(\frac{\partial \zeta}{\partial x}\right)^{-1/2}\varphi_\pm(\zeta) = \bigg((x-\lambda_0)\sqrt{x+2\lambda_0}\bigg)^{-1/2} \exp\bigg(\pm \eta \int_{\lambda_0} ^x S_{-1} \diff x' \bigg) \\ 
        \times\exp\Big(\pm 2\zeta_0\zeta_1\Big)\zeta_0^{\pm E_0/4} \; (1+\mathcal{O}(\eta^{-1/2})).
    \end{split}
    \label{eq:D2}
\end{equation}
Finally, we can compare equations (\ref{eq:D1}) and (\ref{eq:D2}). The ratio between these expressions,
\begin{equation}
    \begin{split}
    C_\pm = \exp\bigg(\pm \eta \int_{-2\lambda_0} ^{\lambda_0} S_{-1} \diff x \bigg)\bigg(\sqrt{x+2\lambda_0}+\sqrt{3\lambda_0}\bigg)^{\mp E_0/2}e^{\mp i \pi E_0/4}\\
    \times \exp\Big(\mp 2\zeta_0\zeta_1\Big) \left(\frac{x-\lambda_0}{\zeta_0}\right)^{\mp E_0/4}(1+\cO(\eta^{-1/2})),
    \end{split}
\end{equation}
must be independent of the coordinate $x$. Hence, we can evaluate the expression at $x = \lambda_0$ which -- using the fact that $\frac{\partial \zeta_0}{\partial x}|_{x=\lambda_0}=-(3\lambda_0)^{1/4}$ -- leads to
\begin{equation}
    C_\pm(t, \eta)= \exp\left(\pm \eta \int_{-2\lambda_0} ^{\lambda_0} S_{-1} \diff x' \right) \left(4(3\lambda_0)^{5/4}\right)^{\mp E_0/4}e^{\mp i \pi E_0/4} \; (1+\mathcal{O}(\eta^{-1/2})).
\end{equation}
This is the expression given in (\ref{eq:Cpm}).


\begin{thebibliography}{99}

\bibitem{PAINLEVE} E.~Picard, {\em Mémoire sur la théorie des fonctions algébriques de deux variables}, J. Math. Pures Appl., \textbf{5} (1889) 135–319.

P.~Painlevé, {\em Mémoire sur les équations différentielles dont l'intégrale générale est uniforme}, Bull. Soc. Math. Fr., \textbf{28} (1900)  201–261.

P.~Painlevé, {\em Sur les équations différentielles du second ordre et d'ordre supérieur dont l'intégrale générale est uniforme}, Acta Math., \textbf{25} (1902) 1–85.

R.~Fuchs, {\em Sur quelques équations différentielles linéaires du second ordre}, Comptes Rendus \textbf{141} (1905) 555–558.

B.~Gambier, {\em Sur les équations différentielles du second ordre et du premier degré dont l'intégrale générale est à points critiques fixes}, Acta Math., \textbf{33} (1910) 1–55.

\bibitem{BOUTROUX} P.~Boutroux, {\em Recherches sur les transcendantes de M. Painlevé et l'étude asymptotique des équations différentielles du second ordre}, Ann. Sci. Ecole Norm. Sup. \textbf{30} (1913) 255-375.

P.~Boutroux, {\em Recherches sur les transcendantes de M. Painlevé et l'étude asymptotique des équations différentielles du second ordre (suite)}, Ann. Sci. Ecole Norm. Sup. \textbf{31} (1914) 99-159. 

\bibitem{Yos} S.~Yoshida, {\em 2-Parameter family of solutions for Painlevé equations(I)--(V) at an irregular singular point}, Funkcialaj Ekvacioj \textbf{28} (1985) 233.

\bibitem{AKT3} T.~Aoki, T.~Kawai and Y.~Takei, {\em WKB analyis of Painlevé Transcendents with a large Parameter II: Multiple-Scale Analysis of Painlevé Transcendents}, RIMS-1038.

\bibitem{Marino:2008ya} M.~Marino, {\em Nonperturbative effects and nonperturbative definitions in matrix models and topological strings}, JHEP \textbf{12}, 114 (2008) [\href{https://arxiv.org/abs/0805.3033}{0805.3033}].

\bibitem{Marino:2008vx} M.~Marino, R.~Schiappa and M.~Weiss, {\em Multi-Instantons and Multi-Cuts}, J. Math. Phys. \textbf{50}, 052301 (2009) [\href{https://arxiv.org/abs0809.2619}{0809.2619}].

\bibitem{GIKM} S.~Garoufalidis, A.~Its, A.~Kapaev and M.~Mariño,
{\em Asymptotics of the instantons of Painlevé~I}, 
Int. Math. Res. Not. (2012) no. 3, 561-606
[\href{https://arxiv.org/abs/1002.3634}{1002.3634}].

\bibitem{ASV1}
I.~Aniceto, R.~Schiappa and M.~Vonk,
{\em The Resurgence of Instantons in String Theory,}
Commun. Num. Theor. Phys. \textbf{6} (2012), 339-496.
[\href{https://arxiv.org/abs/1106.5922}{1106.5922}].

\bibitem{Shim} S.~Shimomura, {\em Series expansions of painlevé transcendents near the point at infinity}, Funkcialaj Ekvacioj \textbf{58} (2015) 277.

\bibitem{Iwa1} K.~Iwaki, {\em 2-Parameter $\tau$-Function for the First Painlevé Equation: Topological Recursion and Direct Monodromy Problem via Exact WKB Analysis.} Communications in Mathematical Physics \textbf{377} (2020): 1047-1098. [\href{https://arxiv.org/abs/1902.06439}{1902.06439}]

\bibitem{Edgar} G.~A.~Edgar, {\em Transseries for beginners}, Real Analysis Exchange \textbf{35} (2010) 253-310. [\href{https://arxiv.org/abs/0801.4877}{0801.4877}]

\bibitem{Bonelli:2016qwg}
G.~Bonelli, O.~Lisovyy, K.~Maruyoshi, A.~Sciarappa and A.~Tanzini,
{\em On Painlev\'e/gauge theory correspondence}, Lett. Math. Phys. \textbf{107}, pages 2359-2413 (2017) [\href{https://arxiv.org/abs/1612.06235}{1612.06235}].

\bibitem{Lis1} O.~Lisovyy and J.~Roussillon, {\em On the connection problem for Painlevé I}, J. Phys. A: Math. Theor. \textbf{50} (2017) 255202 [\href{https://arxiv.org/abs/1612.08382}{1612.08382}].

\bibitem{Compere:2021zfj}
G.~Comp\`ere and L.~K\"uchler,
{\em Asymptotically matched quasi-circular inspiral and transition-to-plunge in the small mass ratio expansion}, [\href{https://arxiv.org/abs/2112.02114}{2112.02114}].

\bibitem{STOKES}G.~G.~Stokes, {\em On the numerical calculation of a class of definite integrals and infinite series}, Transactions of the Cambridge Philosophical Society, \textbf{IX} (1847) (I) 166–189.

G.~G.~Stokes, {\em On the discontinuity of arbitrary constants which appear in divergent developments}, Transactions of the Cambridge Philosophical Society, \textbf{X} (1858) (I) 105–128.

\bibitem{Aniceto:2013fka}
I.~Aniceto and R.~Schiappa,
{\em Nonperturbative Ambiguities and the Reality of Resurgent Transseries,}
Commun. Math. Phys. \textbf{335}, no.1, 183-245 (2015)
[\href{https://arxiv.org/abs/1308.1115}{1308.1115}]

\bibitem{BSSV} S.~Baldino, M.~Schwick, R.~Schiappa and R.~Vega, \textit{Resurgent Stokes data for Painlevé equations and 2d quantum (super-) gravity} [\href{https://arxiv.org/abs/2203.13726}{2203.13726}].


\bibitem{ECALLE} J.~Écalle, {\em Les Fonctions Résurgentes}, Prepub. Math. Université de Paris-Sud \textbf{81-05} (1981), \textbf{81-06} (1981), \textbf{85-05} (1985).


\bibitem{Sauzin} D.~Sauzin, {\em Introduction to 1-summability and resurgence}, in ``Divergent Series, Summability and Resurgence I: Monodromy and Resurgence", Lec. Notes Math. \textbf{2153} (2016) [\href{https://arxiv.org/abs/1405.0356}{1405.0356}].

\bibitem{Dor1} D.~Dorigoni, {\em An Introduction to Resurgence, Trans-series and Alien Calculus},
Annals Phys. \textbf{409} (2019), 167914
[\href{https://arxiv.org/abs/1411.3585}{1411.3585}]. 

\bibitem{ABS1} I. Aniceto, G. Basar and R. Schiappa,
\textit{A Primer on Resurgent Transseries and Their Asymptotics}, 
Phys. Rept. \textbf{809} (2019), 1-135
[\href{https://arxiv.org/abs/1802.10441}{1802.10441}].

\bibitem{Tak1}
Y.~Takei, \emph{On the connection formula for the first Painleve equation: from the viewpoint of the exact WKB analysis}, in ``Painlevé Transcendents and Asymptotic Analysis \textbf{931} (1995) 70.

\bibitem{Tak4} Y.~Takei, \emph{An explicit description of the connection formula for the first Painlevé equation} in "Towards the Exact WKB Analysis of Differential Equations, Linear or Nonlinear" (2000) 271.

\bibitem{AKT2}
T.~Aoki, T.~Kawai and Y.~Takei,
{\em Algebraic analysis of singular perturbations: On exact WKB analysis}. {\em American Mathematical Society (2005)} (Originally published in Japanese by Iwanami Shoten, Publishers, Tokyo, 1998).

\bibitem{KT} T.~Kawai and Y.~Takei, {\em WKB analyis of Painlevé Transcendents with a large Parameter I}. Advances in Mathematics \textbf{118}  (1996) 1-33.

\bibitem{KT3} T.~Kawai and Y.~Takei, {\em Algebraic Analysis of Singular Perturbation Theory}, Trans. of Math. Monographs, \textbf{227} (2005).

\bibitem{Kap1} A.~A.~Kapaev, {\em Asymptotic behavior of the solutions of the Painlevé equation of the first kind}, Differensial'nye Uravneniya, \textbf{24}, no 10 (1988) 1684-1695.

\bibitem{Kap2}  A. A. Kapaev, \textit{Quasi-linear Stokes phenomenon for the Painlevé first equation}, Journal of Physics \textbf{A37} (2004) 11149 [\href{https://arxiv.org/abs/nlin/0404026}{0404026}].

\bibitem{Dav1} F.~David, {\em Nonperturbative effects in Matrix Models and Vacua of Two-Dimensional Gravity}, Phys. Lett. \textbf{B302} (1993) 403.

\bibitem{Cos4} O.~Costin, R.D.~Costin, M.~Huang, {\em A direct method to find Stokes multipliers in closed form for P1 and more general integrable systems},  Trans. Am. Math. Soc. \textbf{368} (2016) 7579-7621. [\href{https://arxiv.org/abs/1205.0775}{1205.0775}]

\bibitem{Cos5} O.~Costin, R.D.~Costin, M.~Huang, {\em Tronquée solutions of the Painlevé equation P1}, Constr. Approx. \textbf{41} (2015), 467–494. [\href{https://arxiv.org/abs/1310.5330}{1310.5330}]

\bibitem{Del} E.~Delabaere, {\em Resurgent methods and the first Painlevé equation}, CIMPA Lecture Notes (2015), hal-01067086.

\bibitem{BerryHowls90}
M. V. Berry and C. J. Howls, \textit{Hyperasymptotics.} Proc. R. Soc. Lond. A430 (1990), 653–668.

\bibitem{JK1} N.~Joshi, M.~Kruskal, {\em The Painlevé Connection Problem: An Asymptotic Approach. I}, Studies in Applied Mathematics (1992).

\bibitem{KK} A.A. Kapaev and A.V. Kitaev, \textit{Connection Formulae for the First Painlevé Transcendent in the Complex Plane}, Letters in Mathematical Physics \textbf{27} (1993) 04.

\bibitem{Vor1} A.~Voros, {\em The return of the quartic oscillator: The complex WKB method}, Ann. Henri Poincaré, \textbf{A39} (1983) 211-338.

\bibitem{JMU1} M.~Jimbo, T.~Miwa and K.~Ueno, \textit{Monodromy preserving deformation of linear ordinary differential equations with rational coefficients: I. General theory and $\tau$-function}, Physica D. Nonlinear Phenomena \textbf{2} (1981) 306-352.

\bibitem{JM} M.~Jimbo and T.~Miwa, {\em Monodromy perserving deformation of linear ordinary differential equations with rational coefficients. II}, Physica D. Nonlinear Phenomena \textbf{2} (1981), 407-448. 

\bibitem{Oka1} K.~Okamoto, {\em On the $\tau$-function of the Painlevé equations}, Phys. D \textbf{2} vol 3 (1981), 525–535.

\bibitem{Iwa2}
K.~Iwaki and T.~Nakanishi, {\em Exact WKB analysis and cluster algebras}. Journal of Physics A: Mathematical and Theoretical, 47(47) (2014), [\href{https://arxiv.org/abs/1401.7094}{1401.7094}].

\bibitem{Gregori:2021tvs}
P.~Gregori and R.~Schiappa,
{\em From Minimal Strings towards Jackiw-Teitelboim Gravity: On their Resurgence, Resonance, and Black Holes}
[\href{https://arxiv.org/abs/2108.11409}{2108.11409}].

\bibitem{YN} V. Yu. Novokshenov, {\em Special Solutions of the First and Second Painlevé Equations and Singularities of the Monodromy Data Manifold} Proc. Steklov Inst. Math. \textbf{281} (2013) 105-117.

\bibitem{SvdP} M.~van der Put and M.~Saito, {\em Moduli spaces for linear differential equations and the Painlevé equations}. [\href{https://arxiv.org/abs/0902.1702}{0902.1702}].

\bibitem{Tak2} Y.~Takei {\em On the instanton-type expansions for Painlevé transcendentss and elliptic functions} in Complex Differential and Difference equations (2020), de Gruyter Proceedings in Mathematics, 365-377.

\bibitem{Tak3} Y.~Takei, {\em Riccati Equations Revisited: Linearization and Analytic Interpretation of Instanton-Type Solutions}, Complex Analysis and Operator Theory (2020).

\bibitem{FW} B.~Fornberg and J.A.C.~Weideman,
\textit{A numerical methodology for the Painlevé equations},
J. Comp. Phys. \textbf{230} 15 (2011).

\bibitem{Cos6} O.~Costin and G.~V.~Dunne, {\em Resurgent extrapolation: rebuilding a function from asymptotic data. Painlev\'e I}, J. Phys. A \textbf{52}, no.44, 445205 (2019) [\href{https://arxiv.org/abs/1904.11593}{1904.11593}].

\bibitem{Cos7} O.~Costin and G.~V.~Dunne, {\em Uniformization and Constructive Analytic Continuation of Taylor Series} [\href{https://arxiv.org/abs/2009.01962}{2009.01962}].

\bibitem{Cos1} O.~Costin, {\em Correlation between pole location and asymptotic behavior for Painlevé I solutions}, Comm. Pure Appl. Math. \textbf{52} (1999) 461 [\href{https://arxiv.org/abs/math/9709223}{math/9709223}].

\bibitem{Cos2} O.~Costin and R.D.~Costin, {\em Singular normal form for the Painlevé equation P1}, Nonlinearity \textbf{11} (1998) 1195 [\href{https://arxiv.org/abs/math/9710209}{math/9710209}].

\bibitem{Cos3} O.~Costin and R.D.~Costin, {\em On the formation of singularities of solutions of nonlinear differential systems in antistokes directions}, Invent. math. \textbf{145} (2001) 425-485, [\href{https://arxiv.org/abs/math/0202234}{math/0202234}] 

\bibitem{ASV2}
I.~Aniceto, R.~Schiappa and M.~Vonk, to appear. See also the talks by I.~Aniceto and M.~Vonk at the KITP conference ``Resurgence in gauge and string theory'' (2017): [\href{online.kitp.ucsb.edu/online/resurgent\_c17/aniceto}{https://online.kitp.ucsb.edu/online/resurgent\_c17/aniceto}] and [\href{online.kitp.ucsb.edu/online/resurgent\_c17/vonk}{https://online.kitp.ucsb.edu/online/resurgent\_c17/vonk}].

\bibitem{talk-Alexander} A.~van Spaendonck, {\em Painlevé 1 and exact WKB}, talk at the Summary Meeting of the workshop ``Applicable resurgent asymptotics: towards a universal theory'', Isaac Newton Institute for Mathematical Sciences, Cambridge, 17 June 2021. [\href{https://www.newton.ac.uk/seminar/33147/}{https://www.newton.ac.uk/seminar/33147/}], slides at \href{https://www.slideshare.net/secret/uFGOMQMWZNmT3j}{https://www.slideshare.net/secret/uFGOMQMWZNmT3j}.

\bibitem{Dun1} J.~L.~Dunham, {\em The Wentzel-Brillouin-Kramers Method of Solving the Wave Equation}, Phys. Rev. \textbf{41} 713 (1932).

\bibitem{FKIN1} A.~Fokas, A.~Its, A.~Kapaev and V. Novokshenov, \textit{Painlevé transcendents}, Mathematical surveys and monographs, vol 128, American Mathematical Society (2006).

\bibitem{KT4} T.~Kawai and Y.~Takei, {\em WKB Analysis of Painlevé Transcendents with a Large Parameter
III. Local Reduction of 2-Parameter Painlevé Transcendents}, Adv. Math. \textbf{134} (1998), 178-218.

\end{thebibliography}
\end{document}